\documentclass[a4paper,11pt]{article}
\usepackage{jheppub} % for details on the use of the package, please see the JINST-author-manual
\usepackage{lineno}
\usepackage{xcolor}
\usepackage{soul}
\usepackage{physics}
\usepackage{booktabs}
\usepackage{array}
\usepackage{multirow}
\usepackage{amsmath}
\usepackage{tikz}
\usepackage{graphicx}
\usepackage{subcaption}
\usetikzlibrary{arrows.meta,decorations.pathmorphing}

\newcolumntype{L}[1]{>{\raggedright\arraybackslash}p{#1}}
\providecommand{\ii}{\mathrm i}

\providecommand{\sech}{\operatorname{sech}}
\providecommand{\ee}{\mathrm e}
\providecommand{\G}{\Gamma}

\providecommand{\Ihat}{\widehat{\mathcal I}}
\arxivnumber{} % if you have one

\title{ 
\begin{center}
{\Large\bfseries
Bootstrapping a Higher-Derivative Cosmological Collider: \\[0.3em]
Borel--Laplace Resummation and Stokes Phenomena
\par}
\end{center}
}

\author[a]{Matheus~C.~Ferreira,}
\author[a,b]{Felipe~T.~Falciano,}

\affiliation[a]{Brazilian Center for Research in Physics (CBPF), \\ Dr. Xavier Sigaud st. 150, zip 22290-180, Rio de Janeiro, RJ, Brazil}
\affiliation[b]{PPGCosmo, CCE, Federal University of Espírito Santo (UFES), \\
Av. Fernando Ferrari, 540, CEP 29.075-910, Vitória, ES, Brazil}

\emailAdd{matheuscurado@cbpf.br}
\emailAdd{ftovar@cbpf.br}

\abstract{We develop a higher-derivative cosmological bootstrap for massive spectators with quartic dispersion, $\omega^2\propto k^4$, revealing a new boundary-value problem for inflationary correlators. The exchange equation becomes fourth order, and the collapsed limit becomes an irregular singular point, where collider and WKB branches coexist and exhibit Stokes phenomena. Since the folded configuration is now an ordinary point, the relativistic criterion of Bunch--Davies regularity at the folded singularity is no longer available. We instead impose an asymptotic boundary condition that removes the WKB branches and matches the surviving collider branches to Schwinger--Keldysh data. We also construct a convergent Borel--Laplace particular solution that resums a factorially divergent expansion and satisfies both bootstrap equations simultaneously; its leading collapsed-limit behavior is reproduced by direct in-in integrals. In the bispectrum, the squeezed-limit collider signal retains its mass-dependent frequency, while its amplitude and phase depend on the dispersion scale; near the equilateral configuration, the full shape is also sensitive to this scale and can change sign. Our results expose a new analytic structure of non-relativistic cosmological correlators and signatures beyond the squeezed limit.}

\begin{document}
\maketitle
\flushbottom

%%%%%%%%%%%%%%%%%%%%%%%%%%%%%%
%%%%%%%%%%%%%%%%%%%%%%%%%%%%%%
%%%%%%%%%%%%%%%%%%%%%%%%%%%%%%
\section{Introduction}\label{sec:intro}

Much of what we know about the analytic structure of inflationary correlators is anchored to relativistic physics. Correlators develop a singularity as the total energy
\begin{equation}
    k_t \equiv k_1+k_2+k_3+\cdots
\end{equation}
approaches zero, and the residue at this singularity reproduces a Lorentz-invariant flat-space scattering amplitude \cite{MaldacenaPimentel_GravitonNonGaussianities_2011,Raju_FlatSpaceLimitAdSCFT_2012,MarolfMorrisonSrednicki_dSSmatrix_2012,GhoshKunduRajuTrivedi_FourPointSlowRoll_2014,CespedesDavisMelville_TimeEvolutionCosmologicalCorrelators_2020,ArmstrongGomezJusinskasLipsteinMei_AdSRecursions_2022,BonifacioGoodhewJoycePajerStefanyszyn_GravitonFourPointdS_2022,Mei_AmplitudeBootstrapAdSdS_2023,MelvillePimentel_dSSmatrixMasses_2023,ChowdhuryLipsteinMeiSachsVanhove_SubtleSimplicity_2023,MelvillePimentel_AmputatedCosmologicalCorrelators_2024,MarottaSkenderisVerma_FlatSpaceSpinningMassiveCFT_2024,CespedesJazayeri_MassiveFlatSpaceLimit_2025}. Inflation, however, does not preserve all of these symmetries. Although the underlying theory is relativistic, the time-dependent inflationary background spontaneously breaks time translations and de Sitter boosts. Boosts are then realized nonlinearly, and the fluctuations need not propagate relativistically \cite{Extra:07090293,Extra:151204100,Extra:240305274}. This raises a basic question: which features of the analytic structure of cosmological correlators are consequences of consistency alone, and which are inherited from unbroken boost invariance?

The question matters because the flat-space limit is not merely a convenient approximation; it is a structural pillar of the cosmological bootstrap. Inspired by the S-matrix bootstrap, this program reconstructs correlators from symmetry, analyticity, and factorization, without relying on a Lagrangian description or on the detailed time evolution in the bulk \cite{ArkaniHamedMaldacena_CosmologicalCollider_2015,ArkaniHamedBenincasaPostnikov_CosmologicalPolytopes_2017,ArkaniHamedBenincasa_ScatteringFacet_2018,Benincasa_FlatSpaceSmatrixWavefunction_2018,ArkaniHamedBaumannLeePimentel_CosmologicalBootstrap_2018}. In practice, the location of the total-energy singularity and the factorization of its residue fix boundary conditions and organize the answer in terms of singular loci and dimensionless kinematic variables. This logic runs through essentially every implementation of the program: boundary differential equations, weight-shifting operators, Mellin-space methods, and spinning correlators \cite{Sleight_MellinSpaceCosmologicalCorrelators_2019,SleightTaronna_BootstrappingInflationaryCorrelatorsMellin_2019,BaumannEtAl_WeightShiftingScalarSeeds_2019,BaumannEtAl_SpinningCorrelators_2020}; cosmological optical theorems, cutting rules, and the classification of singularities \cite{GoodhewJazayeriPajer_CosmologicalOpticalTheorem_2020,BenincasaMcLeodVergu_SteinmannWavefunction_2020,Pajer_BoostlessBootstrapBispectrum_2020,JazayeriPajerStefanyszyn_LocalityUnitarityCosmologicalCorrelators_2021,MelvillePajer_CosmologicalCuttingRules_2021,GoodhewJazayeriLeePajer_CuttingCosmologicalCorrelators_2021,BaumannEtAl_LinkingSingularities_2021}; analytic and nonperturbative approaches \cite{BonifacioPajerWang_AmplitudesToContactCosmologicalCorrelators_2021,Meltzer_InflationaryWavefunctionAnalyticityFactorization_2021,HogervorstPenedonesVaziri_NonPerturbativeCosmologicalBootstrap_2021,DiPietroGorbenkoKomatsu_AnalyticityUnitarityCosmologicalCorrelators_2021,BaumannEtAl_SnowmassCosmologicalBootstrap_2022}; recursive, spectral, and differential-equation methods for multiple exchanges \cite{Extra:240717715,Extra:231205303,Extra:260408658,Extra:250305866,Extra:250514609,Extra:240403716}; cosmological scattering equations, dressing prescriptions, and the relation to Euclidean AdS \cite{Extra:210611903,Extra:250310598,Extra:250909536,Extra:220407217}; and geometric descriptions in terms of cosmological polytopes, triangulations, cohomology, and cluster algebras \cite{Extra:230803753,Extra:230305876,Extra:250313393,Extra:250217564,Extra:250619907,Extra:251214854,Extra:260308670}. In each case, the singularity structure plays the role that Lorentz invariance, locality, and unitarity play for scattering amplitudes, where on-shell methods such as spinor-helicity variables, recursion relations, color--kinematics relations, scattering equations, soft theorems, and positivity bounds exploit the same constraints \cite{Extra:13081697,Extra:170803872,Extra:170509678,Extra:230605395,Extra:HepTh0412308,Extra:HepTh0501052,Extra:170904891,Extra:08053993,Extra:150600974,Extra:13072199,Extra:14123479,Extra:14044091,Extra:13122229,Extra:HepTh0602178,Extra:160706109,Extra:160706110}. In relativistic de Sitter exchange problems, the outcome is particularly clean: the soft, folded, and factorization limits appear as regular singular points of second-order differential equations, whose solutions are familiar special functions.

When boosts are broken, this picture is modified, though not discarded. Amplitude constructions without boosts and their soft limits show how the relation between singularities and Lorentz-invariant amplitudes is relaxed \cite{Extra:200700027,Extra:200914289,Extra:240305459,Extra:200502366,Extra:221016276,Extra:220814544}, and the boostless cosmological bootstrap has extended the program to nonstandard propagation, reduced sound speeds, and resonant settings \cite{PimentelWang_BoostlessCosmologicalColliderBootstrap_2022,JazayeriRenauxPetel_CosmologicalBootstrapSlowMotion_2022,SalcedoLeeMelvillePajer_AnalyticWavefunction_2022,DuasoPueyoPajer_ResonantCosmologicalBootstrap_2023,Extra:250601555}. In these settings, singularities and factorization survive, but the interpretation of the residues and of the natural kinematic variables changes. Most of this work, however, modifies the speed of propagation while keeping the dispersion relation linear at short distances. A qualitatively different dispersion relation can do more: it can change the order of the bootstrap equations and the very nature of their singular points.

The ghost condensate provides the sharpest realization of this possibility \cite{Nima_Arkani_Hamed_2004,Hamed_2004}. In ghost inflation, the inflaton has a time-dependent background that spontaneously breaks time translations, and its fluctuations obey the non-relativistic dispersion relation
\begin{equation}\label{quarticdisp}
    \omega^2 \sim k^4 \,,
\end{equation}
which reflects the spontaneous breaking of boosts by the background. Anisotropic scaling and Lifshitz-type ultraviolet completions provide broader contexts for such dispersion relations \cite{Extra:09013775,Extra:09023657,Extra:180301352}. Correlators in these theories still have singularities and still factorize, but their flat-space limit no longer reproduces a Lorentz-invariant amplitude. This is not a breakdown of consistency but a relaxation of assumptions specific to relativistic theories: the singularities remain meaningful, while their residues encode non-Lorentz-invariant physics.

The consequences are most pronounced for massive exchange, because this is where the analytic structure carries direct physical information. In the cosmological collider program, heavy particles produced during inflation leave nonanalytic imprints in the soft and collapsed limits of correlators, whose frequency and phase encode their masses and spins \cite{ArkaniHamedMaldacena_CosmologicalCollider_2015,Extra:160703735,Extra:11090292,Extra:161006597,Extra:160407841,Extra:161208122}. This picture grew out of quasi-single-field inflation \cite{Extra:09090496,Noumi_2013,Extra:12050160,Extra:171003054} and has been applied to Higgs and neutrino signatures, gauge and spin-two exchanges, standard clocks, and parity-violating observables \cite{Extra:180502656,Extra:190707390,Extra:190901819,Extra:200402887,Extra:220306349,Extra:171103988,Extra:180304412}, as well as to leptogenesis, preheating, and flavor oscillations \cite{Extra:211210793,Extra:210811385,Extra:200512278,Extra:211205710}. The amplitude and phase of these signals are known to be sensitive to the environment of the heavy field, for instance to heavy lifting, chemical potentials, or additional light sectors \cite{Extra:191012876,Extra:201004727,Extra:190811378,Extra:220411869,Extra:240907524,Extra:250722978}. With searches in galaxy surveys and the cosmic microwave background now underway \cite{Extra:161006559,Extra:180107265,Extra:240401894,Extra:240407203,Extra:250402931,Extra:260315728}, it is important to understand how the analytic structure, and hence the signal, changes when the heavy field itself propagates non-relativistically.

In this work, we bootstrap the exchange of a massive spectator with a ghost-condensate dispersion relation. By ``ghost spectator'' we do not mean a negative-norm state, but a stable massive field whose short-distance propagation is governed by the quartic dispersion relation~\eqref{quarticdisp}. In previous work, we showed that such a spectator can weaken the Boltzmann suppression of heavy-particle signals and derived the corresponding bootstrap equations~\cite{Ferreira:2026tyj}. Here we solve these equations, determine the boundary conditions that select the physical solution, and assess their consequences for the full bispectrum shape.

The change in the analytic structure is qualitative rather than perturbative. The quartic term replaces, rather than corrects, the relativistic gradient term in the mode equation, so the de Sitter result is not recovered as a smooth limit of the quartic one. Although the relativistic and quartic spectators share the same late-time weights, $\Delta_\pm=3/2\pm i\mu$, the quartic dispersion promotes the bootstrap equation from second to fourth order, so that the homogeneous problem has four independent solutions instead of two. The singularity structure changes accordingly. In terms of the cross-ratio $u=s/(k_1+k_2)$, the de Sitter equation has regular singular points at the soft limit $u\to0$, at the folded point $u=1$, and at the factorization point $u=-1$, where solutions behave as powers of $u$. For the quartic spectator, $u=0$ becomes an irregular singular point, whereas $u=\pm1$ and $u=\infty$ become ordinary points of the homogeneous equation. Near $u=0$, the solution space contains, besides the two familiar collider branches $u^{\ell+5/2\pm i\mu}$, two WKB branches $\exp[\pm i/(2\gamma u^2)]$, whose relative dominance changes from one sector of the complex $u$-plane to the next. This sectorial structure is the origin of the Stokes phenomenon in the quartic bootstrap.

The disappearance of the folded singularity has an important consequence for the choice of vacuum. In the de Sitter bootstrap, the Bunch--Davies condition is imposed by requiring the absence of the folded singularity at $u=1$. For the quartic spectator, $u=1$ is an ordinary point, and this criterion is no longer available. We show that the vacuum information must instead be supplied by an asymptotic boundary condition at the irregular point: along the physical direction $u\to0^+$, both WKB branches must be absent, and the two surviving collider branches must reproduce the coefficients of the Bunch--Davies-normalized Schwinger--Keldysh integrals. We construct the four independent homogeneous solutions as generalized hypergeometric functions around the ordinary point $u=\infty$ and impose this condition to obtain a global, WKB-free basis of two complex-conjugate solutions, each of which asymptotes to a single collider branch. This prescription selects the physical asymptotic sector without requiring an explicit determination of the Stokes multipliers.

The time-ordered components obey two sourced bootstrap equations, one for each vertex. Near the irregular point, their particular solution is a factorially divergent series, which we resum by a Borel--Laplace transform. Under this transform, the fourth-order operator is intertwined with a second-order Bessel operator, and the resulting Green kernel yields a convergent integral representation that satisfies both sourced equations simultaneously. Direct Schwinger--Keldysh integrals reproduce its leading collapsed-limit behavior. Borel resummation has recently been used to recover cosmological-collider signals from effective expansions of massive inflationary correlators~\cite{LiXianyuResurgence2026}; here it serves a different purpose, namely to select a particular solution of a higher-order bootstrap system. Whether this representative coincides with the full Schwinger--Keldysh seed at finite momenta, rather than differing from it by a homogeneous term, remains an open connection problem.

Finally, we apply these results to the bispectrum. In the squeezed limit, the frequency of the collider oscillations remains fixed by the mass of the spectator, while their amplitude and phase depend on $\gamma=H/M$, and the isolated nonanalytic signal can be substantially enhanced relative to the relativistic case. Away from the squeezed limit, the analytic contributions can partially mask this enhancement, but they carry their own imprint of the quartic dispersion: the equilateral amplitude is sensitive to $\gamma$ and can even change sign. Higher-derivative propagation therefore affects the full momentum dependence of the bispectrum, not only its squeezed limit.

The paper is organized as follows. Section~\ref{sec:setup} introduces the relativistic and quartic spectators and their Schwinger--Keldysh propagators. Section~\ref{sec:dS_bootstrap} reviews the de Sitter scalar-exchange bootstrap, which serves as the relativistic benchmark. Section~\ref{sec:ghost_bootstrap} develops the quartic bootstrap: we classify its singular points, describe the Stokes structure at the irregular point, construct the WKB-free homogeneous basis, and derive the Borel--Laplace particular solution. Section~\ref{sec:pheno} studies the resulting bispectrum in the squeezed and equilateral limits, and Section~\ref{sec:conclusion} summarizes our results and outlines future directions. The appendices collect the asymptotics of generalized hypergeometric functions, the direct collapsed-limit Schwinger--Keldysh integrals, the details of the de Sitter bootstrap, and the mathematical justification of the Borel--Laplace construction.

%%%%%%%%%%%%%%%%%%%%%%%%%%%%%%
%%%%%%%%%%%%%%%%%%%%%%%%%%%%%%
%%%%%%%%%%%%%%%%%%%%%%%%%%%%%%
\section{Theoretical Setup}\label{sec:setup}

This section introduces only the field-theory ingredients that will be used later in the bootstrap construction.  We begin with an ordinary scalar in de Sitter space and then repeat the same logic for a massive spectator with ghost-condensate dispersion.  The purpose is to make clear which change in the microscopic mode equation is responsible for the different analytic structures of the two bootstrap problems.

\subsection{Standard Relativistic Scalar Fields}

During the approximately de Sitter stage of inflation, we use the line element
\begin{equation}
    ds^2 = a(\eta)^2 (-d\eta^2 + d\vec{x}^2), \quad a(\eta) = -\frac{1}{H\eta}, \label{eq:metric}
\end{equation}
where \(H\) is the Hubble scale and \(\eta<0\) is conformal time.  A generic free scalar field \(\sigma\) of mass \(m\) propagating on this fixed background is described by
\begin{equation}
    S_2 = \int d\eta\, d^3x
    \left[
    \frac{1}{2H^2\eta^2}\Big((\partial_\eta \sigma)^2
    -(\nabla \sigma)^2\Big)
    -\frac{m^2}{2H^4\eta^4}\, \sigma^2
    \right].
    \label{eq:free-scalar-action}
\end{equation}
In Fourier space, the field operator is decomposed into mode functions and creation and annihilation operators as $\sigma(\vec{k}, \eta) = \sigma_k(\eta)\, a_{\vec{k}} + \sigma_k^*(\eta)\, a^\dagger_{-\vec{k}} $. The mode function \(\sigma_k(\eta)\) then satisfies
\begin{align}
\left( \Box_{dS} + \frac{m^2}{H^2} \right) \sigma_k(\eta) = 0
    \quad ,\qquad 
    \Box_{dS} \equiv \eta^2 \partial_\eta^2 - 2\eta \partial_\eta + k^2 \eta^2\ .
    \label{eq:mode-eom}
\end{align}
Imposing the Bunch–Davies positive-frequency condition at early times gives, namely, $\sigma_k(\eta \to -\infty) = i \frac{H\eta}{\sqrt{2k}} e^{-ik\eta}$, we obtain the solution
\begin{equation}
\sigma_k(\eta) = -i \frac{H\sqrt{\pi}}{2k^{3/2}}  e^{i\pi/4} e^{-\pi \mu/2} (-k\eta)^{3/2} H^{(1)}_{i\mu}(-k\eta)\quad , \label{eq:sigma-mode}
%\quad \xrightarrow{\eta \to -\infty} \quad i \frac{H\eta}{\sqrt{2k}} e^{-ik\eta}, \label{eq:sigma-mode}
\end{equation}
where
\begin{equation}
    \mu = \sqrt{\frac{m^2}{H^2} - \frac{9}{4}}. \label{eq:mu-def}
\end{equation}
Near the late-time boundary, \(\eta\to0^-\), the same mode separates into the two characteristic powers
\begin{equation}
    \lim_{\eta \to 0} \sigma(k, \eta) = \mathcal{O}_+(k)\, \eta^{\Delta_+} + \mathcal{O}_-(k)\, \eta^{\Delta_-}, \label{eq:late-time-sigma}
\end{equation}
with late-time weights $\Delta_\pm = \frac{3}{2} \pm i\mu$. Two external fields used repeatedly below are the massless scalar \(\phi\), with \(m^2=0\), and the conformally coupled scalar \(\varphi\), with \(m^2=2H^2\).  Their mode functions are
\begin{align}
    \phi_k(\eta) &= \frac{H}{\sqrt{2k^3}} (1 + i c_s k \eta) e^{-i c_s k \eta}, \quad \Delta = 3, \label{eq:phi-mode} \\
    \varphi_k(\eta) &= i \frac{H \eta}{\sqrt{2 k}} e^{-i c_s k \eta}, \quad \Delta = 2. \label{eq:varphi-mode}
\end{align}
Here \(c_s\) denotes the sound speed of the external mode; in the bootstrap analysis below, we set \(c_s=1\).  In the applications below, the massless mode \(\phi\) provides the standard proxy for the scalar curvature fluctuation.

Cosmological correlators are computed using the Schwinger--Keldysh, or in-in, formalism. Diagrammatic and flow-based approaches offer complementary ways to organize the in-in evolution \cite{Chen:2017ryl,Extra:231206559,Extra:230200655,Extra:240415416,Extra:251207941}.  The two propagator types needed here have distinct roles.  Bulk-to-bulk propagators connect two interaction times in the inflationary bulk, while bulk-to-boundary propagators connect an interaction time to the late-time observation surface.  For the massive field \(\sigma\), the four bulk-to-bulk components are
\begin{align}
    G_{++}^\sigma(k, \eta, \eta') &= \sigma_k(\eta)\, \sigma_k^*(\eta')\, \Theta(\eta - \eta') + \sigma_k^*(\eta)\, \sigma_k(\eta')\, \Theta(\eta' - \eta), \nonumber \\
    G_{+ -}^\sigma(k, \eta, \eta') &= \sigma_k^*(\eta)\, \sigma_k(\eta'), \nonumber \\
    G_{- +}^\sigma(k, \eta, \eta') &= \sigma_k(\eta)\, \sigma_k^*(\eta'), \nonumber \\
    G_{--}^\sigma(k, \eta, \eta') &= \sigma_k(\eta)\, \sigma_k^*(\eta')\, \Theta(\eta' - \eta) + \sigma_k^*(\eta)\, \sigma_k(\eta')\, \Theta(\eta - \eta'). \label{eq:bulk-to-bulk}
\end{align}
The time-ordered components satisfy the inhomogeneous equation
\begin{equation}
    \left( \Box_{dS} + \frac{m^2}{H^2} \right) G_{\pm\pm}^\sigma(k, \eta, \eta') = \mp i \eta^2 \eta'^2H^2\delta(\eta - \eta'), \label{eq:inhom-eom}
\end{equation}
whereas \(G_{\pm\mp}^\sigma\) satisfy the corresponding homogeneous equation.  For an external massless field, the bulk-to-boundary propagators are
\begin{equation}
    K^{\phi}_+(k, \eta) = \phi_k(\eta_0)\, \phi_k^*(\eta), \quad K^{\phi}_-(k, \eta) = \phi_k^*(\eta_0)\, \phi_k(\eta), \label{eq:bulk-to-boundary}
\end{equation}
where \(\eta_0\to0^-\) denotes the late-time boundary.  The same construction applies to \(\varphi\).  These external propagators satisfy
\begin{equation}
    \left( \Box_{dS} + \frac{m^2}{H^2} \right) K_\pm = 0, \quad \text{with} \quad K_+ = K_-^*.
\end{equation}
The distinction between the two propagator roles reflects the different late-time behavior of the fields. A heavy field decays as $(-\eta)^{3/2}$, see Eq.~\eqref{eq:late-time-sigma}, and therefore does not survive until the end of inflation; it exists only as an intermediate state in the bulk and enters correlators through the bulk-to-bulk propagator. Light fields, by contrast, decay slowly enough, or freeze out entirely in the massless case, and persist until the late-time boundary $\eta_0\to0^-$, where they carry the imprint of the heavy exchange. The massive field is therefore never observed directly, but only through its effect on the correlators of the light fields. For this reason, the dynamics of the exchanged field, and hence the bootstrap equations derived below, originate from the equation of motion of the bulk-to-bulk propagator.

%%%%%%%%%%%%%%%%%%%%%%%%%%%%%%%%%%%%%%%%
%%%%%%%%%%%%%%%%%%%%%%%%%%%%%%%%%%%%%%%%
%%%%%%%%%%%%%%%%%%%%%%%%%%%%%%%%%%%%%%%%
%%%%%%%%%%%%%%%%%%%%%%%%%%%%%%%%%%%%%%%%
\subsection{Ghost Condensate Scalar Field}
\label{subsec:ghost_condensate_scalar}

Originally, ghost inflation \cite{Nima_Arkani_Hamed_2004} was proposed as an inflationary phase driven by a scalar field whose time-dependent background spontaneously breaks time translations. In this work, we use the same low-energy structure in a different way: the exchanged particle is treated as a massive spectator degree of freedom with ghost-condensate dispersion \cite{Ferreira:2026tyj}. Thus, the word ``ghost'' refers to the quartic dispersion relation inherited from the ghost-condensate effective theory, not to a negative-norm state. The quadratic Lagrangian for the spectator fluctuation may be written schematically,
\begin{equation}
    \mathcal{L}_2
    =
    \frac{1}{2}M^4\dot{\chi}^2
    -
    \frac{1}{2}M^2(\nabla^2\chi)^2
    -
    \frac{1}{2}M^4m^2_{\rm gh}\chi^2 ,
\end{equation}
where \(M\) denotes the cutoff scale of the ghost effective theory and \(m_{\rm gh}\) is the spectator mass. Its key feature is that the usual gradient term \((\nabla\chi)^2\) is absent and replaced by the higher-derivative term \((\nabla^2\chi)^2\). The resulting dispersion relation is
\begin{equation}
    \omega^2 = m_{\rm gh}^2 + \frac{k_{\rm phys}^4}{M^2},
    \qquad k_{\rm phys}=\frac{k}{a},
    \label{eq:ghost_dispersion}
\end{equation}
so the quartic term controls the short-distance propagation, while the mass term controls the late-time behavior. From this point onward, we write \(m\equiv m_{\rm gh}\) and \(\gamma\equiv H/M\). In conformal time, the mode function obeys
\begin{equation}
    \left(
    \Box_{\rm ghost}
    +
    \frac{m^2}{H^2}
    \right)\chi_k(\eta)=0,
    \qquad
    \Box_{\rm ghost}
    \equiv
    \eta^2\partial_\eta^2
    -2\eta\partial_\eta
    +
    \gamma^2k^4\eta^4.
    \label{eq:mode-eom2}
\end{equation}
Compared with Eq.~\eqref{eq:mode-eom}, the only change is in the gradient term,
\begin{equation}
    k^2\eta^2=\frac{k_{\rm phys}^2}{H^2}
    \quad\longrightarrow\quad
    \gamma^2k^4\eta^4=\frac{k_{\rm phys}^4}{M^2H^2},
\end{equation}
which is precisely the quartic term of Eq.~\eqref{eq:ghost_dispersion}; the mass term is unchanged. To fix the normalization, consider first the far past, \(\eta\to-\infty\). There, the quartic term dominates over the mass term, and in terms of the canonical variable \(f_k\equiv a\chi_k=-\chi_k/(H\eta)\), Eq.~\eqref{eq:mode-eom2} reduces to
\begin{equation}
    f_k''+\gamma^2k^4\eta^2 f_k\simeq0 .
\end{equation}
The mode is adiabatic, with conformal frequency \(\omega_k(\eta)=\gamma k^2(-\eta)>0\), which grows toward the past. We define the vacuum by the positive-frequency WKB solution,\footnote{Note there is a  sign typo in Eq.~(2.11) of Ref.~\cite{Ferreira:2026tyj} in the exponent of the WKB phase; the correct positive-frequency phase is the one in Eq.~\eqref{eq:ghost_WKB}.}
\begin{equation}
    f_k(\eta\to-\infty)\to 
    \frac{1}{\sqrt{2\omega_k}}\exp\!\left(-i\int^{\eta}\omega_k\,d\eta'\right)
    =\frac{e^{i\gamma k^2\eta^2/2}}{k\sqrt{2\gamma(-\eta)}} .
    \label{eq:ghost_WKB}
\end{equation}
This is a Bunch--Davies vacuum in the adiabatic sense appropriate to a \(k^4\) dispersion relation, rather than the standard relativistic one. To find the exact solution, we write \(\chi_k=(-\eta)^{3/2}\mathcal Z(z)\) with \(z=\gamma k^2\eta^2/2\), which turns Eq.~\eqref{eq:mode-eom2} into Bessel's equation of order \(i\mu/2\). Of its two independent solutions, only \(H^{(1)}_{i\mu/2}(z)\sim e^{iz}\) reproduces the phase of Eq.~\eqref{eq:ghost_WKB}, and matching the amplitude fixes the normalization,
\begin{equation}
    \chi_k^{\mathrm{BD}}(\eta)
     =
     i\frac{H\sqrt{\pi}}{\sqrt8 k^{3/2}}
     e^{-i\pi/4}e^{-\pi\mu/4}
     (-k\eta)^{3/2}
     H_{i\mu/2}^{(1)}
     \left(\frac{\gamma k^2\eta^2}{2}\right)
    %\xrightarrow{\eta\to-\infty}     \frac{H}{k}\sqrt{\frac{-\eta}{2\gamma}}    e^{i\gamma k^2\eta^2/2},
    \label{eq:sigma-mode2}
\end{equation}
with $\mu$ defined in Eq.~\eqref{eq:mu-def}. The factor \(e^{-\pi\mu/4}\) compensates for the growth \(e^{+\pi\mu/4}\) of the Hankel function at large arguments, so that the early-time amplitude is independent of the mass, as it must be in the regime where the quartic term dominates. The order of the Hankel function is half the relativistic one because \(z\propto\eta^2\). However, at late times, this is compensated, since \((-\eta)^{3/2}z^{\pm i\mu/2}\propto(-\eta)^{3/2\pm i\mu}\). The ghost spectator, therefore, has the same late-time weights, \(\Delta_\pm=3/2\pm i\mu\), as the relativistic field
\begin{equation}
    \lim_{\eta\to0}\chi(k,\eta)
    =
    \mathcal O_+(k)\eta^{\Delta_+}
    +
    \mathcal O_-(k)\eta^{\Delta_-}.
    \label{eq:late-time-sigma2}
\end{equation}

The effective description requires \(H<M\) and \(m<M\). We also take the spectator to be heavier than the inflaton fluctuations, whose mass in single-field slow-roll inflation is parametrically \(m_{\rm inf}^2\sim\eta_V H^2\), with \(\eta_V\equiv M_{\rm Pl}^2V_{,\phi\phi}/V\sim10^{-2}\) \cite{Baumann:2022mni,Planck:2018jri}. For a characteristic Hubble scale \(H\sim10^{13}\,{\rm GeV}\sim10^{-5}M_{\rm Pl}\) and, for illustration, a cutoff \(M\sim10^{-2}M_{\rm Pl}\), these conditions give the parametric window
\begin{equation}
10^{-1}
\lesssim
\frac{m}{H}
\lesssim
10^{3},
\end{equation}
and \(\gamma=H/M\sim10^{-3}\). We focus on the principal-series regime, \(m/H>3/2\), where \(\mu\) is real, and the exchanged field produces the oscillatory nonanalytic behavior characteristic of the cosmological collider. The Schwinger--Keldysh propagators of \(\chi\) are obtained from Eq.~\eqref{eq:bulk-to-bulk} by replacing \(\sigma_k\to\chi_k\),
\begin{equation}
    G^\chi_{ab}(k,\eta,\eta')=G^\sigma_{ab}(k,\eta,\eta')\Big|_{\sigma_k\to\chi_k},
    \qquad a,b=\pm,
    \label{eq:bulk-to-bulkghost}
\end{equation}
and the time-ordered components satisfy
\begin{equation}
    \left( \Box_{\rm ghost} + \frac{m^2}{H^2} \right) G_{\pm\pm}^\chi(k, \eta, \eta') = \mp i \eta^2 \eta'^2H^2\delta(\eta - \eta'), \label{eq:inhom-eomghost}
\end{equation}
whereas \(G_{\pm\mp}^\chi\) satisfy the corresponding homogeneous equation. This distinction is important because the fourth-order bootstrap operator derived below originates from the equation of motion of the internal propagator.

%%%%%%%%%%%%%%%%%%%%%%%%%%%%%%%%%%%%
%%%%%%%%%%%%%%%%%%%%%%%%%%%%%%%%%%%%
%%%%%%%%%%%%%%%%%%%%%%%%%%%%%%%%%%%%
\subsection{Late-time weights and de Sitter representations}

\begin{figure}[t]
\centering
\begin{tikzpicture}[
x=1.55cm,
y=1.15cm,
>=Latex,
every node/.style={font=\small}
]

% -------------------------------------------------
% Axes
% -------------------------------------------------

\draw[->, line width=0.65pt]
(-0.25,0) -- (3.45,0)
node[right=2pt] {$\operatorname{Re}\Delta$};

\draw[->, line width=0.65pt]
(0,-2.25) -- (0,2.25)
node[above=2pt] {$\operatorname{Im}\Delta$};

% -------------------------------------------------
% Complementary series
% -------------------------------------------------

\draw[
line width=1.6pt,
red!75!black,
line cap=round
]
(0,0) -- (3,0);

% -------------------------------------------------
% Principal series
% -------------------------------------------------

\draw[
line width=0.95pt,
densely dashed,
black
]
(1.5,-2.05) -- (1.5,2.05);

% -------------------------------------------------
% Massive-field weights
% -------------------------------------------------

\fill[green!50!black]
(1.5,1.50) circle (2.0pt);

\fill[green!50!black]
(1.5,-1.50) circle (2.0pt);

% -------------------------------------------------
% Tick marks
% -------------------------------------------------

\draw[line width=0.55pt] (0,-0.065) -- (0,0.065);
\draw[line width=0.55pt] (1.5,-0.065) -- (1.5,0.065);
\draw[line width=0.55pt] (3,-0.065) -- (3,0.065);

\draw[line width=0.55pt] (-0.055,1.50) -- (0.055,1.50);
\draw[line width=0.55pt] (-0.055,-1.50) -- (0.055,-1.50);

% -------------------------------------------------
% Tick labels
% -------------------------------------------------

\node[
below=5pt,
fill=white,
inner sep=1.2pt
] at (0,0) {$0$};

\node[
below=5pt,
fill=white,
inner sep=1.2pt
] at (1.5,0) {$3/2$};

\node[
below=5pt,
fill=white,
inner sep=1.2pt
] at (3,0) {$3$};

\node[
left=7pt,
fill=white,
inner sep=1.2pt
] at (0,1.50) {$\mu$};

\node[
left=7pt,
fill=white,
inner sep=1.2pt
] at (0,-1.50) {$-\mu$};

\end{tikzpicture}

\caption{Scalar late-time weights in the complex \(\Delta\) plane.
The red segment denotes the complementary series of scalar de Sitter representations, while the vertical dashed line at
\(\operatorname{Re}\Delta=3/2\) corresponds to the principal series.
The green markers indicate the two late-time weights
\(\Delta_{\pm}=3/2\pm i\mu\) of a massive principal-series field.
The relativistic massive scalar and the quartic ghost spectator share these same late-time weights; their distinction arises instead from their different short-distance dispersion relations.
}
\label{fig:ds_series_delta_plane}
\end{figure}
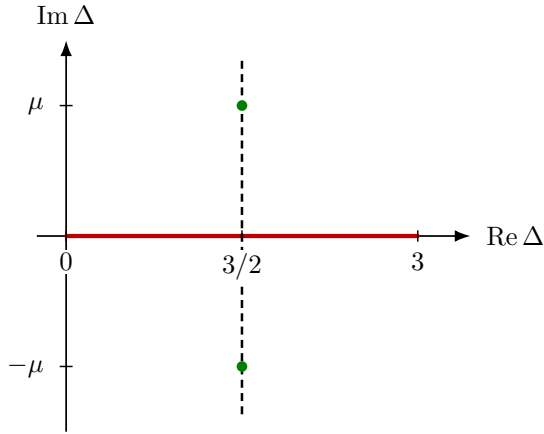

The late-time behavior of scalar fields in de Sitter space can be organized in terms of the unitary representations of the de Sitter isometry group \(SO(1,4)\). For a massive scalar, the two independent late-time weights are
\begin{equation}
\Delta_{\pm}
=
\frac{3}{2}\pm i\mu,
\qquad
\Delta_{+}\equiv\Delta,
\qquad
\Delta_{-}=3-\Delta,
\label{anomalousdimen}
\end{equation}
with \(\mu\) defined in Eq.~\eqref{eq:mu-def}. The two weights \(\Delta_{+}\) and \(\Delta_{-}\) correspond to the two independent late-time branches of the same bulk field. When a single symbol \(\Delta\) is needed below, we use \(\Delta\equiv\Delta_{+}\), so that the second branch is \(3-\Delta\). For scalar fields, the unitary de Sitter representations relevant here separate into the complementary and principal series, see Figure~\ref{fig:ds_series_delta_plane}.

\textbf{Complementary series}: for sufficiently light scalars in de Sitter, the two weights are real and lie in the range \begin{equation} \Delta_{\pm} \in (0,3), \quad 0 < \frac{m^2}{H^2} < \frac{9}{4}. \end{equation} This is the regime in which the late-time behavior is power-law rather than oscillatory.

\textbf{Principal series}: for \(m>3H/2\), the weights form a complex-conjugate pair, \begin{equation}\label{principalseries} \Delta_{\pm} = \frac32 \pm i\mu, \quad \frac{m^2}{H^2} > \frac{9}{4}. 
\end{equation}
The corresponding late-time branches therefore contain logarithmic oscillations. Their interpretation as heavy-particle signals motivates studies of standard-model backgrounds, loop-induced signatures, and prospects for cosmological collider searches \cite{Extra:161006597,Extra:160407841,Extra:161208122,Extra:180502656,Extra:190707390}.~These non-analytic oscillatory contributions are the characteristic imprint of massive-particle exchange in cosmological correlators and provide the basic signal exploited in the cosmological collider program \cite{ArkaniHamedMaldacena_CosmologicalCollider_2015}. The two branches should not be interpreted as distinct representations; rather, \(\Delta_{+}\) and \(\Delta_{-}=3-\Delta_{+}\) are the two late-time weights associated with the same principal-series bulk field. The important point is that the relativistic massive scalar and the ghost spectator have the same late-time scaling dimensions, \(\Delta_\pm=3/2\pm i\mu\), even though their short-distance propagation is qualitatively different. For the relativistic field, the early-time mode behaves as
\begin{equation}
\sigma_{k}(\eta)
\sim
e^{-ik\eta},
\end{equation}
whereas for the quartic ghost spectator,
\begin{equation}
\chi_k(\eta)\sim
    e^{\!+\frac{i}{2}\gamma k^2\eta^2}.
\end{equation}
Both are positive-frequency phases, \(\exp[-i\int^\eta\omega\,d\eta']\). The opposite sign of the exponent arises because the conformal frequency of the quartic mode, \(\omega=\gamma k^2(-\eta)\), grows toward the past, whereas that of the relativistic mode, \(\omega=k\), is constant.
The modification of the bootstrap problem therefore does not originate from a change in the infrared de Sitter weights but from the quartic ultraviolet dispersion relation. As we shall see, this difference promotes the standard second-order de Sitter bootstrap equation to a fourth-order equation and qualitatively changes the nature of its singular points.

%%%%%%%%%%%%%%%%%%%%%%%%%%%%%%%%%
%%%%%%%%%%%%%%%%%%%%%%%%%%%%%%%%%
%%%%%%%%%%%%%%%%%%%%%%%%%%%%%%%%%
%%%%%%%%%%%%%%%%%%%%%%%%%%%%%%%%%
\section{The de Sitter Bootstrap}\label{sec:dS_bootstrap}

Before turning to the ghost case, we review the standard de Sitter bootstrap for the exchange of a massive scalar. This provides the relativistic benchmark against which the higher-derivative problem will be compared. The bootstrap equations may be understood from the conformal symmetries of de Sitter space and the associated Ward identities
\cite{ArkaniHamedMaldacena_CosmologicalCollider_2015,ArkaniHamedBaumannLeePimentel_CosmologicalBootstrap_2018,BaumannEtAl_WeightShiftingScalarSeeds_2019,PimentelWang_BoostlessCosmologicalColliderBootstrap_2022}.
For our purposes, however, it will be more convenient to derive the differential equation directly from the equation of motion of the exchanged propagator. This formulation makes the comparison with the ghost spectator particularly transparent.

We consider a scalar-exchange four-point function in the channel
\begin{equation}
s
\equiv
\left|\mathbf{k}_1+\mathbf{k}_2\right|
=
-\left|\mathbf{k}_3+\mathbf{k}_4\right|,
\end{equation}
with
\begin{equation}
k_{12}\equiv k_1+k_2,
\qquad
k_{34}\equiv k_3+k_4.
\end{equation}
After stripping off the kinematic prefactors associated with the external legs, the dynamical information can be organized into a family of reduced Schwinger--Keldysh seed functions. Following the conventions of
\cite{Aoki:2023wdc,Qin:2023ejc}, we define
\begin{align}\label{Seedgeralzero}
    \hat{\mathcal{I}}^{\,\ell_1,\ell_2}_{ab} 
    \equiv 
    -ab\, s^{5+\ell_1+\ell_2}
    \int_{-\infty}^{0} d\eta_1\, d\eta_2\, 
    (-\eta_1)^{\ell_1}(-\eta_2)^{\ell_2}
    e^{ai k_{12}\eta_1 + bi k_{34} \eta_2}
    G_{ab}(s,\eta_1,\eta_2) \quad a,b=\pm 1 .
\end{align}
The prefactor $s^{5+\ell_1+\ell_2}$ defines the normalized seed used throughout this work, so that its remaining dependence is expressed entirely in terms of dimensionless momentum ratios. The parameters $\ell_1$ and $\ell_2$ encode the time dependence associated with the two interaction vertices. In particular, $\ell=0$ corresponds to the standard massless external weight, while $\ell=-2$ corresponds to the conformally coupled case.~Mixed choices, such as $(\ell_1,\ell_2)=(0,-2)$, are also useful for relating the seed to inflationary observables.

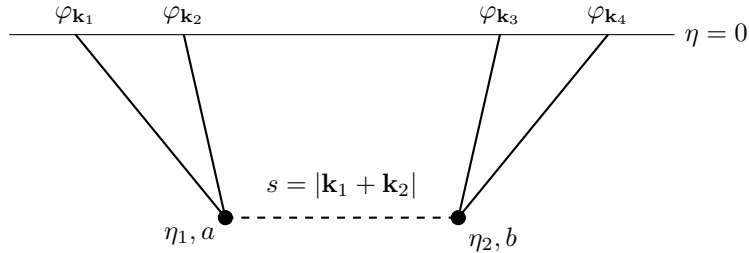
\begin{figure}[t]
\centering
\begin{tikzpicture}[scale=1.1, every node/.style={font=\small}]

    % Boundary
    \draw[thin] (-4.0,2.2) -- (4.0,2.2);
    \node[right] at (4.0,2.2) {$\eta=0$};

    % Vertices
    \coordinate (V1) at (-1.4,0);
    \coordinate (V2) at ( 1.4,0);

    % Boundary attachment points
    \coordinate (B1) at (-3.2,2.2);
    \coordinate (B2) at (-1.9,2.2);
    \coordinate (B3) at ( 1.9,2.2);
    \coordinate (B4) at ( 3.2,2.2);

    % External legs
    \draw[thick] (V1) -- (B1);
    \draw[thick] (V1) -- (B2);
    \draw[thick] (V2) -- (B3);
    \draw[thick] (V2) -- (B4);

    % Internal exchanged line
    \draw[thick,dashed] (V1) -- (V2);

    % Vertices as dots
    \filldraw[black] (V1) circle (2.4pt);
    \filldraw[black] (V2) circle (2.4pt);

    % Vertex labels
    \node[below left]  at (V1) {$\eta_1,a$};
    \node[below right] at (V2) {$\eta_2,b$};

    % Momentum labels
    \node[above] at (B1) {$\varphi_{\mathbf{k}_1}$};
    \node[above] at (B2) {$\varphi_{\mathbf{k}_2}$};
    \node[above] at (B3) {$\varphi_{\mathbf{k}_3}$};
    \node[above] at (B4) {$\varphi_{\mathbf{k}_4}$};

    % Internal momentum label
    \node[above] at (0,0.1) {$s=|\mathbf{k}_1+\mathbf{k}_2|$};

\end{tikzpicture}
\caption{
Schematic Schwinger--Keldysh exchange diagram contributing to the scalar four-point function. The two bulk vertices, located at conformal times \(\eta_1\) and \(\eta_2\), may lie on either branch \(a,b=\pm\) of the in-in contour. The four external legs end on the late-time boundary \(\eta=0\) and carry boundary momenta \(\mathbf{k}_1,\mathbf{k}_2,\mathbf{k}_3,\mathbf{k}_4\). The dashed internal line denotes the exchanged field in the \(s\)-channel, with \(s=|\mathbf{k}_1+\mathbf{k}_2|\).
}
\label{fig:sk_exchange_seed}
\end{figure}

The Schwinger--Keldysh structure of Eq.~\eqref{Seedgeralzero} is illustrated in Fig.~\ref{fig:sk_exchange_seed}. The two bulk vertices are located at conformal times $\eta_1$ and $\eta_2$ and can independently lie on either branch $a,b=\pm$ of the in-in contour, while the internal line carries the exchange momentum $s$. The complete correlator is obtained by summing over the corresponding contour assignments. For the bootstrap construction, the important point is that the dependence on the exchanged field is entirely contained in the bulk-to-bulk propagator $G_{ab}$. Acting with its equation of motion inside Eq.~\eqref{Seedgeralzero} therefore converts the bulk dynamics into a differential equation for the reduced seed.

From the diagrammatic perspective, in appropriate limits, the non-analytic part of the four-point function is inherited from the cubic subdiagrams attached to the internal propagator. Thus, much of the local singular behavior can already be anticipated from the analytic structure of a single vertex integral. However, the full singularity structure of the four-point seed function requires combining both vertices with the internal propagator and performing the Schwinger--Keldysh sum over contour assignments. This is the reason why, before moving to the ghost and higher-derivative cases, it is useful to review the singularity structure of the standard de Sitter seed equation. Once this structure is understood in the ordinary scalar-exchange problem, we can ask how it is modified when the exchanged field has a non-linear dispersion relation. In this sense, the study of singularities provides the bridge between the diagrammatic Schwinger--Keldysh computation of cosmological correlators and the bootstrap description in terms of seed functions.
There are related ways of imposing the physical constraints, including cosmological tree and cutting theorems, positivity and reality properties, and the reconstruction of correlators from physical cuts \cite{Extra:230800680,Extra:241109695,Extra:230907769,Extra:240706258,Extra:241113636,Extra:230413295}.

\subsection{Singularities and cross-ratios}

Let us first recall how the singularities appear directly from bulk time integrals. Consider a contact interaction involving conformally coupled external fields. Up to an overall power of conformal time fixed by the vertex, the corresponding seed integral has the schematic form
\begin{equation}
    \mathcal I_{\rm contact}
    \sim
    \int_{-\infty}^{0} d\eta\,(-\eta)^n e^{i k_t\eta},
    \qquad
    k_t\equiv k_1+k_2+k_3+k_4 .
\end{equation}
With the usual in--in \(i\epsilon\) prescription, the far past part of the contour is slightly rotated so that the oscillatory integral is damped. The integral then gives
\begin{equation}
    \mathcal I_{\rm contact}
    \sim
    e^{-i\pi(n+1)/2}
    \frac{\Gamma(n+1)}{k_t^{n+1}} .
\end{equation}
Thus, the total-energy singularity at \(k_t\to0\) is already visible in the elementary bulk integral. In a relativistic theory, this is the cosmological version of the flat-space limit: the residue of the total-energy pole is related to a flat-space scattering amplitude. For an exchange diagram, the same logic appears locally at each cubic vertex. Let the internal momentum flowing through the exchange channel be $ s\equiv \lvert \boldsymbol{k}_1+\boldsymbol{k}_2\rvert    = - \lvert\boldsymbol{k}_3+\boldsymbol{k}_4\rvert$. The two time integrals contain phases of the schematic form
\begin{equation}
    e^{i(k_1+k_2)\eta_1}e^{\pm i s\eta_1},
    \qquad
    e^{i(k_3+k_4)\eta_2}e^{\pm i s\eta_2}.
\end{equation}
The elementary energy denominators are therefore controlled by combinations such as $ k_1+k_2\pm s$, $ k_3+k_4\pm s$. This motivates the dimensionless ratios
\begin{equation}
    u\equiv \frac{s}{k_1+k_2},
    \qquad
    v\equiv \frac{s}{k_3+k_4}.
\end{equation}
In terms of \(u\), the singular loci associated with the left vertex are
\begin{equation}
    k_1+k_2+s=0
    \quad\Longleftrightarrow\quad
    u=-1,
    \qquad
    k_1+k_2-s=0
    \quad\Longleftrightarrow\quad
    u=1.
\end{equation}
These two limits have different physical interpretations. The first one, \(u=-1\), is the partial-energy or factorization singularity after analytic continuation. It is a physical singularity: its residue is fixed by factorization into lower-point data. The second one, \(u=1\), is the folded configuration. This is not the physical factorization channel. In a Bunch--Davies state, one does not want an independent folded singularity. Such folded singularities are characteristic of non--Bunch-Davies initial states, where negative-frequency components can generate denominators of the form \(k_1+k_2-s\). Therefore, in the Bunch--Davies seed, the would-be singular behavior at \(u=1\) must cancel. 

The soft exchange limit is, instead,
\begin{equation}
    s\to0
    \quad\Longleftrightarrow\quad
    u\to0,
\end{equation}
and this is the limit where the characteristic cosmological collider scaling of a massive exchanged field appears. The bootstrap equation makes this picture precise: the kinematic limits \(u=0,\pm1\) become singular points of a differential equation in the variable \(u\).

\subsection{Differential structure}

The homogeneous equation for a massive scalar propagator in de Sitter is
\begin{equation}
    \left(
    \eta_1^2\partial_{\eta_1}^2
    -2\eta_1\partial_{\eta_1}
    +s^2\eta_1^2
    +\frac{m^2}{H^2}
    \right)
    G_{ab}(\eta_1,\eta_2;s)=0,\qquad a\neq b .
    \label{eq:dS_prop_eom_hom}
\end{equation}
After the change of integration variables
\begin{equation}
    z_1=(k_1+k_2)\eta_1,
    \qquad
    z_2=(k_3+k_4)\eta_2,
\end{equation}
the exchange seed can be written as
\begin{align}
    \hat{\mathcal I}^{\ell_1,\ell_2}_{ab}(u,v)
    =&
    -ab\,
    u^{1+\ell_1}v^{1+\ell_2}
    \int_{-\infty}^{0} dz_1\,(-z_1)^{\ell_1}e^{ia z_1}
    \int_{-\infty}^{0} dz_2\,(-z_2)^{\ell_2}e^{ib z_2}
    \hat G_{ab}(uz_1,vz_2),
    \label{eq:dS_seed_general}
\end{align}
where \(a,b=\pm1\), and
\begin{equation}
    \hat G_{ab}\equiv s^3G_{ab}.
\end{equation}
The powers \(\ell_1,\ell_2\) encode the time dependence of the two interaction vertices. The key observation is that the propagator depends on the products \(uz_1\) and \(vz_2\). Therefore derivatives with respect to bulk time can be traded for derivatives with respect to the cross-ratios. The required integration-by-parts identities are
\begin{align}
    \int_{-\infty}^{0} dz\,(-z)^\ell e^{iaz}\,z\,f(uz)
    &=
    ia\left(\ell+1+u\partial_u\right)
    \int_{-\infty}^{0} dz\,(-z)^\ell e^{iaz}f(uz),
    \\
    \int_{-\infty}^{0} dz\,(-z)^\ell e^{iaz}\,z^2\,f(uz)
    &=
    -\left(u\partial_u+\ell+2\right)
     \left(u\partial_u+\ell+1\right)
    \int_{-\infty}^{0} dz\,(-z)^\ell e^{iaz}f(uz).
    \label{eq:integration_identities_dS}
\end{align}
These identities allow us to commute the propagator equation of motion through the time integrals. At an intermediate stage the differential operator acts on the stripped combination
\begin{equation}
    u^{-1-\ell_1}v^{-1-\ell_2}
    \hat{\mathcal I}^{\ell_1,\ell_2}_{ab}(u,v),
\end{equation}
rather than directly on the full seed. After restoring the prefactor \(u^{1+\ell_1}v^{1+\ell_2}\), the equation can be written directly in terms of \(\hat{\mathcal I}\). For the Wightman components, one obtains
\begin{equation}
    \mathcal D_u^{\ell_1}
    \hat{\mathcal I}^{\ell_1,\ell_2}_{ab}(u,v)=0,\qquad a\neq b,
    \label{eq:dS_bootstrap_hom}
\end{equation}
where the de Sitter bootstrap operator in the usual variable \(u\) is
\begin{equation}
    \mathcal D_u^\ell
    \equiv
    \left(u^2-u^4\right)\partial_u^2
    -
    \left[(4+2\ell)u+2u^3\right]\partial_u
    +\mu^2+\left(\ell+\frac52\right)^2\ ,
\label{eq:dS_bootstrap_operator_usual}
\end{equation}
recalling that $\mu\equiv \sqrt{m^2/H^2-9/4}$. The contact term in the propagator equation gives
\begin{equation}
    \mathcal D_u^{\ell_1}
    \hat{\mathcal I}^{\ell_1,\ell_2}_{aa}(u,v)
    =
    H^2 e^{-a i\pi\ell_{12}/2}
    \Gamma(5+\ell_{12})
    \left(\frac{uv}{u+v}\right)^{5+\ell_{12}}, \quad   \ell_{12}\equiv \ell_1+\ell_2.
    \label{eq:dS_bootstrap_source}
\end{equation}
Let us now study the singular points of the homogeneous equation
\begin{equation}
    \mathcal D_u^\ell Y(u)=0.
    \label{eq:dS_homogeneous_Y_usual}
\end{equation}
Dividing by the coefficient of \(Y''\), the equation becomes
\begin{equation}
    Y''+P_\ell(u)Y'+Q_\ell(u)Y=0,
\end{equation}
with
\begin{equation}
    P_\ell(u)
    =
    -\frac{(4+2\ell)u+2u^3}{u^2(1-u^2)},
    \qquad
    Q_\ell(u)
    =
    \frac{
    \mu^2+\left(\ell+\frac52\right)^2
    }{
    u^2(1-u^2)
    }.
    \label{eq:PQ_dS_usual}
\end{equation}
The finite singular points are therefore
\begin{equation}
    u=0,
    \qquad
    u=1,
    \qquad
    u=-1.
\end{equation}
They are all regular singular points\footnote{Around \(u=u_\star\), a second-order equation has a regular singular point if
$(u-u_\star)P_\ell(u)$ and $(u-u_\star)^2Q_\ell(u)$ are analytic near \(u_\star\). This criterion is satisfied at \(u=0,\pm1\).}. At \(u=0\), one finds $ uP_\ell(u)\to -(4+2\ell)$ and $u^2Q_\ell(u)\to  \mu^2+\left(\ell+5/2\right)^2$. The indicial equation is
\begin{equation}
    \rho(\rho-1)
    -(4+2\ell)\rho
    +
    \mu^2+\left(\ell+\frac52\right)^2=0,
\end{equation}
whose roots are $\rho_\pm=\ell+\frac52\pm i\mu $. Thus, near the soft exchange limit,
\begin{equation}
    Y(u)\sim
    c_+u^{\ell+\frac52+i\mu}
    +
    c_-u^{\ell+\frac52-i\mu}.
\end{equation}
This is the usual cosmological collider behavior. The two branches are the boundary imprint of the two late-time falloffs of a massive de Sitter field. At \(u=1\), we have $(u-1)P_\ell(u)\to \ell+3$ and $    (u-1)^2Q_\ell(u)\to 0$. The indicial equation is
\begin{equation}
    \rho(\rho-1)+(\ell+3)\rho=0,
\end{equation}
and therefore $\rho=0$ and $\rho=-\ell-2$. The same exponents are obtained at \(u=-1\). In the commonly used case \(\ell=-2\), the two exponents coincide and the local solutions contain logarithms,
\begin{equation}
    Y(u)\sim c_1+c_2\log(u\mp1),
    \qquad
    u\to\pm1,
\end{equation}

\begin{figure}[t]
\centering
\begin{tikzpicture}[
    >=Latex,
    scale=1.05,
    every node/.style={font=\footnotesize},
    axis/.style={gray, line width=0.6pt},
    sing/.style={circle, fill=blue!65!black, inner sep=1.5pt},
    origin/.style={circle, fill=blue!65!black, inner sep=1.2pt},
    folded/.style={circle, draw=red!70!black, fill=white, line width=0.9pt, inner sep=2.0pt},
    leader/.style={gray, line width=0.5pt}
]

% axes
\draw[axis] (-3.0,0) -- (3.0,0) node[below ] {$\mathrm{Re}\,u$};
\draw[axis] (0,-1.8) -- (0,1.8) node[above] {$\mathrm{Im}\,u$};

% |u|=1 circle
\draw[line width=1.0pt] (0,0) circle (1.1);

% wavy segment from -1 to 0
\draw[blue!65!black, line width=1.2pt,
      decorate, decoration={snake, amplitude=0.8pt, segment length=4.5pt}]
      (-2.0,0) -- (0,0);

% singular / special points
\node[sing] at (-2.0,0) {};
\node[below=2pt] at (-2.0,0) {$-1$};

%\node[sing] at (-1.15,0) {};
\node[below=2pt] at (-1.35,0) {$-v$};

\node[origin] at (0,0) {};
\node[below right=1pt] at (0,0) {$0$};

% optional folded point
\node[folded] at (2.0,0) {};
\node[below=2pt] at (2.0,0) {$1$};

% optional labels
\draw[leader] (-2.0,0.08) -- (-2.0,1.0);
\node[align=center, font=\scriptsize] at (-2.0,1.22) {factorization};

\draw[leader] (2.0,0.08) -- (2.0,1.0);
\node[align=center, font=\scriptsize] at (2.0,1.22) {folded};

\end{tikzpicture}
\caption{
Analytic structure of the ordinary de Sitter bootstrap in the complex
$u$-plane. The point $u=0$ is a regular singular point associated with
the soft cosmological-collider limit. The point $u=-1$ corresponds to
the physical factorization singularity, while $u=1$ is the folded
configuration. For fixed $v$, the point $u=-v$ lies on the negative
real axis and is relevant to the inhomogeneous structure.
}
\label{fig:ds_analytic_structure}
\end{figure}
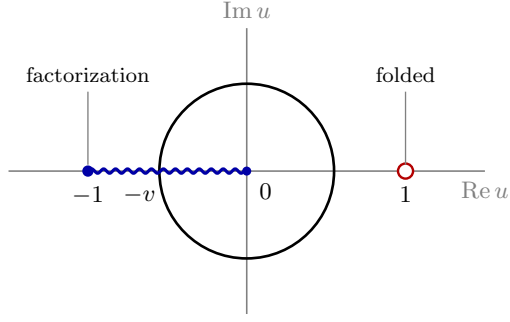

The interpretation of these two regular singular points is different. The point \(u=1\) is the folded configuration. For the Bunch--Davies answer this singularity must be absent. The point \(u=-1\) is the partial-energy factorization singularity. It should not be removed; its coefficient is fixed by factorization.

This is the basic analytic structure of the ordinary de Sitter bootstrap equation in the usual variable, as illustrated in Fig.~\ref{fig:ds_analytic_structure}. The soft point \(u=0\), the folded point \(u=1\), and the factorization point \(u=-1\) are all regular singular points. This is the structure we want to compare later with the fourth-order ghost equation.

%%%%%%%%%%%%%%%%%%%%%%%%%%%%%%%%
%%%%%%%%%%%%%%%%%%%%%%%%%%%%%%%%
%%%%%%%%%%%%%%%%%%%%%%%%%%%%%%%%
\subsection{Bootstrap solutions}
\label{subsec:ds_bootstrap_solutions}

The technical Frobenius derivation of the homogeneous and non-homogeneous solutions is given in Appendix~\ref{app:ds_bootstrap_details}. The special-function approach is complementary to modern methods for massive exchanges, including families of exchange diagrams and more general backgrounds \cite{Extra:230910849,Extra:241207843,Extra:250902684,Extra:251108677,Extra:240307050,Extra:240409547}.  Here we keep the derivation at the level needed to follow the physical result.  The Schwinger--Keldysh equations naturally separate into homogeneous mixed components and sourced same-sign components.

%%%%%%%%%%%%%%%%%%%%%%%%%%%%%%
%%%%%%%%%%%%%%%%%%%%%%%%%%%%%%
%%%%%%%%%%%%%%%%%%%%%%%%%%%%%%
\subsection*{Homogeneous Solution}

The mixed Schwinger--Keldysh components satisfy the homogeneous equation in Eq.~\eqref{eq:dS_homogeneous_Y_usual}.  Up to an overall normalization, its two independent solutions can be written as
\begin{equation}
    \mathcal Y^\ell (u)= C_{+}\mathcal{Y}^{l}_{+}(u) + C_{-}\mathcal{Y}^{l}_{-}(u),
\end{equation}
where,
\begin{equation}
    \mathcal Y^\ell_\pm(u)
    =
    2^{\mp i\mu}
    (iu)^{\ell+\frac52\pm i\mu}
    {}_2F_1
    \left(
    \frac{\ell+\frac52\pm i\mu}{2},
    \frac{\ell+\frac72\pm i\mu}{2};
    1\pm i\mu;
    u^2
    \right) .
\end{equation}
The relative normalization of the two branches can be fixed directly from the folded limit. Using the hypergeometric representation at $ u^2\to 1$ (Appendix~\ref{app:generalized_hypergeometric}), the combination of the two branches is regular provided their coefficients are chosen as
\begin{equation}
C_\pm
=
e^{\pm\pi\mu/2}\,
\Gamma\left(\ell+\frac52 \pm  i\mu\right)
\Gamma(\mp i\mu).
\end{equation}
Indeed, since
\begin{equation}
(iu)^{\ell+\frac52+\pm i\mu}
=
i^{\ell+\frac52}
e^{\mp\pi\mu/2}
u^{\ell+\frac52\pm i\mu},
\end{equation}
the exponential factor in $C_\pm$ cancels, while the remaining factor
$i^{\ell+5/2}$ is common to both branches and can be absorbed into the
overall normalization. The overall normalization is fixed by the convention used for the full de Sitter seed. From this point onward, \(\mathcal Y^\ell_\pm\) denotes the normalized basis,
\begin{equation}
\mathcal Y^\ell_\pm(u)
=
2^{\mp i\mu}
 u^{\ell+\frac52 \pm i\mu}
\Gamma\!\left(\ell+\frac52 \pm i\mu\right)
\Gamma(\mp i\mu)
\,{}_2F_1\!\left(
\begin{matrix}
\frac{\ell}{2}+\frac54 \pm \frac{ i\mu}{2},
\frac{\ell}{2}+\frac74 \pm \frac{ i\mu}{2}\\
1 \pm i\mu
\end{matrix}
;u^2
\right),
\label{eq:Y_sigma_normalized_main}
\end{equation}
With this convention, the complete non-time-ordered Schwinger--Keldysh sector, fixed by the boundary conditions, can be written entirely in terms of these normalized functions. In particular, they determine the $(ab)$ components of the Schwinger--Keldysh correlator with $a\neq b$. By contrast, the time-ordered $(aa)$ components also contain the non-homogeneous contribution and must therefore be treated separately.

%%%%%%%%%%%%%%%%%%%%%%%%%%%%%%
%%%%%%%%%%%%%%%%%%%%%%%%%%%%%%
%%%%%%%%%%%%%%%%%%%%%%%%%%%%%%
\subsection*{Non-homogeneous solution}

For the time-ordered Schwinger--Keldysh components, the particular solution
is determined by the non-homogeneous bootstrap equation
\begin{equation}
\mathcal D_u^{\ell_1}
\widehat{\mathcal I}^{\,\ell_1\ell_2}_{\pm \pm,\mathrm{part}}(u,v)
=
H^2 e^{\mp i\pi(\ell_1+\ell_2)/2}
\Gamma(5+\ell_1+\ell_2)
\left(\frac{uv}{u+v}\right)^{5+\ell_1+\ell_2},
\label{eq:dS_particular_equation}
\end{equation}
with
\begin{equation}
\mathcal D_u^\ell
=
u^2(1-u^2)\partial_u^2
-2u\left(\ell+2+u^2\right)\partial_u
+\left(\ell+\frac52\right)^2+\mu^2 .
\label{eq:dS_operator}
\end{equation}

In the region $u<v$, the source can be expanded directly as
\begin{equation}
\begin{aligned}
\left(\frac{uv}{u+v}\right)^{5+\ell_1+\ell_2}
&=u^{5+\ell_1+\ell_2}\sum_{n=0}^{\infty}
(-1)^n\frac{(5+\ell_1+\ell_2)_n}{n!}
\left(\frac uv\right)^n .
\end{aligned}
\label{eq:dS_source_series}
\end{equation}
We therefore take the particular solution in the double-series form
\begin{equation}
\begin{aligned}
\widehat{\mathcal I}^{\,\ell_1\ell_2}_{\pm \pm ,\mathrm{part}}(u,v)
={}&
H^2 e^{\mp i\pi(\ell_1+\ell_2)/2}\Gamma(5+\ell_1+\ell_2)
%\\&
\times
\sum_{m,n=0}^{\infty}
c_{mn}^{(\ell_1,\ell_2)}
\,u^{5+\ell_1+\ell_2+2m}
\left(\frac uv\right)^n .
\end{aligned}
\label{eq:dS_particular_ansatz}
\end{equation}
For a fixed pair $(m,n)$, the total power of $u$ is $\beta=5+\ell_1+\ell_2+n+2m$. At $m=0$, matching the source gives
\begin{equation}
c_{0n}^{(\ell_1,\ell_2)}
=
(-1)^n
\frac{(5+\ell_1+\ell_2)_n}
{n!\left[
\left(n+\ell_2+\frac52\right)^2+\mu^2
\right]} \ .
\label{eq:dS_c0n}
\end{equation}
For $m\geq1$, the coefficient of each monomial
$u^{5+\ell_1+\ell_2+n+2m}v^{-n}$ must vanish. This gives
\begin{equation}
c_{mn}^{(\ell_1,\ell_2)}
=
\frac{
(3+\ell_1+\ell_2+n+2m)(4+\ell_1+\ell_2+n+2m)
}{
\left(n+2m+\ell_2+\frac52\right)^2+\mu^2
}
c_{m-1,n}^{(\ell_1,\ell_2)} .
\label{eq:dS_recurrence}
\end{equation}
The first terms are
\begin{align}
c_{0n}^{(\ell_1,\ell_2)}
&=
(-1)^n\frac{(5+\ell_1+\ell_2)_n}
{n!\,A_n} \ ,
&&
c_{1n}^{(\ell_1,\ell_2)}
=
(-1)^n\frac{(5+\ell_1+\ell_2)_{n+2}}
{n!\,A_nA_{n+2}}\ ,
\\
c_{2n}^{(\ell_1,\ell_2)}
&=
(-1)^n\frac{(5+\ell_1+\ell_2)_{n+4}}
{n!\,A_nA_{n+2}A_{n+4}} \ ,
&&
c_{3n}^{(\ell_1,\ell_2)}
=
(-1)^n\frac{(5+\ell_1+\ell_2)_{n+6}}
{n!\,A_nA_{n+2}A_{n+4}A_{n+6}}\ ,
\end{align}
where, only in these four lines, $A_k\equiv
\left(k+\ell_2+5/2\right)^2+\mu^2 $. The general coefficient is therefore
\begin{equation}
c_{mn}^{(\ell_1,\ell_2)}
=
(-1)^n
\frac{(5+\ell_1+\ell_2)_{n+2m}}
{n!\displaystyle\prod_{r=0}^{m}
\left[
\left(n+2r+\ell_2+\frac52\right)^2+\mu^2
\right]} .
\label{eq:dS_cmn_general}
\end{equation}
Hence, the particular solution, kept explicitly as a series, is
\begin{align}
&\widehat{\mathcal I}^{\,\ell_1\ell_2}_{\pm \pm,\mathrm{part}}(u,v)
={}
H^2 e^{\mp i\pi(\ell_1+\ell_2)/2}\Gamma(5+\ell_1+\ell_2)
\label{eq:dS_particular_series}\\
&\qquad \times
\sum_{m,n=0}^{\infty}
(-1)^n
\frac{(5+\ell_1+\ell_2)_{n+2m}}
{n!\displaystyle\prod_{r=0}^{m}
\left[
\left(n+2r+\ell_2+\frac52\right)^2+\mu^2
\right]}
%\\&\hspace{1.2cm}
\times
u^{5+\ell_1+\ell_2+2m}
\left(\frac uv\right)^n,
\qquad u<v \ . \notag
\end{align}
The full time-ordered seed is the sum of a homogeneous contribution and the
particular solution, 
\begin{equation}
    \hat{\mathcal I}^{\ell_1\ell_2}_{\pm \pm}(u,v)
    =
    \hat{\mathcal I}^{\ell_1\ell_2}_{\pm \pm ,\mathrm{hom}}(u,v)
    +
    \hat{\mathcal I}^{\ell_1\ell_2}_{\pm \pm ,\mathrm{inh}}(u,v).
\end{equation}
Before imposing the Schwinger--Keldysh boundary conditions, the homogeneous
sector should be written as the most general linear combination of products
of homogeneous solutions. By the Bunch--Davies condition, the requirement that the singular part of the particular solution be canceled in the folded limit (Appendix \ref{app:ds_bootstrap_details}). For the time-ordered Bunch--Davies correlator, this choice leads to the combination
\begin{equation}
    e^{\pi\mu}\mathcal Y^{\ell_1}_{+}(u)
    +
    e^{-\pi\mu}\mathcal Y^{\ell_1}_{-}(u)
\end{equation}
in the \(u\)-channel. Therefore, the full solution may be written as,
\begin{equation}
\begin{aligned}
&\hat{\mathcal I}^{\ell_1\ell_2}_{\pm\mp}(u,v)
=
\frac{e^{\mp i\pi(\ell_1-\ell_2)/2}}{4\pi}
\left[\mathcal Y^{\ell_1}_+(u)+\mathcal Y^{\ell_1}_-(u)\right]
\left[\mathcal Y^{\ell_2}_+(v)+\mathcal Y^{\ell_2}_-(v)\right],\\
&\hat{\mathcal I}^{\ell_1\ell_2}_{\pm\pm}(u,v)
=
\frac{\pm i\,e^{\mp i\pi\ell_{12}/2}}{4\pi}
\left[
e^{\pi\mu}\mathcal Y^{\ell_1}_{\pm}(u)
+
e^{-\pi\mu}\mathcal Y^{\ell_1}_{\mp}(u)
\right]
\left[
\mathcal Y^{\ell_2}_{+}(v)
+
\mathcal Y^{\ell_2}_{-}(v)
\right]
\\
& \qquad \qquad \quad + \hat{\mathcal I}^{\ell_1\ell_2}_{\pm\pm,\mathrm{inh}}(u,v).
\end{aligned}
\end{equation}
In the collapsed limit, the non-analytic part of the total
Schwinger--Keldysh seed reproduces the usual de Sitter
cosmological-collider behavior. For \(\ell_1=\ell_2=-2\), one finds
\begin{align}
\begin{aligned}
\widehat{\mathcal I}_{\rm seed,\,non\text{-}an}^{-2,-2}(u,v)
\Big|_{\rm coll}
={}&
\frac{\pi\,\operatorname{sech}(\pi\mu)}{2\mu}
(uv)^{\frac12-i\mu}
\left[
u^{2i\mu}\left(\operatorname{csch}(\pi\mu)+i\right)
+
v^{2i\mu}\left(\operatorname{csch}(\pi\mu)-i\right)
\right]
\\[1mm]
&+
\frac{
2^{-1-2i\mu}
\left(1+i\sinh(\pi\mu)\right)
\Gamma(-i\mu)^2
\Gamma\!\left(\frac12+i\mu\right)^2
}{
\pi
}
(uv)^{\frac12+i\mu}
\\[1mm]
&+
\frac{
2^{-1+2i\mu}
\left(1-i\sinh(\pi\mu)\right)
\Gamma(i\mu)^2
\Gamma\!\left(\frac12-i\mu\right)^2
}{
\pi
}
(uv)^{\frac12-i\mu}.
\end{aligned}
\label{eq:dS_seed_collapsed_limit}
\end{align}
Equivalently, the first line can be written as
\begin{equation}
\frac{\pi\,\operatorname{sech}(\pi\mu)}{2\mu}
(uv)^{\frac12}
\left[
\left(\frac{u}{v}\right)^{i\mu}
\left(\operatorname{csch}(\pi\mu)+i\right)
+
\left(\frac{u}{v}\right)^{-i\mu}
\left(\operatorname{csch}(\pi\mu)-i\right)
\right],
\end{equation}
which makes the usual collapsed-limit oscillatory dependence explicit. The
remaining two terms are the two conjugate scaling branches,
\((uv)^{1/2+i\mu}\) and \((uv)^{1/2-i\mu}\), with the expected thermal
coefficients. Thus, the non-analytic part of the seed recovers the standard
de Sitter cosmological-collider structure in the collapsed limit.

%%%%%%%%%%%%%%%%%%%%%%%%%%%%%%
%%%%%%%%%%%%%%%%%%%%%%%%%%%%%%
%%%%%%%%%%%%%%%%%%%%%%%%%%%%%%
\section{The Non-linear Bootstrap}\label{sec:ghost_bootstrap}

We now turn to the bootstrap approach for the pure ghost spectator field. The field-theory setup was already introduced in the subsection~\ref{subsec:ghost_condensate_scalar} on the ghost-condensate scalar field. The exchanged field is the massive mode \(\chi\) whose quadratic equation, Bunch–-Davies solution, and late-time weights were given in Eqs.~\eqref{eq:mode-eom2}--\eqref{eq:sigma-mode2}. We shall not repeat those mode functions here. The only property that we would like to stress here is the early-time ghost phase contained in Eq.~\eqref{eq:sigma-mode2}.
\begin{equation}
    \chi_k(\eta)\sim
    e^{\!+\frac{i}{2}\gamma k^2\eta^2}
    \label{eq:ghost_early_phase_recall}
\end{equation}
Here \(\gamma=H/M\) is the parameter already introduced with the ghost mode above. This early-time behavior contrasts with ordinary de Sitter exchange fields. In the relativistic case, the early-time phases are linear in \(s\eta\), and the bootstrap operator is second order. In the pure ghost case, the phase is quadratic in \(s\eta\), and the corresponding seed equation becomes fourth order.

%%%%%%%%%%%%%%%%%%%%%%%%%%%%%%
%%%%%%%%%%%%%%%%%%%%%%%%%%%%%%
%%%%%%%%%%%%%%%%%%%%%%%%%%%%%%
\subsection{Cross-ratios and singular limits}
\label{subsec:ghost_cross_ratios_singularities}

As in the de Sitter section, we work in the exchange channel $s\equiv \lvert \boldsymbol{k}_1+\boldsymbol{k}_2\rvert = - \lvert\boldsymbol{k}_3+\boldsymbol{k}_4\rvert $ and use the cross ratios $ u={s}{/k_{12}}$ and $v= {s}/{k_{34}}$. The rescaling is the same, \(z_1=k_{12}\eta_1\) and \(z_2=k_{34}\eta_2\), which makes the argument of the ghost mode function take the form $\gamma u^2z_1^2$ and $\gamma v^2z_2^2$. Thus, the same dimensionless variables \((u,v)\) remain natural, even though the analytic structure of these variables changes. This change is already evident at the level of a single vertex. The left vertex is controlled by the elementary integral
\begin{equation}
    \mathcal V_\nu(k_{12},s)
    =
    \int_0^\infty dx\,x^\nu
    \exp\!\left[-ik_{12}x-\frac{i}{2}\gamma s^2x^2\right].
    \label{eq:ghost_vertex_integral_x_short}
\end{equation}
Using the standard parabolic-cylinder integral, one obtains
\begin{equation}
    \mathcal V_\nu(k_{12},s)
    =
    k_{12}^{-\nu-1}\,
    \Gamma(\nu+1)
    (i\gamma u^2)^{-\frac{\nu+1}{2}}
    \exp\!\left[\frac{i}{4\gamma u^2}\right]
    D_{-\nu-1}\!\left(
    \frac{e^{i\pi/4}}{\sqrt\gamma\,u}
    \right) .
\label{eq:ghost_vertex_integral_exact_short}
\end{equation}
The soft exchange limit \(u\to0\) corresponds to \(s to 0\) with \(k_{12}\neq 0\), and the large-argument asymptotic of \(D_{-\nu-1}\) cancels the exponential prefactor to give
\begin{equation}
    \mathcal V_\nu(k_{12},s)
    \xrightarrow{u\to0}
    e^{-i\pi(\nu+1)/2}
    \frac{\Gamma(\nu+1)}{k_{12}^{\nu+1}} .
    \label{eq:ghost_vertex_soft_limit_short}
\end{equation}
By contrast, the limit \(k_{12}\to0\) at fixed \(s\) corresponds to \(u\to\infty\), in which case
\begin{equation}
    \mathcal V_\nu(0,s)
    =
    \frac12
    \left(\frac{i\gamma s^2}{2}\right)^{-\frac{\nu+1}{2}}
    \Gamma\!\left(\frac{\nu+1}{2}\right)
    \propto s^{-\nu-1} .
    \label{eq:ghost_vertex_Kzero_limit_short}
\end{equation}
Therefore, distinct from the relativistic de Sitter singular loci \(u=\pm1\), which encode the folded and factorization channels, in the pure ghost case, the nontrivial endpoints suggested by the vertex analysis are located at \(u=0\) and \(u=\infty\).

%%%%%%%%%%%%%%%%%%%%%%%%%%%%%%
%%%%%%%%%%%%%%%%%%%%%%%%%%%%%%
%%%%%%%%%%%%%%%%%%%%%%%%%%%%%%
\subsection{Differential structure}
\label{subsec:ghost_differential_structure}

The seed is defined by the same Schwinger--Keldysh integral as in Eq.~\eqref{Seedgeralzero}, but with the pure ghost bulk-to-bulk propagator. Acting with the ghost equation of motion on the left endpoint and commuting the result through the time integral gives a differential equation in \(u\). It is useful to introduce the Euler operator
\begin{equation}
    \theta\equiv u\partial_u .
\end{equation}
The integration-by-parts relation can then be written directly in terms of the function that enters the seed, this will give us some vertex function $Y(u)$ multiplied by some momentum ratio $u$,
\begin{equation}
\int_{-\infty}^{0}dz\, (-z)^\ell e^{iaz}z^4 F\!\left(\frac{\gamma u^2z^2}{2}\right)
= \displaystyle\prod_{k=1}^{4}  (\theta+\ell+k)
    \left[u^{-\ell-1}Y(u)\right].
\end{equation}
After restoring the seed prefactor, the pure ghost bootstrap operator is
\begin{equation}
    \mathcal G_u^\ell
    =
    \left[
    \theta-\left(\ell+\frac52\right)
    \right]^2
    +
    \mu^2
    +
    \gamma^2u^4\theta(\theta+1)(\theta+2)(\theta+3).
\label{eq:ghost_bootstrap_operator_theta}
\end{equation}
Equivalently,
\begin{equation}
\begin{aligned}
    \mathcal G_u^\ell
    ={}&
    \gamma^2\left(u^8\partial_u^4+12 u^7 \partial_u^3 +24\gamma^2u^5 \partial_u\right)
    +\left(u^2+36\gamma^2u^6\right)\partial_u^2
    %\\&
    -2(\ell+2)u \partial_u
    +\mu^2+\left(\ell+\frac52\right)^2 .
\end{aligned}
\label{eq:ghost_bootstrap_operator_derivative}
\end{equation}
The Wightman components obey
\begin{equation}
    \mathcal G_u^{\ell_1}
    \hat{\mathcal I}^{\ell_1\ell_2}_{ab}(u,v)=0,\qquad a\neq b,
    \label{eq:ghost_wightman_bootstrap}
\end{equation}
For the time-ordered components, the source depends on the sum of the two external weights.  Since this combination occurs only in the following equation in the present subsection, we keep it explicit rather than introducing another abbreviation:
\begin{equation}
    \mathcal G_u^{\ell_1}
    \hat{\mathcal I}^{\ell_1\ell_2}_{aa}(u,v)
    =
    H^2 e^{-a i\pi(\ell_1+\ell_2)/2}
    \Gamma(5+\ell_1+\ell_2)
    \left(\frac{uv}{u+v}\right)^{5+\ell_1+\ell_2}.
    \label{eq:ghost_time_ordered_bootstrap}
\end{equation}
For the same-sign components it is useful to separate the homogeneous freedom from one particular solution of the sourced equation,
\begin{equation}
\hat{\mathcal I}^{\ell_1\ell_2}_{aa}(u,v)
=
\hat{\mathcal I}^{\ell_1\ell_2}_{aa,{\rm hom}}(u,v)
+
\hat{\mathcal I}^{\ell_1\ell_2}_{aa,{\rm part}}(u,v) .
\label{eq:k4_source_decomposition}
\end{equation}
The first term obeys the homogeneous fourth-order equation, whereas the second obeys Eq.~\eqref{eq:ghost_time_ordered_bootstrap}.  
The singularity classification follows directly from the derivative form of the fourth-order equation.  At any finite point $u_0\neq0$, the coefficient of the highest derivative is $\gamma^2u_0^8$, which is nonzero. Therefore, every finite point away from the origin is an ordinary point; in particular, the de Sitter folded and factorization loci $u=1$ and $u=-1$ are ordinary points of the quartic homogeneous equation. The normalized equation is obtained by dividing Eq.~\eqref{eq:ghost_bootstrap_operator_derivative} by $\gamma^2u^8$, and we identify the following polynomials
\begin{equation}
    P_3(u)=\frac{12}{u}
    \ , \quad
    P_2(u)=\frac{36}{u^2}+\frac{1}{\gamma^2u^6}
    \ , \quad
    P_1(u)=\frac{24}{u^3}-\frac{2(\ell+2)}{\gamma^2u^7}
    \ , \quad
    P_0(u)=\frac{\mu^2+(\ell+5/2)^2}{\gamma^2u^8} .
\end{equation}
For a fourth-order equation, a regular singular point requires \(u^{4-k}P_k(u) \), for $k=0,1,2,3$, to be analytic at $u=0$. We conclude that $u=0$ is an irregular singular point. Finally, the point at infinity is classified by setting
\begin{equation}
    x\equiv\frac1u,
    \qquad
    \vartheta\equiv x\partial_x=-\theta .
    \label{eq:k4_x_vartheta}
\end{equation}
The homogeneous operator becomes
\begin{equation}
\mathcal G_x^\ell
=\gamma^2\partial_x^4+x^2\partial_x^2+(2\ell+6)x\partial_x
+\left(\ell+\frac52\right)^2+\mu^2 .
\label{eq:k4_hom_x_ordinary}
\end{equation}
The coefficient of the highest derivative at $x=0$ is the nonzero constant
$\gamma^2$, and all other coefficients are analytic. Hence,
\begin{equation}
    u=0\quad\text{is an irregular singular point,}
    \qquad
    u=\pm1\ \text{and}\ u=\infty\quad\text{are ordinary points.}
\label{eq:ghost_singularity_summary}
\end{equation}
This is the central analytic distinction between the pure ghost and
ordinary de Sitter bootstrap equations. The irregular singularity also gives rise to the sectorial asymptotic
structure illustrated in Fig.~\ref{fig:k4_stokes}. As shown below,
in addition to the two algebraic collider branches, the homogeneous
equation admits two WKB branches proportional to
$u^{-\ell-1/2}\exp[\pm i/(2\gamma u^2)]$. Writing
$u=|u|e^{i\theta}$, for $\gamma>0$, their exponential factors satisfy
\begin{equation}
\left|\exp\left(\pm\frac{i}{2\gamma u^2}\right)\right|
=\exp\left(\pm\frac{\sin(2\theta)}{2\gamma|u|^2}\right).
\end{equation}
The red and blue regions therefore correspond to alternating WKB
growth and decay. Both exponential functions are oscillatory along the coordinate axes, while the dashed diagonals indicate directions of maximal exponential dominance (Stokes rays in the convention adopted here). More generally,
analytic continuation between sectors can change the asymptotic
representation of a solution by revealing exponentially subdominant
contributions. This is the Stokes phenomenon
\cite{OlverAsymptotics1974}.

The sectorial structure does not, by itself, impose a physical boundary
condition or determine the Stokes multipliers. Along the physical
direction $u\to0^+$, both WKB branches are oscillatory; for $\ell=-2$
, they are even sub-leading to the collider power laws. Consequently,
matching the leading collider scaling does not guaranty the absence
of WKB contributions. The boundary condition imposed below is stronger:
it sets both WKB coefficients to zero in the positive-real-axis
asymptotic expansion, after which the collider coefficients are fixed
by the Bunch--Davies-normalized mode functions.

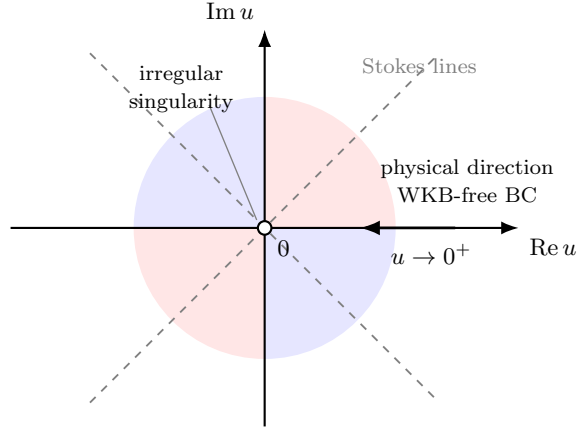
\begin{figure}[t]
\centering
\begin{tikzpicture}[
    >=Latex,
    scale=1.05,
    every node/.style={font=\footnotesize},
    axis/.style={->, line width=0.8pt},
    stokes/.style={dashed, line width=0.7pt, gray},
    sectorlabel/.style={align=center, font=\scriptsize},
    leader/.style={gray, line width=0.5pt},
    openpoint/.style={circle, draw=black, fill=white, line width=0.8pt, inner sep=1.8pt}
]

%--- shaded sectors ---
\fill[red!10]   (0,0) -- (1.65,0) arc[start angle=0,end angle=90,radius=1.65] -- cycle;
\fill[blue!10]  (0,0) -- (0,1.65) arc[start angle=90,end angle=180,radius=1.65] -- cycle;
\fill[red!10]   (0,0) -- (-1.65,0) arc[start angle=180,end angle=270,radius=1.65] -- cycle;
\fill[blue!10]  (0,0) -- (0,-1.65) arc[start angle=270,end angle=360,radius=1.65] -- cycle;

%--- axes ---
\draw[axis] (-3.2,0) -- (3.2,0) node[below right] {$\mathrm{Re}\,u$};
\draw[axis] (0,-2.5) -- (0,2.5) node[above left] {$\mathrm{Im}\,u$};

%--- Stokes lines ---
\draw[stokes] (-2.2,-2.2) -- (2.2,2.2);
\draw[stokes] (-2.2,2.2) -- (2.2,-2.2);

%--- origin ---
\draw[line width=0.8pt] (0,0);
\node[openpoint] at (0,0) {};
\node[below right=1pt] at (0,0) {$0$};

%--- label for origin ---
\draw[leader] (-0.10,0.10) -- (-0.70,1.55);
\node[sectorlabel] at (-1.05,1.75) {irregular\\singularity};

%--- physical positive real direction ---
%\draw[line width=1.2pt] (0,0) -- (2.55,0);
%\fill (2.55,0) circle (1.5pt);
%\node[openpoint] at (2.55,0) {};
\draw[->, line width=1.0pt] (2.4,0) -- (1.2,0);
\node[below=2pt] at (2.1,0) {$u\to0^+$};
%--- label for physical direction / BC ---
%\draw[leader] (2.55,0.08) -- (2.55,0.8);
\node[sectorlabel] at (2.55,0.6) {physical direction\\WKB-free BC};

%--- sector labels ---
\node[sectorlabel, text=red!70!black] at (1.05,0.95)
{};

\node[sectorlabel, text=blue!70!black] at (-1.05,0.95)
{%$\exp\!\left(-\dfrac{i}{2\gamma u^2}\right)$ grows\\
%$\exp\!\left(+\dfrac{i}{2\gamma u^2}\right)$ decay
};

\node[sectorlabel, text=red!70!black] at (-1.05,-0.95)
{%$e^{\frac{i}{2\gamma u^2}}$ \\
 %$e^{-\frac{i}{2\gamma u^2}}$
};

\node[sectorlabel, text=blue!70!black] at (1.05,-0.95)
{%$e^{\frac{i}{2\gamma u^2}}$ \\
%$e^{-\frac{i}{2\gamma u^2}}$
};

%--- optional small label for dashed lines ---
\node[font=\scriptsize, gray] at (1.95,2.05) {Stokes lines};
\node[font=\scriptsize, gray, rotate=-45] at (2.0,-1.95) {};

\end{tikzpicture}

\caption{
Stokes structure near the irregular singular point $u=0$
for quartic dispersion ($\gamma>0$).
The red and blue regions represent alternating WKB sectors.
In the red sectors, the branch
$e^{+i/(2\gamma u^2)}$ is exponentially enhanced,
while $e^{-i/(2\gamma u^2)}$ is suppressed;
the opposite occurs in the blue sectors.
The dashed diagonals indicate the Stokes lines,
along which the exponential dominance is maximal,
whereas the coordinate axes correspond to oscillatory
directions, where both branches have unit modulus.
Along the physical direction $u\to0^+$,
our boundary condition requires the absence of both
WKB contributions, selecting the purely
collider-type asymptotic behavior.
}
\label{fig:k4_stokes}

\end{figure}

\subsection{Homogeneous bootstrap solution}
\label{subsec:k4_homogeneous_solution}

From the vertex analysis in Eq.~\eqref{eq:ghost_vertex_Kzero_limit_short}, we see that the irregular singular point corresponds to the asymptotic limit of the mathematical representation of the vertex function. The nonlinear dispersion relation therefore appears to invert the usual role of the cross-ratio limits and, at the same time, makes the higher-point correlators more regular by removing the finite singular points that characterize the relativistic de Sitter case. The appropriate boundary condition must then be reformulated accordingly, with the analog of the Bunch-Davies prescription imposed at the asymptotic endpoint selected by the quartic dispersion relation. We now study the homogeneous equation
\begin{equation}
\mathcal G_u^\ell \, \mathcal F(u)=0
\end{equation}
in the limit \(u\to0\). Since the differential equation is fourth order, four independent asymptotic behaviors are expected. We first look for power-law solutions,
\begin{equation}
\mathcal F(u)\sim u^\rho .
\end{equation}
Keeping only leading order terms in \(u\), the homogeneous equation becomes
\begin{equation}
\left[
\rho(\rho-1)
-2(\ell+2)\rho
+\mu^2
+\left(\ell+\frac52\right)^2
\right]u^\rho=0 .
\end{equation}
Hence, \( \rho=\ell+\frac52\pm i\mu \) and the first two independent asymptotic behaviors are
\begin{equation}
\mathcal F(u)\sim
u^{\ell+\frac52+i\mu},
\qquad
\mathcal F(u)\sim
u^{\ell+\frac52-i\mu}.
\end{equation}
These two power-law solutions do not exhaust the four independent solutions of the fourth-order differential equation. The remaining two solutions have an essential exponential behavior around \(u=0\). To see this, consider the ansatz
\begin{equation}
\mathcal F(u)\sim
u^\rho
\exp\left(\frac{A}{u^2}\right).
\end{equation}
The two most singular contributions behave as
\begin{equation}
u^2\partial_u^2\mathcal F
\sim
4A^2u^{-4}\mathcal F
\quad , \qquad \gamma^2u^8\partial_u^4\mathcal F
\sim
16\gamma^2A^4u^{-4}\mathcal F .
\end{equation}
Hence, the leading \(u^{-4}\) contribution to the differential equation gives \( 16\gamma^2A^4+4A^2=0 \). For the nontrivial exponential branches, \(A\neq0\), this implies \(A=\pm{i}/{2\gamma} \), which fixes the exponential behavior of the solution. To determine the power-law prefactor, we substitute
\begin{equation}
\mathcal F(u)\sim
u^\rho
\exp\left(
\pm\frac{i}{2\gamma u^2}
\right)
\end{equation}
into the differential equation and consider the next order after the leading \(u^{-4}\) cancellation. The coefficient of the \(u^{-2}\) contribution gives \( 4\rho+4\ell+2=0 \), and hence \( \rho=-(\ell+1/2) \). The two WKB-type asymptotic solutions are therefore
\begin{equation}
\mathcal F(u)\sim
u^{-\ell-\frac12}
\exp\left(
+\frac{i}{2\gamma u^2}
\right)
\quad , \quad
\mathcal F(u)\sim
u^{-\ell-\frac12}
\exp\left(
-\frac{i}{2\gamma u^2}
\right) \quad .
\end{equation}

Consequently, the four independent asymptotic behaviors around \(u=0\) are
\begin{equation}
u^{\ell+\frac52+i\mu},
\qquad
u^{\ell+\frac52-i\mu},
\qquad
u^{-\ell-\frac12}
\exp\left(
+\frac{i}{2\gamma u^2}
\right),
\qquad
u^{-\ell-\frac12}
\exp\left(
-\frac{i}{2\gamma u^2}
\right) \, 
\end{equation}
which contains two power-law branches, together with two rapidly oscillating WKB branches. The appearance of the exponential function \( \exp\left( \pm {i}/{2\gamma u^2} \right) \) is the characteristic asymptotic behavior associated with the irregular singular point at \(u=0\). This behavior is not present at the collapsed limit of the four-point function \cite{Ferreira:2026tyj}; hence, we must set the boundary conditions to recover only the collider branches at this asymptotic limit. Since \(u=0\) corresponds to an asymptotic limit and is an irregular singular point, it is convenient to construct the global homogeneous solution from the opposite endpoint. In the variable \(x=1/u\), the point \(u=\infty\), corresponding to \(x=0\), is an ordinary point of the differential equation. The homogeneous solution can therefore be organized as an ordinary Taylor expansion around \(x=0\), providing a convenient global basis in which the physical boundary conditions can subsequently be imposed. 

In terms of \(\vartheta=x\partial_x\), the homogeneous equation reads
\begin{equation}
\left\{
\gamma^2(-\vartheta)(1-\vartheta)(2-\vartheta)(3-\vartheta)
+x^4\left[\left(\vartheta+\ell+\frac52\right)^2+\mu^2\right]
\right\}Y(x)=0 .
\label{eq:k4_hom_x}
\end{equation}
Assuming a solution of the form
\begin{equation}
    Y(x)=\sum_{m=0}^{\infty}c_mx^m \ ,
    \label{eq:k4_taylor_infty}
\end{equation}
and substitution into Eq.~\eqref{eq:k4_hom_x} we get
\begin{equation}
    \gamma^2(m+4)(m+3)(m+2)(m+1)c_{m+4}
    +\left[\left(m+\ell+\frac52\right)^2+\mu^2\right]c_m=0 \ .
    \label{eq:k4_taylor_rec_general}
\end{equation}
The equation couples coefficients that differ in the fourth power. Therefore, we have four independent congruence classes associated with the four Taylor data $c_0,c_1,c_2,c_3$.  Another interesting manner to express these solutions is to use the ansatz
\begin{equation}
    Y_r^{(\ell)}(x)=x^r\sum_{n=0}^{\infty}a_n^{(r,\ell)}x^{4n},
    \qquad
    r=0,1,2,3,
    \qquad
    a_0^{(r,\ell)}=1 \ ,
\end{equation}
which, for $n\geq1$, gives 
\begin{equation}
\frac{a_n^{(r,\ell)}}{a_{n-1}^{(r,\ell)}}
=-\frac{\left(4n+r+\ell-\frac32\right)^2+\mu^2}
{\gamma^2(r+4n)(r+4n-1)(r+4n-2)(r+4n-3)} .
\label{eq:k4_taylor_class_rec_infty}
\end{equation}
This recurrence relation can be resumed into four ${}_2F_3$ functions. Returning to the $u=1/x$ variable, we find
\begin{equation}
\begin{aligned}
 h_0^{(\ell)}(u)={}&{}_2F_3\!\left(
 \begin{matrix}\frac{\ell+\frac52}{4}+\frac{\ii\mu}{4},\frac{\ell+\frac52}{4}-\frac{\ii\mu}{4}\\
 \frac14,\frac12,\frac34\end{matrix};-
 \frac{1}{16\gamma^2u^4}\right),\\
 h_1^{(\ell)}(u)={}&\frac1u{}_2F_3\!\left(
 \begin{matrix}\frac{\ell+\frac72}{4}+\frac{\ii\mu}{4},\frac{\ell+\frac72}{4}-\frac{\ii\mu}{4}\\
 \frac12,\frac34,\frac54\end{matrix};-
 \frac{1}{16\gamma^2u^4}\right),\\
 h_2^{(\ell)}(u)={}&\frac1{u^2}{}_2F_3\!\left(
 \begin{matrix}\frac{\ell+\frac92}{4}+\frac{\ii\mu}{4},\frac{\ell+\frac92}{4}-\frac{\ii\mu}{4}\\
 \frac34,\frac54,\frac32\end{matrix};-
 \frac{1}{16\gamma^2u^4}\right),\\
 h_3^{(\ell)}(u)={}&\frac1{u^3}{}_2F_3\!\left(
 \begin{matrix}\frac{\ell+\frac{11}{2}}{4}+\frac{\ii\mu}{4},\frac{\ell+\frac{11}{2}}{4}-\frac{\ii\mu}{4}\\
 \frac54,\frac32,\frac74\end{matrix};-
 \frac{1}{16\gamma^2u^4}\right).
\end{aligned}
\label{eq:k4_hr}
\end{equation}
The label $r$ is not an indicial exponent of a singular point; it labels the four ordinary Taylor classes modulo four.  Equivalently, the four leading behaviors at $u=\infty$ are $1,u^{-1},u^{-2},u^{-3}$.

%%%%%%%%%%%%%%%%%%%%%%%%%%%%%%%
%%%%%%%%%%%%%%%%%%%%%%%%%%%%%%%
%%%%%%%%%%%%%%%%%%%%%%%%%%%%%%%
\subsubsection{Asymptotic boundary condition}

The fourth-order nature of the bootstrap equation implies that its general homogeneous solution is a linear combination of four linearly independent solutions, with four undetermined coefficients to be fixed by appropriate boundary conditions. Since \(u=0\) is an irregular singular point, the asymptotic expansion of Eq.~\eqref{eq:k4_hr} is not exhausted by power-law solutions. We use the large-argument asymptotics of ${}_2F_3$
collected in Appendix~\ref{app:hypergeometric_q_less_p}, especially
Eq.~\eqref{eq:q_p_minus_one_asymptotic}, to separate the
oscillatory and algebraic sectors. The four-dimensional solution space separates into
two oscillatory WKB branches and two nonanalytic collider branches,
\begin{equation}
\mathcal{Y}^{\ell}(u)
\underset{u\to0}{\sim}
\mathcal{Y}_{\rm WKB}^{(+)}(u)
+
\mathcal{Y}_{\rm WKB}^{(-)}(u)
+
\mathcal{Y}_{\rm coll}^{(+)}(u)
+
\mathcal{Y}_{\rm coll}^{(-)}(u),
\end{equation}
where the first two pieces contain the rapidly oscillating factors
\begin{equation}
\exp\left(\frac{i}{2\gamma u^2}\right),
\qquad
\exp\left(-\frac{i}{2\gamma u^2}\right),
\end{equation}
whereas the collider sector is characterized by
\begin{equation}
u^{\ell+\frac52+i\mu},
\qquad
u^{\ell+\frac52-i\mu}.
\end{equation}

The crucial challenge is to understand how we can restrict the space of solutions to only the physical solution by identifying the correct boundary condition. The differential equation in Eq.~\eqref{eq:ghost_bootstrap_operator_theta} does not, by itself, retain the information associated with the choice of mode functions or vacuum state. Rather, it follows from the equation of motion satisfied by the Schwinger--Keldysh propagator, Eq.~\eqref{eq:inhom-eomghost}, while the vacuum information must be supplied through an appropriate boundary condition.

In the standard de Sitter bootstrap, this information is encoded in the behavior of the correlator at the folded configuration. For modes prepared in the Bunch--Davies vacuum, the time-integrated correlator must be free of the unphysical folded singularity, and imposing its absence selects the corresponding physical solution. In the nonlinear case, the mode functions are normalized according to the same Bunch--Davies prescription, Eq.~\eqref{eq:sigma-mode2}. However, the higher-derivative structure modifies the singularity structure of the bootstrap equation, so that the folded-limit prescription no longer provides the natural way to formulate the boundary condition.

Nevertheless, the late-time behavior of the time-integrated nonlinear modes contains the same cosmological-collider data associated with the normalization of the underlying mode functions \cite{Ferreira:2026tyj}. The vacuum information is therefore still encoded in the asymptotic collider limit, although it appears through a different analytic structure. Matching the four-point function to this collider data consequently fixes the relative normalization of the global solutions and selects the physical solution of the bootstrap equation.

The appropriate asymptotic boundary condition for the \(k^4\) bootstrap problem is therefore to select the linear combination of global solutions for which both WKB branches are absent, while the remaining collider branches reproduce the coefficients dictated by the Bunch--Davies-normalized mode functions. Explicitly, we impose

\begin{equation}
\operatorname{Coeff}
\left[
\mathcal{Y}^{\ell}(u),
e^{-\frac{i}{2\gamma u^2}}
\right]
=0,
\qquad
\operatorname{Coeff}
\left[
\mathcal{Y}^{\ell}(u),
e^{+\frac{i}{2\gamma u^2}}
\right]
=0.
\label{eq:k4_wkb_boundary_condition}
\end{equation}
Since the asymptotic decomposition is sector-dependent, these WKB-free conditions are imposed along the physical direction \(u\to0^+\), with the branches of the hypergeometric asymptotics fixed accordingly.
These two conditions remove the two asymptotic solutions associated with the irregular singular point and project the four-dimensional homogeneous solution space onto its two-dimensional WKB-free subspace. They determine two of the four integration constants, which may be taken to be \(C_{3}\) and \(C_{4}\), in terms of \(C_{1}\) and \(C_{2}\). Substituting the resulting relations back into the asymptotic expansion gives
\begin{equation}
\mathcal{Y}^{\ell}_{\rm WKB}(u)=0
\end{equation}
identically. The remaining two constants are then fixed independently by matching the coefficients of the two nonanalytic collider branches to the late-time data inherited from the Bunch--Davies-normalized mode functions. The boundary prescription therefore has two distinct ingredients: cancellation of the two WKB branches and normalization of the two surviving collider branches.

With the normalization adopted throughout this work, the resulting asymptotic late-time solution becomes
\begin{align}
\mathcal{Y}^{\ell}_{\pm,\mathrm{hom}}(u)
\underset{u\to0}{\sim}
{}&
\frac{e^{\mp \pi(2i \ell + i3)/4}}{\sqrt{2\pi}}\,
\bigg\{ 2^{-1 + i \mu} \gamma^{-\frac{i\mu}{2}}\,
e^{\mp\frac{\pi\mu}{4}}
u^{\ell+5/2-i\mu} \Gamma\left(\frac{i\mu}{2}\right)
\Gamma\left(\ell+\frac52-i\mu\right)
\nonumber\\
&+\,2^{-1 - i \mu}\gamma^{\frac{i\mu}{2}}e^{\pm \frac{\pi \mu}{4}}u^{\ell+5/2+i\mu}
\Gamma\left(-\frac{i\mu}{2}\right)
\Gamma\left(\ell+\frac52+i\mu\right)
\bigg\}.
\label{eq:k4_collider_asymptotic}
\end{align}
The asymptotic formulas in Appendix~\ref{app:hypergeometric_q_less_p}
provide two linear constraints from the WKB coefficients and two from the
collider coefficients. Their solution is most economically expressed by
constructing a global basis in which each of the two collider powers is
normalized separately. The coefficients of the original four Taylor
blocks are displayed below in Eq.~\eqref{eq:phi-master}.
The global branch functions are fixed by the WKB-free conditions and the
Bunch--Davies collider coefficients, independently of any integral
representation of an individual time vertex (see
Appendix~\ref{app:sk_collapsed} for their direct Schwinger--Keldysh
normalization). We therefore write

\begin{equation}
\mathcal{Y}^{\ell}_{\pm,\mathrm{hom}}(u)
=
\frac{e^{\mp\pi(2i \ell + i3)/4}}{2\sqrt{2\pi}}\,
\bigg\{ e^{\mp \pi \mu /4}\,
\Phi_{-}^{\ell}(u)
+
e^{ \pm \pi \mu /4}\,
\Phi_{+}^{\ell}(u) \bigg\},
\label{eq:k4_pure_collider_asymptotic}
\end{equation}
where the two branch functions must be normalized such that
\begin{equation}
\Phi_{\pm}^{\ell}(u)
\underset{u\to0}{\sim}
2^{\mp i \mu} \gamma^{\pm i \mu / 2} u^{5/2 + \ell \pm i \mu}
\Gamma\left(\frac{5}{2} + \ell \pm i \mu\right)
\Gamma\left(\mp \frac{i \mu}{2} \right).
\label{eq:Phi_asymptotic_definition}
\end{equation}
Each asymptotic condition isolates a single collider exponent. Imposing these two conditions separately, therefore, promotes the two late-time branches to exact global homogeneous solutions,
\begin{equation}
\begin{aligned}
\Phi_{\pm}^{\ell}(u)
={}&
\frac{
(-1)^{1/8}2^\ell
e^{\frac{\pi}{4}(-i\ell+\mu)}
\gamma^{-\frac{11}{4}-\frac{\ell}{2}}
}
{
3\left(e^{\pi\mu}-1\right)u^3
}
\sum_{r=0}^{3}
C^{\pm}_r(\ell)\,
\left(u\sqrt{\gamma}\right)^{3-r}\Gamma\left(
\frac{5+2r+2\ell-2i\mu}{8}
\right)
\\
&\times
\Gamma\left(
\frac{5+2r+2\ell+2i\mu}{8}
\right)
{}_2F_3\left(
\begin{matrix}
\frac{5+2r+2\ell-2i\mu}{8},
\frac{5+2r+2\ell+2i\mu}{8}
\\
\boldsymbol{\beta}_r
\end{matrix};
-\frac{1}{16\gamma^2u^4}
\right).
\end{aligned}
\label{eq:phi-master}
\end{equation}
where
\begin{equation}
\begin{aligned}
\boldsymbol{\beta}_0&=
\left(\frac14,\frac12,\frac34\right),
&
\boldsymbol{\beta}_1&=
\left(\frac12,\frac34,\frac54\right),
\\
\boldsymbol{\beta}_2&=
\left(\frac34,\frac54,\frac32\right),
&
\boldsymbol{\beta}_3&=
\left(\frac54,\frac32,\frac74\right).
\end{aligned}
\end{equation}
The coefficients of the global homogeneous solutions are
determined by imposing four independent asymptotic
conditions. We first expand the four hypergeometric
blocks around $u\to 0^+$, using the large-argument
formulas collected in Appendix~A.1. Requiring the
coefficients of both oscillatory WKB branches,
$e^{\pm i/(2\gamma u^2)}$, to vanish yields two linear
relations among the four integration constants.
The remaining two constants are then fixed by
requiring that one collider branch vanishes while
the other reproduces the Bunch--Davies normalization
obtained from the Schwinger--Keldysh integrals in
Appendix~B. Repeating this procedure for the two
collider exponents determines the global solutions
$\Phi_\pm^\ell(u)$, with coefficients $C_r^\pm(\ell)$ given by
\begin{alignat}{2}
C^-_0(\ell) &=
-3i\left[
\sqrt{2}\,e^{\frac{i\pi\ell}{2}}
+(1-i)e^{\frac{\pi\mu}{2}}
\right],
\qquad &
C^+_0(\ell) &=
3\left[
(1+i)
+i\sqrt{2}\,e^{\frac{\pi}{2}(i\ell+\mu)}
\right],
\label{eq:phi-masterCpm0}\\[6pt]
C^-_1(\ell) &=
6\left[
(-1+i)e^{\frac{i\pi\ell}{2}}
+\sqrt{2}\,e^{\frac{\pi\mu}{2}}
\right],
\qquad &
C^+_1(\ell) &=
-6\left[
\sqrt{2}
-(1-i)e^{\frac{\pi}{2}(i\ell+\mu)}
\right],
\label{eq:phi-masterCpm1}\\[6pt]
C^-_2(\ell) &=
6\left[
\sqrt{2}\,e^{\frac{i\pi\ell}{2}}
-(1-i)e^{\frac{\pi\mu}{2}}
\right],
\qquad &
C^+_2(\ell) &=
-6\left[
(-1+i)
+\sqrt{2}\,e^{\frac{\pi}{2}(i\ell+\mu)}
\right],
\label{eq:phi-masterCpm2}\\[6pt]
C^-_3(\ell) &=
-4\left[
(1+i)e^{\frac{i\pi\ell}{2}}
+i\sqrt{2}\,e^{\frac{\pi\mu}{2}}
\right],
\qquad &
C^+_3(\ell) &=
4\left[
i\sqrt{2}
+(1+i)e^{\frac{\pi}{2}(i\ell+\mu)}
\right] \ .
\label{eq:phi-masterCpm3}
\end{alignat}

The resulting coefficients simultaneously cancel both WKB sectors and isolate a single normalized collider branch in each global solution. The two solutions thus form a natural basis for implementing the physical boundary conditions. The coefficients $C_r^\pm(\ell)$ defined Eqs.~\eqref{eq:phi-masterCpm0}-\eqref{eq:phi-masterCpm3} are the only coefficients needed for the global homogeneous solutions  Eq.~\eqref{eq:phi-master}.  Inserting that expression into Eq.~\eqref{eq:k4_pure_collider_asymptotic} fixes the four Taylor-block coefficients without introducing a second set of constants.  The asymptotic conditions and the normalized SK integrals are collected in Appendices~\ref{app:hypergeometric_q_less_p} and \ref{app:sk_collapsed}, respectively.

The basis \(\bigl(\Phi_{+}^{\ell},\Phi_{-}^{\ell}\bigr)\) has an additional useful reality property. For real \(u>0\), \(\gamma>0\), \(\mu\), and \(\ell\), and with the branch convention \((-1)^{1/8}=e^{i\pi/8}\) used above, the hypergeometric blocks and the products of Gamma functions appearing in Eq.~\eqref{eq:phi-master} are real. The coefficients of the two branches obey
\begin{equation}
C_r^{+}(\ell)
=
e^{\frac{i\pi}{4}(2\ell-1)}
\left[C_r^{-}(\ell)\right]^{*},
\qquad r=0,1,2,3,
\label{eq:Cpm_conjugation}
\end{equation}
while the common prefactor in Eq.~\eqref{eq:phi-master} carries precisely the compensating phase. Consequently,
\begin{equation}
\Phi_{+}^{\ell}(u)
=
\left[\Phi_{-}^{\ell}(u)\right]^{*},
\qquad
\Phi_{-}^{\ell}(u)
=
\left[\Phi_{+}^{\ell}(u)\right]^{*}.
\label{eq:Phi_hermitian_relation}
\end{equation}
Thus, the two functions form a complex-conjugate, or Hermitian, pair. This relation is already manifest in their late-time normalization since the two powers \(u^{\ell+5/2\pm i\mu}\) and their Gamma-function coefficients are mapped into one another by complex conjugation. In particular,
\begin{equation}
\Phi_{+}^{\ell}(u)+\Phi_{-}^{\ell}(u)
=2\,\mathrm{Re}\,\Phi_{+}^{\ell}(u),
\end{equation}
so the sum appearing repeatedly in the seed is real for real kinematics. Together with the fact that each \(\Phi_{\pm}^{\ell}\) is WKB-free and asymptotes to only one nonanalytic collider branch, Eq.~\eqref{eq:Phi_hermitian_relation} makes \(\bigl(\Phi_{+}^{\ell},\Phi_{-}^{\ell}\bigr)\) a particularly convenient basis for the physical homogeneous sector.

These functions constitute the building blocks of the factorized mixed-sign seed and, at the same time, provide the natural homogeneous basis for constructing the ordered solutions of the inhomogeneous bootstrap equation. The ordered same-sign solution contains two contributions: a particular solution fixed by the source term and a homogeneous contribution,
\begin{equation}
    \widehat{\mathcal I}_{\pm\pm}^{\ell_1\ell_2}(u,v) =\widehat{\mathcal I}_{\pm\pm,\mathrm{hom}}^{\ell_1\ell_2}(u,v)+\widehat{\mathcal I}_{\pm\pm,\mathrm{part}}^{\ell_1\ell_2,\mathrm{BL}}(u,v).
\end{equation}
The homogeneous term is fixed by the asymptotic boundary condition in Eq.~\eqref{eq:k4_wkb_boundary_condition} together with the collider data $\widehat{\mathcal I}_{\pm\pm,\mathrm{hom}}^{\ell_1\ell_2}(u,v) = \mathcal{Y}^{\ell_1}_{\pm \pm }(u)\mathcal{Y}^{\ell_2}_{\pm \mp }(v)$. The ordered Schwinger--Keldysh components carry an additional relative weight between the two collider branches, fixed by the Bunch--Davies normalization of the underlying mode functions, 
\begin{align}
\mathcal{Y}^{\ell}_{\pm \mp}(u)
\underset{u\to0}{\sim}
{}&
\frac{e^{\mp\pi(2i \ell + i3)/4}}{\sqrt{2\pi}}\,
\bigg\{ 2^{-1 + i \mu} \gamma^{-\frac{i\mu}{2}}\,
e^{\mp\frac{\pi\mu}{4}}e^{\pm\frac{b\pi\mu}{2}}
u^{\ell+5/2-i\mu} \Gamma\left(\frac{i\mu}{2}\right)
\Gamma\left(\ell+\frac52-i\mu\right)
\nonumber\\
&+\,2^{-1 - i \mu}\gamma^{\frac{i\mu}{2}}e^{\frac{\pm \pi \mu}{4}}e^{\frac{\mp \pi\mu}{2}}u^{\ell+5/2+i\mu}
\Gamma\left(-\frac{i\mu}{2}\right)
\Gamma\left(\ell+\frac52+i\mu\right)
\bigg\}.
\end{align}
Rewriting the same result in the branch-diagonal basis, we obtain
\begin{align}
\mathcal{Y}^{\ell}_{a,b}(u)
\underset{u\to0}{\sim}
{}&
\frac{e^{-a\pi(2i \ell + i3)/4}}{2\sqrt{2\pi}}\,
\bigg\{ e^{-\frac{a\pi\mu}{4}}e^{-\frac{b\pi\mu}{2}}
\Phi_{-}^{\ell}(u)+ e^{\frac{a\pi \mu}{4}}e^{\frac{b\pi\mu}{2}}
\Phi_{+}^{\ell}(u) 
\bigg\}, \qquad a,b = \pm .
\end{align}
Within the basis $(\Phi^{\ell}_{+},\Phi^{\ell}_{-})$, the homogeneous ordered solution therefore takes a form directly analogous to the de Sitter result: the dynamical information is carried by two universal conjugate branch functions, while the Schwinger--Keldysh prescription appears only through their relative coefficients. The remaining task is to determine the particular solution generated by the contact source, which we now construct in the ordered region $u<v$.

%%%%%%%%%%%%%%%%%%%%%%%%%%%%%%%%%%%%
%%%%%%%%%%%%%%%%%%%%%%%%%%%%%%%%%%%%
%%%%%%%%%%%%%%%%%%%%%%%%%%%%%%%%%%%%
\subsection{Particular solution from Borel--Laplace}
\label{sec:k4_borel_particular}

The equal-sign Schwinger--Keldysh components obey two independent sourced
bootstrap equations.  In contrast to an inverse-power construction at one
vertex, we seek a particular representative that solves \emph{both}
constraints at once.  For arbitrary real vertex weights $\ell_1,\ell_2$, the equations to be solved are
\begin{equation}
 \mathcal G_u^{\ell_1}\widehat{\mathcal I}^{\,\ell_1\ell_2}_{\pm \pm,\mathrm{part}}
 =\mathcal G_v^{\ell_2}\widehat{\mathcal I}^{\,\ell_1\ell_2}_{\pm \pm,\mathrm{part}}
 =H^2e^{\mp i\pi(\ell_1+\ell_2)/2}\Gamma(5+\ell_1+\ell_2)
 \left(\frac{uv}{u+v}\right)^{5+\ell_1+\ell_2}.
\label{eq:k4_bootstrap_source}
\end{equation}
We first assume $\gamma>0$, $\mu>0$, $u,v>0$, $\ell_1+\ell_2>-5$, and
$\ell_2>-5/2$; in particular, these conditions include the phenomenological weights $\ell_i\in\{-2,0\}$.  The role of these sufficient hypotheses and
possible analytic continuations is clarified in
Appendix~\ref{app:k4_borel_laplace}. Introduce $x=u^{-1}$ and $z=v^{-1}$.  Since
$u\partial_u=-x\partial_x$ and
$u^4\prod_{j=0}^3(u\partial_u+j)=\partial_x^4$,
the fourth-order operator in Eq.~\eqref{eq:ghost_bootstrap_operator_theta}
becomes
\begin{equation}
 \mathcal L_x^\ell
 =\gamma^2\partial_x^4+x^2\partial_x^2
 +2(\ell+3)x\partial_x
 +\mu^2+\left(\ell+\frac52\right)^2.
\label{eq:k4_bl_x_operator}
\end{equation}
The source simplifies to
$H^2e^{-a i\pi(\ell_1+\ell_2)/2}\Gamma(5+\ell_1+\ell_2)(x+z)^{-(5+\ell_1+\ell_2)}$. The Gamma--Laplace representation
\begin{equation}
 \frac{\Gamma(5+\ell)}{(x+z)^{5+\ell}}
 =\int_0^\infty dt\,e^{-(x+z)t}t^{4+\ell}
 \ , \quad \operatorname{Re}(5+\ell)>0\ ,
 \label{eq:k4_bl_source_laplace}
\end{equation}
motivates the integral ansatz
\begin{equation}
 \widehat{\mathcal I}^{\,\ell_1\ell_2,\mathrm{BL}}_{\pm \pm,\mathrm{part}}(u,v)
 =H^2e^{\mp i\pi(\ell_1+\ell_2)/2}
 \int_0^\infty dt\,e^{-t/u}B(t,1/v).
 \label{eq:k4_bl_ansatz}
\end{equation}
For a kernel with the endpoint behavior derived below, repeated integration
by parts (without boundary terms) gives
\begin{equation}
 \mathcal L_x^{\ell_1}\int_0^\infty dt\,e^{-xt}B(t,z)
 =\int_0^\infty dt\,e^{-xt}\mathcal P_t^{\ell_1}B(t,z),
 \label{eq:k4_bl_intertwining}
\end{equation}
where
\begin{equation}
 \mathcal P_t^\ell=t^2\partial_t^2-2(\ell+1)t\partial_t
 +\gamma^2t^4+\mu^2+\left(\ell+\frac32\right)^2.
 \label{eq:k4_bl_kernel_operator}
\end{equation}
It is therefore sufficient to find the solution of the \emph{second-order}
inhomogeneous equation
\begin{equation}
\mathcal P_t^{\ell_1}B(t,z)=t^{4+\ell_1+\ell_2}e^{-zt}.
\label{eq:k4_bl_kernel_equation}
\end{equation}
Complementary approaches to the analytic reconstruction of massive cosmological correlators include spectral and dispersive bootstrap methods, as well as loop-level differential and integral representations \cite{Extra:221103810,Extra:240712299,Extra:260602686,Extra:240902072,Extra:241113636,Extra:230814802,Extra:210914635}.
The Borel--Laplace construction is useful precisely because it replaces
an irregular fourth-order equation with a second-order equation for a kernel whose regular branch can be selected locally.  The Borel reconstruction of massive inflationary correlators in Ref.~\cite{LiXianyuResurgence2026} is closely related in spirit: there, Borel-plane singularities and contour deformations are used to recover cosmological-collider signals, whereas here the regular positive-axis kernel selects a particular representative, and the physical homogeneous matching remains separate.  For general background on divergent asymptotic series and Laplace reconstruction, see
Refs.~\cite{OlverAsymptotics1974,BalserDivergent1994,CostinBorel2008,BenderOrszag1999}.

The homogeneous equation associated with
Eq.~\eqref{eq:k4_bl_kernel_equation} is
$\mathcal P_t^\ell g(t)=0$.
To solve it, we introduce
\begin{equation}
    y=\frac{\gamma t^2}{2},
    \qquad
    g(t)=t^{\ell+3/2}f(y).
\end{equation}
Substituting into the homogeneous equation gives
\begin{equation}
    y^2 f''(y)+y f'(y)
    +\left(y^2+\frac{\mu^2}{4}\right)f(y)=0.
\end{equation}
This is the standard Bessel equation with imaginary
order $\nu=\pm i\mu/2$. Its two independent solutions
can be written in terms of Bessel functions of the
first kind, $f_\pm(y)=J_{\pm i\mu/2}(y)$.
Returning to the original variable $t$, we obtain
the following pair of homogeneous solutions:
\begin{equation}
    g_\pm^\ell(t)
    =t^{\ell+3/2}
    J_{\pm i\mu/2}\!
    \left(\frac{\gamma t^2}{2}\right).
    \label{eq:k4_bl_g_bessel}
\end{equation}
To construct the particular solution, we also need
their Wronskian,
\begin{equation}
\begin{aligned}
W[g_+^\ell,g_-^\ell](t)
=-\frac{4i\sinh(\pi\mu/2)}{\pi}
  t^{2\ell+2}.
\end{aligned}
\label{eq:k4_bl_wronskian}
\end{equation}
Variation of parameters then yields the inhomogeneous kernel selected by
the absence of independent homogeneous contributions at $t=0$,
\begin{equation}
\begin{aligned}
B(t,z)=\frac{\pi}{4i\sinh(\pi\mu/2)}\bigg[
 &g_+^{\ell_1}(t)
  \int_0^t ds\,s^{\ell_2-\ell_1}g_-^{\ell_1}(s)e^{-zs}
 \\
-&g_-^{\ell_1}(t)
  \int_0^t ds\,s^{\ell_2-\ell_1}g_+^{\ell_1}(s)e^{-zs}
 \bigg].
\end{aligned}
\label{eq:k4_bl_kernel_green}
\end{equation}
The fractional powers of the individual Bessel solutions cancel in this
combination.  More precisely, the kernel takes the Frobenius form
$B=t^{4+\ell_1+\ell_2}F(t,z)$, with $F$ analytic at $t=0$ and uniquely determined
by the source.  In particular, Eq.~\eqref{eq:k4_bl_kernel_green} is a
well-defined, single-valued function on the positive $t$ axis under our
assumptions.

Then, we have constructed a particular solution that satisfies both bootstrap equations. Its explicit form is, 
\begin{equation}
\begin{aligned}
\widehat{\mathcal I}^{\,\ell_1\ell_2,\mathrm{BL}}_{aa,\mathrm{part}}(u,v)
={}&\frac{\pi H^2e^{-a i\pi(\ell_1+\ell_2)/2}}{4i\sinh(\pi\mu/2)}
\int_0^\infty dt\,e^{-t/u}
\bigg[
 g_+^{\ell_1}(t)
 \int_0^t ds\,s^{\ell_2-\ell_1}g_-^{\ell_1}(s)e^{-s/v}
 \\
&\hspace{12mm}-
 g_-^{\ell_1}(t)
 \int_0^t ds\,s^{\ell_2-\ell_1}g_+^{\ell_1}(s)e^{-s/v}
\bigg] , \qquad a = \pm.
\end{aligned}
\label{eq:k4_bl_particular}
\end{equation}
For $\ell_1=\ell_2=-2$, $H=1$, this reduces exactly to the
Bessel--Laplace particular solution for the source $uv/(u+v)$.
For $z>0$, the inner integrals are finite and
$g_\pm^{\ell_1}(t)=O(t^{\ell_1+1/2})$ at large positive $t$, so the outer
Laplace integral converges for every $u>0$.  The detailed endpoint
analysis is given in Appendix~\ref{app:k4_borel_laplace}.

The simultaneous compatibility does not rely on a matching argument.
Indeed, writing $\alpha=\ell_1+\ell_2+4$, one checks directly that
\begin{equation}
 \mathcal P_t^{\ell_1}[t^\alpha e^{-zt}]
 =\mathcal L_z^{\ell_2}[t^\alpha e^{-zt}].
 \label{eq:k4_bl_source_intertwining}
\end{equation}
Since $\mathcal L_z^{\ell_2}$ commutes with $\mathcal P_t^{\ell_1}$,
the difference
$\mathcal L_z^{\ell_2}B-t^\alpha e^{-zt}$
solves the homogeneous kernel equation.  It also belongs to the same
$t^\alpha$-analytic class as $B$.  The homogeneous equation has no
nonzero solution in that class for real $\mu>0$ and $\ell_2>-5/2$;
therefore
\begin{equation}
\begin{aligned}
\mathcal L_z^{\ell_2}B&=t^{4+\ell_1+\ell_2}e^{-zt},\\
\mathcal G_u^{\ell_1}\widehat{\mathcal I}^{\mathrm{BL}}_{\mathrm{part}}
&=\mathcal G_v^{\ell_2}\widehat{\mathcal I}^{\mathrm{BL}}_{\mathrm{part}}
% \\&
=H^2e^{-a i\pi(\ell_1+\ell_2)/2}\Gamma(5+\ell_1+\ell_2)
\left(\frac{uv}{u+v}\right)^{5+\ell_1+\ell_2}.
\end{aligned}
 \label{eq:k4_bl_compatibility}
\end{equation}
The complete coefficient proof, including the absence of nonzero analytic
homogeneous terms, appears in Appendix~\ref{app:k4_borel_laplace}.

For comparison with the irregular-point asymptotics, define
$D_k^{(\ell)}\equiv (k+\ell+5/2)^2+\mu^2$.  The analytic kernel has the
convergent expansion
\begin{equation}
 B(t,z)=t^{4+\ell_1+\ell_2}\sum_{n,m=0}^\infty
 \frac{(-1)^{n+m}\gamma^{2m}z^nt^{n+4m}}{
  n!\prod_{j=0}^{m}D_{n+4j}^{(\ell_2)}}.
 \label{eq:k4_bl_kernel_series}
\end{equation}
Applying Watson's lemma at fixed $z>0$ generates the formal expansion
\begin{equation}
\begin{aligned}
\widehat{\mathcal I}^{\,\ell_1\ell_2,\mathrm{BL}}_{aa,\mathrm{part}}(u,v)
\sim{}&H^2e^{-a i\pi(\ell_1+\ell_2)/2}
\sum_{n,m=0}^\infty
 \frac{(-1)^{n+m}\gamma^{2m}\Gamma(5+\ell_1+\ell_2+n+4m)}
 {n!\prod_{j=0}^{m}D_{n+4j}^{(\ell_2)}}
 \\
 &\times u^{5+\ell_1+\ell_2+n+4m}v^{-n}.
\end{aligned}
\label{eq:k4_bl_asymptotic}
\end{equation}
This is an \emph{asymptotic}, generally divergent expansion; the convergent
object is the Laplace integral in Eq.~\eqref{eq:k4_bl_particular}.
The Gamma-factorial growth must not be confused with a convergent power
series at the irregular point.  The Borel interpretation and the uniform
fixed-ratio collapsed limit are discussed in the appendix.  In particular,
for fixed $0<u/v<1$, the leading algebraic term is
\begin{equation}
\begin{aligned}
 \widehat{\mathcal I}^{\,\ell_1\ell_2,\mathrm{BL}}_{aa,\mathrm{part}}(u,v)
 ={}&H^2e^{-a i\pi(\ell_1+\ell_2)/2}\Gamma(5+\ell_1+\ell_2)u^{5+\ell_1+\ell_2}\\[-1mm]
 &\times\sum_{n=0}^\infty
 \frac{(-u/v)^n(5+\ell_1+\ell_2)_n}{n![(n+\ell_2+\frac52)^2+\mu^2]}
 +O(u^{9+\ell_1+\ell_2}),\qquad 0<u/v<1.
\end{aligned}
 \label{eq:k4_bl_collapsed_fixed_ratio}
\end{equation}
The leading series can be identified directly in the Schwinger--Keldysh
integrals.  In particular, Appendix~\ref{app:sk_collapsed} evaluates all four
contour components with the positive-frequency mode of
Eq.~\eqref{eq:sigma-mode2}.  Its time-ordering correction gives exactly
Eq.~\eqref{eq:k4_bl_collapsed_fixed_ratio}, including its overall phase,
normalization and dependence on the fixed ratio $u/v$.  This establishes
that the Borel--Laplace representative has the correct leading
\emph{algebraic} collapsed behavior. 

The same direct Schwinger--Keldysh calculation also fixes the \emph{leading
collider} coefficients of the factorized pieces.~All the components of the seed function with the proper normalization is the following,
\begin{equation}
\begin{aligned}
\widehat{\mathcal I}_{\pm\mp}^{\ell_1\ell_2}(u,v)
&=
\frac{H^2e^{\mp i\pi(\ell_1-\ell_2)/2}}{8\pi}
\left[
e^{\mp\pi\mu/4}\Phi_-^{\ell_1}(u)
+e^{\pm\pi\mu/4}\Phi_+^{\ell_1}(u)
\right]
\\[-1mm]
&\quad\times
\left[
e^{\pm\pi\mu/4}\Phi_-^{\ell_2}(v)
+e^{\mp\pi\mu/4}\Phi_+^{\ell_2}(v)
\right],
\\[6pt]
\widehat{\mathcal I}_{\pm\pm}^{\ell_1\ell_2}(u,v)
&=
\frac{\pm iH^2e^{\mp i\pi(\ell_1+\ell_2)/2}}{8\pi}
\left[
e^{\mp3\pi\mu/4}\Phi_-^{\ell_1}(u)
+e^{\pm3\pi\mu/4}\Phi_+^{\ell_1}(u)
\right]
\\[-1mm]
&\quad\times
\left[
e^{\mp\pi\mu/4}\Phi_-^{\ell_2}(v)
+e^{\pm\pi\mu/4}\Phi_+^{\ell_2}(v)
\right]
+\widehat{\mathcal I}_{\pm\pm,\mathrm{part}}^{\ell_1\ell_2,\mathrm{BL}}(u,v).
\end{aligned}
\label{eq:k4_complete_seed}
\end{equation}
The coefficients $e^{a\pi\mu/4}$, $e^{3a\pi\mu/4}$ and the
factor $1/(8\pi)$ are inherited from the explicitly normalized
Schwinger--Keldysh integrals in Appendix~\ref{app:sk_collapsed}; they
are not inferred from the differential equation alone.  The particular
integral is given exactly in Eq.~\eqref{eq:k4_bl_particular}.  These
expressions specify a WKB-free global bootstrap continuation compatible
with \emph{all leading collapsed terms}.  Identifying that continuation
with the complete finite-$s$ Schwinger--Keldysh seed requires a separate
global connection check: leading collapsed data alone do not exclude
subleading WKB contributions or other homogeneous shifts.

Substituting Eq.~\eqref{eq:phi-master} into the two components
gives the seed directly in terms of the four hypergeometric blocks
$h_r^{(\ell)}$ of Eq.~\eqref{eq:k4_hr} and their original
coefficients $C_r^\pm(\ell)$:
\begin{align}
\widehat{\mathcal I}_{\pm\mp}^{\ell_1\ell_2}(u,v)
&=\frac{H^2 2^{\ell_1+\ell_2}}{72\pi(e^{\pi\mu}-1)^2}
 e^{\frac{i\pi}{4}+\frac{\pi}{4}(2\mu-i\ell_1-i\ell_2)
 \mp\frac{i\pi}{2}(\ell_1-\ell_2)}
 \sum_{r,s=0}^{3}
 \frac{h_r^{(\ell_1)}(u)h_s^{(\ell_2)}(v)}
 {\gamma^{(\ell_1+\ell_2+5+r+s)/2}}
 \notag\\[-1mm]
&\quad\times
 \prod_{c=\pm1}
 \Gamma\!\left(\frac{\ell_1+\frac52+r+c i\mu}{4}\right)
 \Gamma\!\left(\frac{\ell_2+\frac52+s+c i\mu}{4}\right)
 \notag\\[-1mm]
&\quad\times\left[e^{\mp\pi\mu/4}C_r^-(\ell_1)
 +e^{\pm\pi\mu/4}C_r^+(\ell_1)\right]
 \left[e^{\pm\pi\mu/4}C_s^-(\ell_2)
 +e^{\mp\pi\mu/4}C_s^+(\ell_2)\right].
\label{eq:k4_complete_seed_fully_expanded_mixed}
\end{align}

\begin{align}
\widehat{\mathcal I}_{\pm\pm}^{\ell_1\ell_2}(u,v)
&=\frac{\pm iH^2 2^{\ell_1+\ell_2}}{72\pi(e^{\pi\mu}-1)^2}
 e^{\frac{i\pi}{4}+\frac{\pi}{4}(2\mu-i\ell_1-i\ell_2)
 \mp\frac{i\pi}{2}(\ell_1+\ell_2)}
 \sum_{r,s=0}^{3}
 \frac{h_r^{(\ell_1)}(u)h_s^{(\ell_2)}(v)}
 {\gamma^{(\ell_1+\ell_2+5+r+s)/2}}
 \notag\\[-1mm]
&\quad\times
 \prod_{c=\pm1}
 \Gamma\!\left(\frac{\ell_1+\frac52+r+c i\mu}{4}\right)
 \Gamma\!\left(\frac{\ell_2+\frac52+s+c i\mu}{4}\right)
 \notag\\[-1mm]
&\quad\times\left[e^{\mp3\pi\mu/4}C_r^-(\ell_1)
 +e^{\pm3\pi\mu/4}C_r^+(\ell_1)\right]
 \left[e^{\mp\pi\mu/4}C_s^-(\ell_2)
 +e^{\pm\pi\mu/4}C_s^+(\ell_2)\right]
 \notag\\[1mm]
&\quad+\frac{\pi H^2e^{\mp i\pi(\ell_1+\ell_2)/2}}{4i\sinh(\pi\mu/2)}
\int_0^\infty\!dt\,t^{\ell_1+3/2}e^{-t/u}
\int_0^t\!ds\,s^{\ell_2+3/2}e^{-s/v}
\notag\\[-1mm]
&\qquad\times\Bigg[
 J_{i\mu/2}\!\left(\frac{\gamma t^2}{2}\right)
 J_{-i\mu/2}\!\left(\frac{\gamma s^2}{2}\right)
 -J_{-i\mu/2}\!\left(\frac{\gamma t^2}{2}\right)
 J_{i\mu/2}\!\left(\frac{\gamma s^2}{2}\right)
 \Bigg].
\label{eq:k4_complete_seed_fully_expanded}
\end{align}
The coefficients $C_r^\pm(\ell)$ are those fixed by the homogeneous
boundary conditions, and the double Bessel integral is the particular
solution of Eq.~\eqref{eq:k4_bl_particular}. The finite-momentum
qualification stated above applies unchanged.

For $\ell_1=\ell_2=-2$, the fixed-ratio collapsed contribution
of the particular is given by Eq.~\eqref{eq:k4_bl_collapsed_fixed_ratio}.
Appendix~\ref{app:sk_collapsed} derives the same time-ordering correction
directly from the Schwinger--Keldysh integrals; it is retained along
with the factorized collider terms in the full soft limit. After taking the collapsed limit $u,v\to0$ at fixed $u/v$, the non-analytic part reduces to
\begin{align}
\boxed{
\begin{aligned}
\widehat{\mathcal I}^{-2,-2}_{\rm seed,\,non\text{-}an}(u,v)
\Big|_{\rm coll}
={}&\frac{\pi H^2\sech(\pi\mu)}{2\mu}
(uv)^{\frac12-\ii\mu}
\left[
 u^{2\ii\mu}
 \left(\coth\!\frac{\pi\mu}{2}+\ii\right)
+v^{2\ii\mu}
 \left(\coth\!\frac{\pi\mu}{2}-\ii\right)
\right]
\\[1mm]
&+\frac{H^2}{4\pi}
\left(\frac{\gamma}{4}\right)^{\ii\mu}
\left(1+\ii\sinh\pi\mu\right)
\Gamma\!\left(-\frac{\ii\mu}{2}\right)^2
\Gamma\!\left(\frac12+\ii\mu\right)^2
(uv)^{\frac12+\ii\mu}
\\[1mm]
&+\frac{H^2}{4\pi}
\left(\frac{\gamma}{4}\right)^{-\ii\mu}
\left(1-\ii\sinh\pi\mu\right)
\Gamma\!\left(\frac{\ii\mu}{2}\right)^2
\Gamma\!\left(\frac12-\ii\mu\right)^2
(uv)^{\frac12-\ii\mu}.
\end{aligned}}
\label{eq:k4_summary_collapsed}
\end{align}
The seed reproduces the proper late-time strucuture of the four-point function, with the expected non-analytic scaling in $u$ and $v$ and the correct $\mu$-dependent coefficients \cite{Ferreira:2026tyj}. The first line corresponds to the leading non-analytic contribution, while the second and third lines represent subleading contributions that are also consistent with the expected behavior of the correlator in this limit.

The physical picture is therefore cleanly separated.  The quartic
operator admits four local asymptotic sectors, including two WKB
branches.  The chosen global homogeneous pair $\Phi_\pm^\ell$ is
WKB-free and Hermitian, and the direct Schwinger--Keldysh calculation
fixes the leading coefficients in its two collider branches.  The
Borel--Laplace particular term carries the analytic time-ordering
correction at leading collapsed order.  These statements are exact at
the level specified above; the finite-momentum connection problem
remains distinct from the leading soft matching.

%%%%%%%%%%%%%%%%%%%%%%%%%%%%%%%%%%%%
%%%%%%%%%%%%%%%%%%%%%%%%%%%%%%%%%%%%
%%%%%%%%%%%%%%%%%%%%%%%%%%%%%%%%%%%%
\section{Phenomenology}\label{sec:pheno}

Recent analyzes of full momentum-dependent collider shapes, together with the development of observational templates, have emphasized that the nonanalytic collider signal cannot be assessed in isolation from the analytic contributions \cite{Extra:260407434,Extra:240401894,Extra:240407203,Extra:250402931,Extra:250405389}. We therefore study both, first isolating the nonanalytic signal in the squeezed limit and then restoring the complete shape.

The bispectrum is generated by combining a linear mixing of the inflaton fluctuation and the massive spectator with the cubic interaction involving two inflaton fluctuations.
\begin{equation}
\delta\mathcal L_{2}
\supset
c_2 a^{3} \partial_{\eta}\varphi \chi,
\qquad
\delta\mathcal L_{3}
\supset
c_3 a^{2}
\left(\partial_{\eta}\varphi\right)^{2}\chi,
\label{eq:bispectrum_interactions}
\end{equation}
where $c_2$ and $c_3$ are time-independent couplings. As in the trispectrum discussion, we suppress the overall product of interaction couplings and numerical vertex-normalization factors in what follows since they only determine the overall amplitude of the final shape. These interactions generate the tree-level exchange contribution
\begin{equation}
\left\langle
\varphi_{\mathbf k_1}
\varphi_{\mathbf k_2}
\varphi_{\mathbf k_3}
\right\rangle_{\chi}^{\prime}
={}
-\frac{c_2 c_3 H}{4k_1k_2k^4_3}
\sum_{a,b=\pm1}
\widehat{\mathcal I}_{ab}(u,1)
+2\,{\rm perm.},
\label{eq:bispectrum_exchange_integral}
\end{equation}
It is useful to separate the three exchange channels as
\begin{equation}
\left\langle
\varphi_{\mathbf k_1}
\varphi_{\mathbf k_2}
\varphi_{\mathbf k_3}
\right\rangle_{\chi}^{\prime} =
\mathcal C_{12|3}
+
\mathcal C_{23|1}
+
\mathcal C_{31|2}.
\label{eq:bispectrum_three_channels}
\end{equation}
In the squeezed configuration $k_3\ll k_1\simeq k_2$, the channel $\mathcal C_{12|3}$ is the collider channel. The remaining two channels carry hard internal momenta and contribute to the analytic squeezed-limit background. The soft channel is directly related to the exchange seed introduced above. For the interactions in \eqref{eq:bispectrum_interactions}, the time dependence of the two vertices selects $\ell_1=0,\ell_2=-2$.

To connect with the primordial bispectrum, we use $\zeta=\varphi/(\sqrt{2\epsilon}M_{\rm Pl})$ and $\mathcal P_\zeta=H^2/(8\pi^2\epsilon M_{\rm Pl}^2)$. We define the dimensionless shape function by
\begin{equation}
\left\langle
\zeta_{\mathbf k_1}
\zeta_{\mathbf k_2}
\zeta_{\mathbf k_3}
\right\rangle^{\prime} \equiv (2\pi)^4
\frac{\mathcal P_{\zeta}^{2}}
{(k_1k_2k_3)^2}
S(k_1,k_2,k_3).
\label{eq:bispectrum_shape_definition}
\end{equation}
Consequently, the momentum dependence of the soft exchange channel is entirely determined by the single-folded seed,
\begin{equation}
S_{12|3}(k_1,k_2,k_3)
\propto -
\frac{k_1k_2}{k_3^2}
\sum_{a,b=\pm1}
\widehat{\mathcal I}^{0,-2}_{ab}
\left(
\frac{k_3}{k_1+k_2},1
\right).
\label{eq:bispectrum_shape_seed}
\end{equation}
Thus, the squeezed collider contribution to the bispectrum is completely encoded in the single-folded $\ell_1=0$, $\ell_2=-2$ seed. The complete bispectrum shape is obtained by summing the three exchange channels,
\begin{equation}
\begin{aligned}
S_{\rm Total}(k_1,k_2,k_3)
&= -\mathcal N_B\Bigg[
\frac{k_1k_2}{k_3^2}\,
\widehat{\mathcal I}_{\rm Total}^{\,0,-2}
\left(\frac{k_3}{k_1+k_2},1\right)
\\
&+\frac{k_2k_3}{k_1^2}\,
\widehat{\mathcal I}_{\rm Total}^{\,0,-2}
\left(\frac{k_1}{k_2+k_3},1\right)
%\\&
+\frac{k_3k_1}{k_2^2}\,
\widehat{\mathcal I}_{\rm Total}^{\,0,-2}
\left(\frac{k_2}{k_3+k_1},1\right)
\Bigg],
\end{aligned}
\label{eq:full_bispectrum_shape}
\end{equation}
where $\mathcal N_B$ is a momentum-independent normalization factor fixed by the background and the interaction couplings. These momenta permutations are usually related to the Mandelstam variables as follows:
\begin{equation}
    s : \frac{k_3}{k_1 + k_2}, \quad t : \frac{k_1}{k_3 + k_2},\quad u : \frac{k_2}{k_3 + k_1}
\end{equation}
the s, u, and t channels. This is an important description for the cosmological collider because the non-analytical effects can differ in these channels.

%%%%%%%%%%%%%%%%%%%%%%%%%%
%%%%%%%%%%%%%%%%%%%%%%%%%%
%%%%%%%%%%%%%%%%%%%%%%%%%%
\subsection{Squeezed Limit}

In the squeezed limit, the momenta hierarchy is $k_3\ll k_1 \sim k_2 $; this allows us to re-write the shape function as a one-dimensional family of configurations.
\begin{equation}
k_1=k_2\equiv k,
\qquad
x\equiv\frac{k_3}{k},
\qquad
0<x\leq 1.
\end{equation}
The three exchange channels then reduce to two independent contributions,
\begin{equation}
u_{12|3}=\frac{x}{2},
\qquad
u_{23|1}=u_{31|2}=\frac{1}{1+x},
\end{equation}
and the complete shape becomes
\begin{equation}
\begin{aligned}
S_{\rm Total}(x)
=
-\mathcal N_B
\Bigg\{
\frac{1}{x^2}
\widehat{\mathcal I}_{\rm dS}^{\,0,-2}
\left(
\frac{x}{2},1
\right)
+
2x \,
\widehat{\mathcal I}_{\rm dS}^{\,0,-2}
\left(
\frac{1}{1+x},1
\right) \Bigg\}
\end{aligned}
\label{eq:full_bispectrum_shapx}
\end{equation}
The two characteristic kinematic regimes are then contained in a single variable,
\begin{equation}
x\rightarrow0
\quad\text{(squeezed)},
\qquad
x=1
\quad\text{(equilateral)}.
\end{equation}
It is also convenient to introduce $\kappa\equiv x^{-1}=k/k_3$ when focusing on the squeezed regime. In terms of $\kappa$, the equilateral point is at $\kappa=1$, while the deep squeezed limit corresponds to $\kappa\gg1$.

%%%%%%%%%%%%%%%%%%%%%%%%%%%%%%%
%%%%%%%%%%%%%%%%%%%%%%%%%%%%%%%
%%%%%%%%%%%%%%%%%%%%%%%%%%%%%%%
\subsubsection*{Non-analytic sector}
Oscillatory mass-dependent non-analyticity, as well as its possible enhancement or phase modification, has been investigated through chemical-potential effects and transient backgrounds \cite{Extra:220501692,Extra:220501107,Ferreira:2026rsl}. Loop correlators can also carry nonanalytic collider signals, with different analytic structures from tree-level exchange \cite{Extra:230413295,Extra:230814802,Extra:210914635,Extra:260521419,Extra:260909288}.

We first isolate the non-analytic contribution of the $s$-channel,
$\mathcal C_{12|3}$, which carries the massive-particle signal in the
squeezed limit. By excluding the analytic background and the two hard
exchange channels, we can directly compare the oscillatory signal of the
quartic spectator with its relativistic de Sitter counterpart. The
non-analytic terms retain the characteristic logarithmic oscillations,
whose leading squeezed behavior may be written schematically as
\begin{equation}
S_{12|3}^{\mathrm{non\text{-}an}}(\kappa)
\underset{\kappa\gg1}{\sim}
\kappa^{-1/2}\left[
\mathcal A_+(\mu,\gamma)\,\kappa^{i\mu}
+\mathcal A_-(\mu,\gamma)\,\kappa^{-i\mu}
\right].
\label{eq:k4_nonanalytic_squeezed_shape}
\end{equation}
The oscillation frequency is therefore controlled by the mass parameter
$\mu$, whereas the amplitudes and phases are sensitive to the dispersion
parameter $\gamma$.

\begin{figure}[ht]
    \centering
    \begin{subfigure}[b]{0.48\textwidth}
        \centering
        \includegraphics[width=\linewidth]{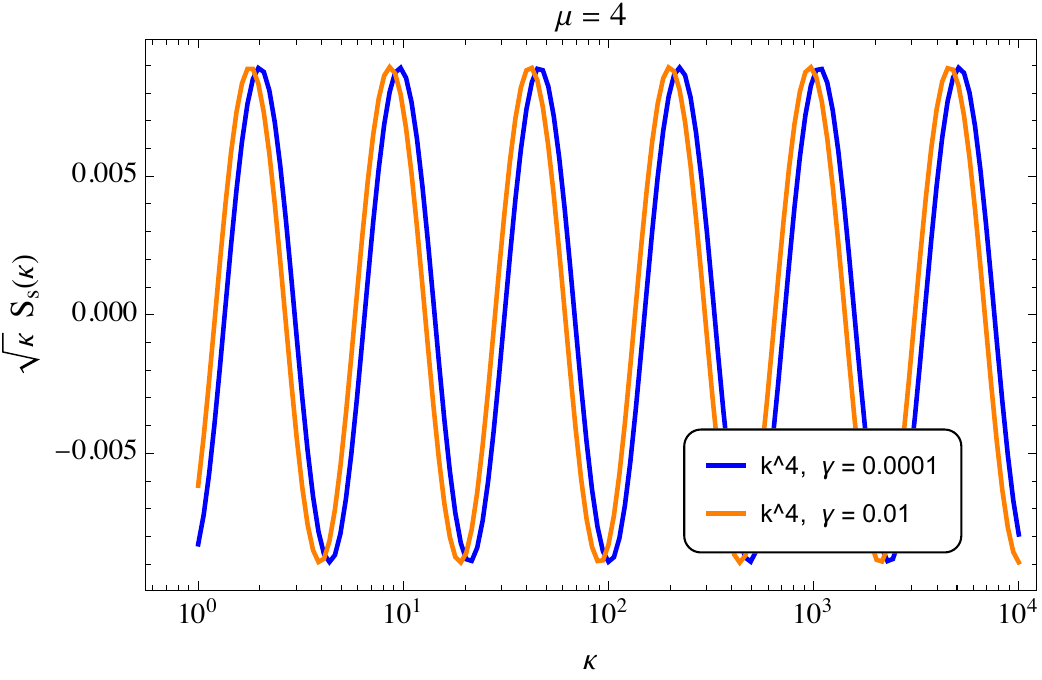}
        \caption{}
        \label{fig:nonanalytic_gamma1}
    \end{subfigure}
    \hfill
    \begin{subfigure}[b]{0.48\textwidth}
        \centering
        \includegraphics[width=\linewidth]{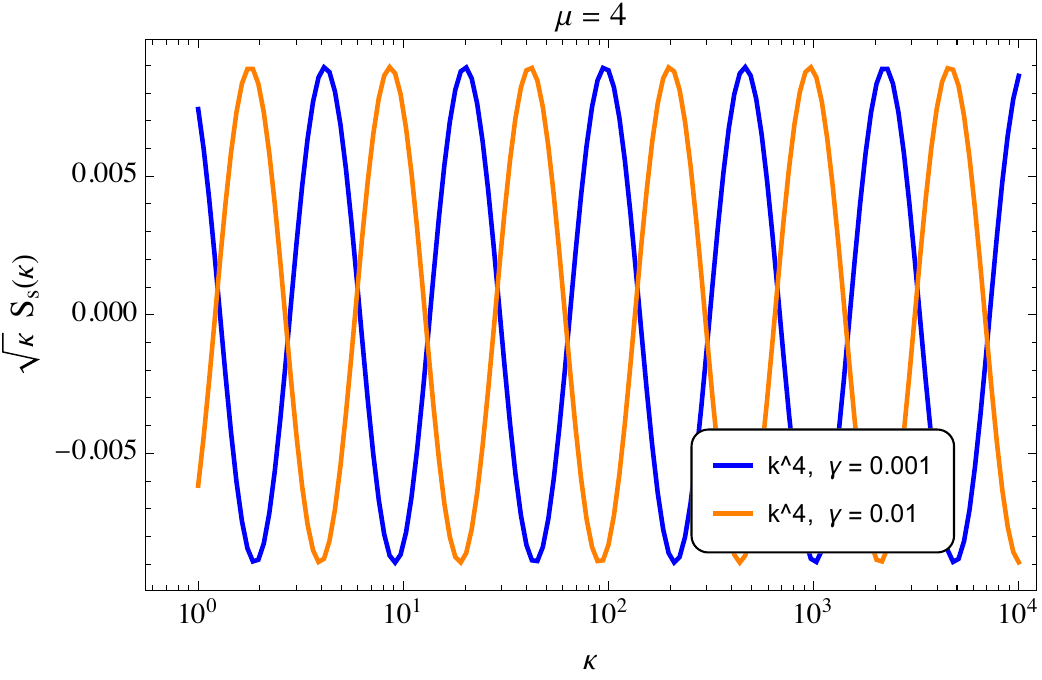}
        \caption{}
        \label{fig:nonanalytic_gamma2}
    \end{subfigure}
    \caption{Non-analytic $s$-channel contribution, displayed as
    $\mathrm{Re}[\sqrt{\kappa}\,S(\kappa)]$ for $\mu=4$.
    The left panel compares $\gamma=10^{-4}$ and $10^{-2}$; the right
    panel compares $\gamma=10^{-3}$ and $10^{-2}$. The logarithmic
    oscillations retain a fixed frequency while their phases depend
    on the higher-derivative parameter.}
    \label{fig:nonanalytic_gamma}
\end{figure}

\begin{figure}[ht]
    \centering
    \begin{subfigure}[b]{0.48\textwidth}
        \centering
        \includegraphics[width=\linewidth]{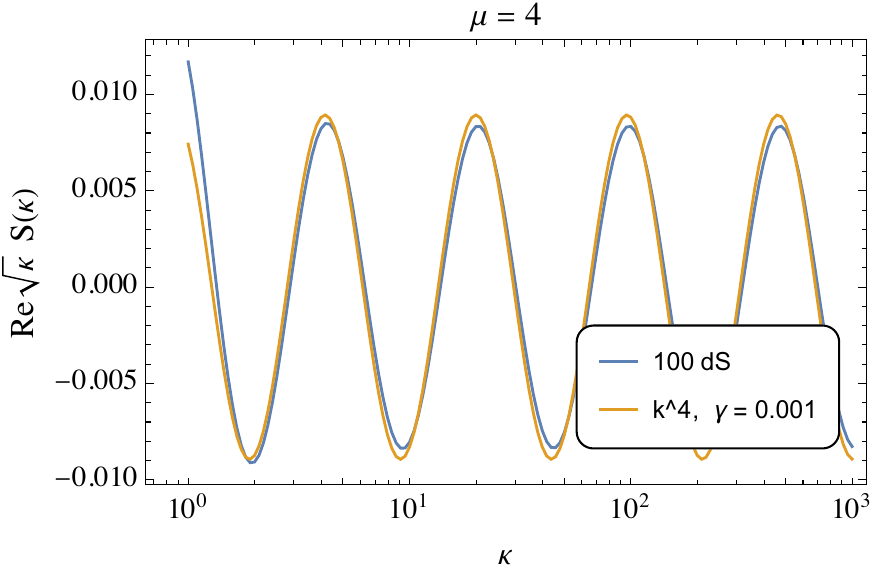}
        \caption{}
        \label{fig:nonanalytic_mu4}
    \end{subfigure}
    \hfill
    \begin{subfigure}[b]{0.48\textwidth}
        \centering
        \includegraphics[width=\linewidth]{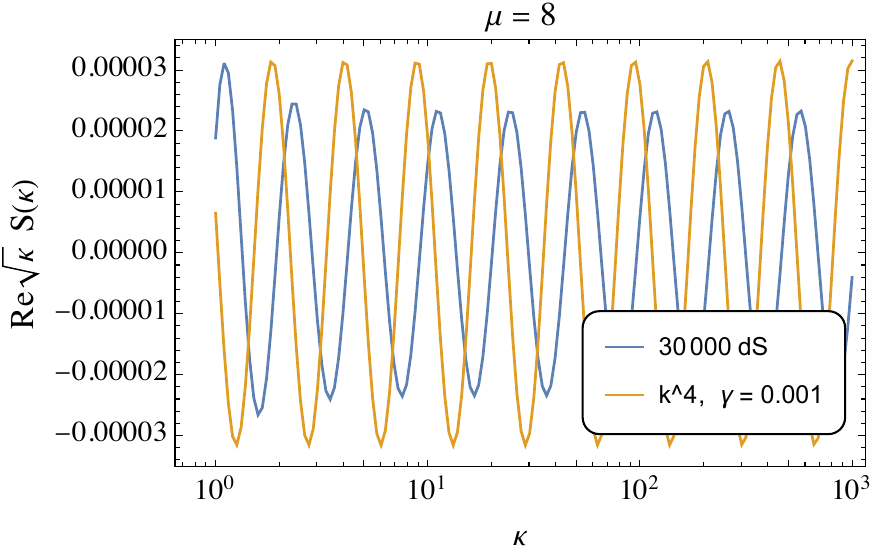}
        \caption{}
        \label{fig:nonanalytic_mu8}
    \end{subfigure}
    \caption{Non-analytic $s$-channel signal for quartic dispersion
    compared with the relativistic de Sitter result.
    Left: $\mu=4$, $\gamma=10^{-3}$; right: $\mu=8$,
    $\gamma=10^{-3}$. For visualization only, the de Sitter curves
    are multiplied by $10^2$ and $3\times10^4$, respectively.
    Both panels show $\mathrm{Re}[\sqrt{\kappa}\,S(\kappa)]$.}
    \label{fig:nonanalytic_mu}
\end{figure}

Figure~\ref{fig:nonanalytic_gamma} illustrates this dependence for
$\mu=4$. Changing $\gamma$ alters the oscillatory pattern, in particular
its phase, without changing the characteristic logarithmic frequency.
More importantly, Fig.~\ref{fig:nonanalytic_mu} shows that the
non-analytic signal can be substantially larger than its relativistic
counterpart. The de Sitter curves have been multiplied by $10^2$ for
$\mu=4$ and by $3\times10^4$ for $\mu=8$ merely to display them on
comparable vertical scales. Thus, when the analytic contributions are
removed, the quartic theory exhibits the same qualitative enhancement
of massive cosmological-collider oscillations found in our earlier
analysis~\cite{Ferreira:2026tyj}. This enhancement is a property of the
isolated non-analytic exchange signal; its imprint on the observable
full shape must be assessed after restoring all contributions.

%%%%%%%%%%%%%%%%%%%%%%%%
%%%%%%%%%%%%%%%%%%%%%%%%
%%%%%%%%%%%%%%%%%%%%%%%%
%%%%%%%%%%%%%%%%%%%%%%%%
\subsubsection*{Full shape function}

We now restore the complete exchange contribution, including the
analytic terms, the inhomogeneous Borel--Laplace particular solution,
and all three momentum channels. These pieces are essential for the
full bispectrum, even though they are not part of the isolated
non-analytic signal discussed above. In particular, the two hard
channels contribute an analytic squeezed-limit background, while the
particular solution introduces additional, nontrivial dependence on
$\gamma$. Consequently, an enhancement of the non-analytic $s$-channel
need not appear as a uniform enhancement of the complete shape.

\begin{figure}[t]
    \centering
    \begin{subfigure}[b]{0.48\textwidth}
        \centering
        \includegraphics[width=\linewidth]{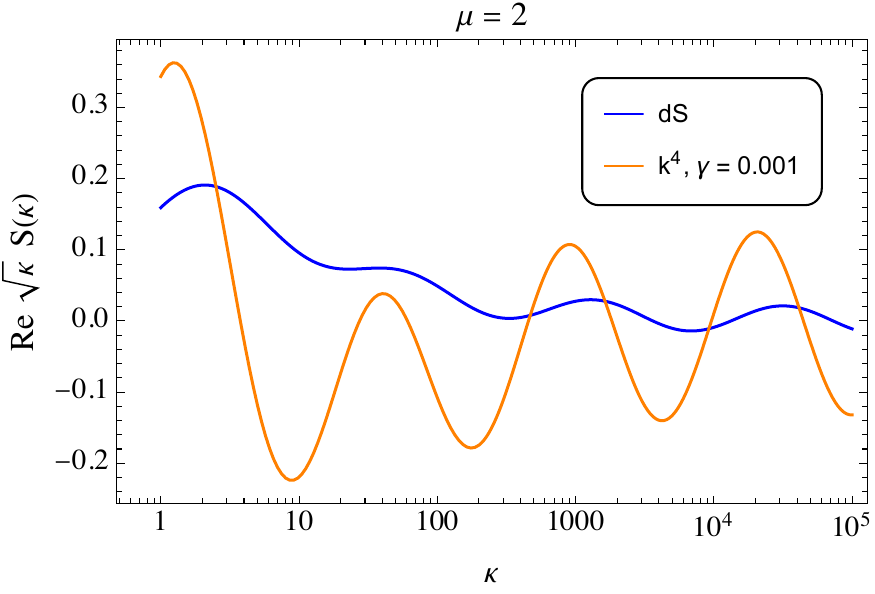}
        \caption{$\mu=2$, $\gamma=10^{-3}$.}
        \label{fig:full_mu2_gamma001}
    \end{subfigure}
    \hfill
    \begin{subfigure}[b]{0.48\textwidth}
        \centering
        \includegraphics[width=\linewidth]{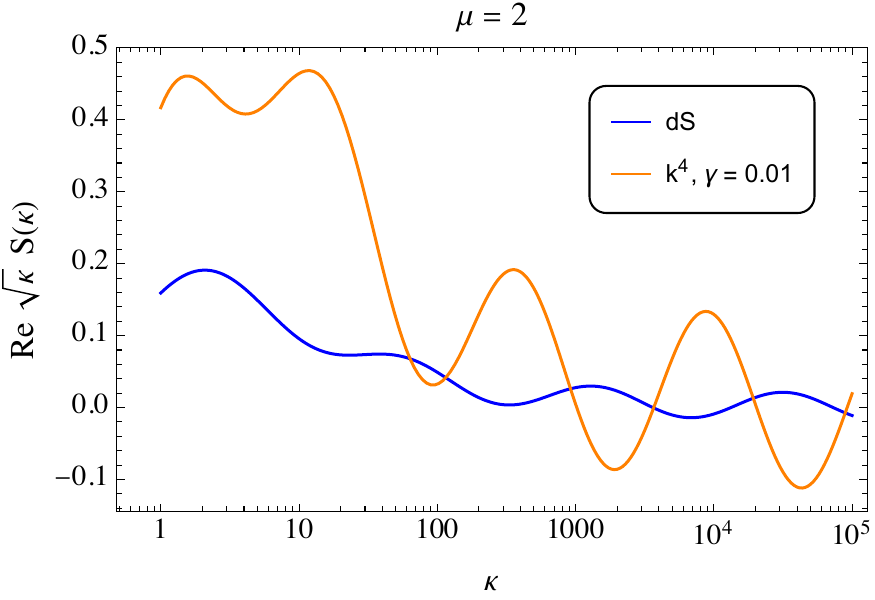}
        \caption{$\mu=2$, $\gamma=10^{-2}$.}
        \label{fig:full_mu2_gamma01}
    \end{subfigure}
    \vspace{0.3cm}
    \begin{subfigure}[b]{0.48\textwidth}
        \centering
        \includegraphics[width=\linewidth]{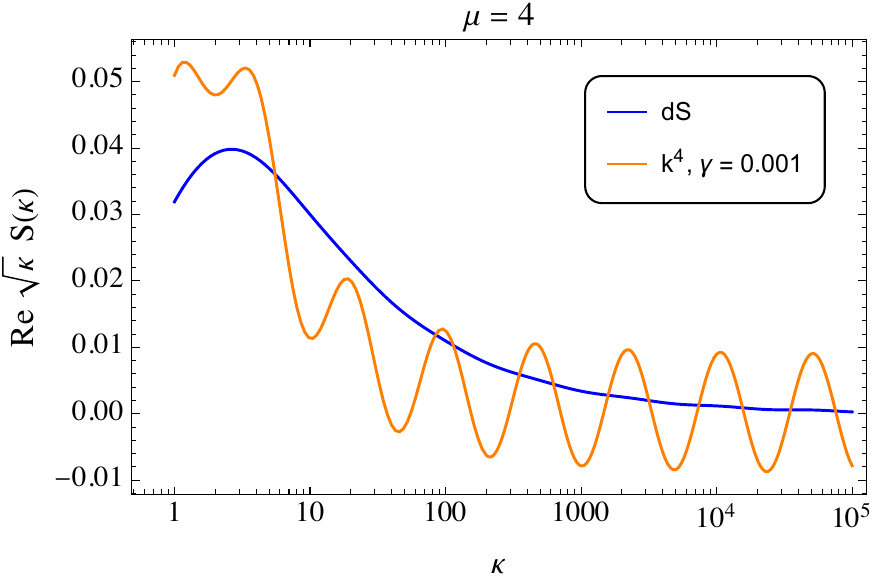}
        \caption{$\mu=4$, $\gamma=10^{-3}$.}
        \label{fig:full_mu4_gamma001}
    \end{subfigure}
    \hfill
    \begin{subfigure}[b]{0.48\textwidth}
        \centering
        \includegraphics[width=\linewidth]{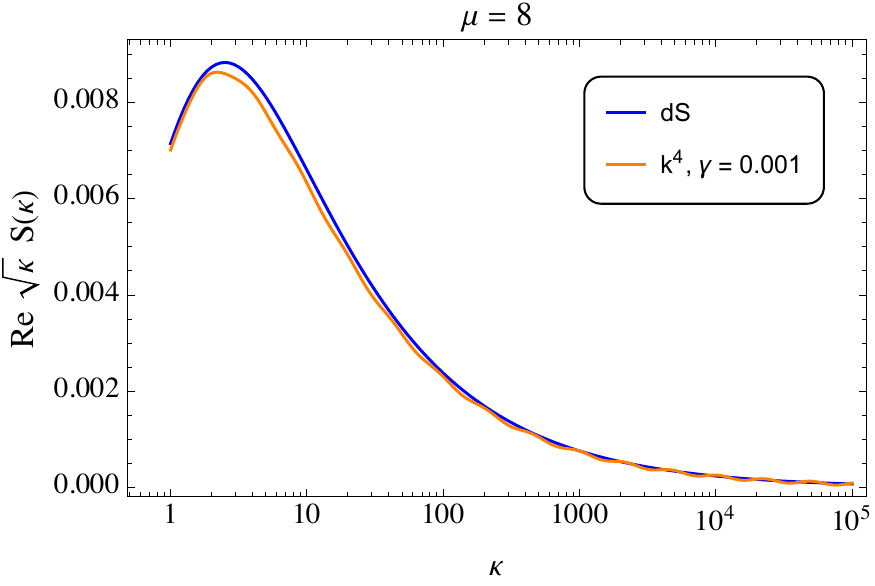}
        \caption{$\mu=8$, $\gamma=10^{-3}$.}
        \label{fig:full_mu8_gamma001}
    \end{subfigure}
    \caption{Full bispectrum shape
    $\mathrm{Re}[\sqrt{\kappa}\,S(\kappa)]$ for the quartic spectator
    (orange) and the standard relativistic de Sitter result (blue),
    including the analytic and non-analytic contributions and all
    momentum permutations. Panels (a) and (b) compare two values
    of $\gamma$ at $\mu=2$; panels (c) and (d) show $\mu=4$ and
    $\mu=8$ for $\gamma=10^{-3}$. Near-equilateral enhancement and
    oscillatory deviations at lower masses give way to near agreement
    of the full shapes for $\mu=8$.}
    \label{fig:full_shape_comparison}
\end{figure}

The contrast is apparent in Fig.~\ref{fig:full_shape_comparison}.
For $\mu=2$, the quartic theory develops a pronounced enhancement and
a different momentum dependence near $\kappa\simeq1$, the equilateral
end of the isosceles family of configurations. The enhancement is more
prominent for $\gamma=10^{-2}$ than for $\gamma=10^{-3}$. Since
$\gamma=H/M$, increasing $\gamma$ corresponds, at fixed $H$, to
lowering the cutoff $M$: the two values shown correspond to
$M=10^2H$ and $M=10^3H$, respectively. Moving the cutoff farther
above the Hubble scale therefore seems to reduce this near-equilateral effect.
The explicit $\gamma$ dependence of the Borel--Laplace particular
solution also provides a source of this sensitivity, as is clear
from Eq.~\eqref{eq:k4_bl_particular}. We stress that this
near-equilateral enhancement should be distinguished from the
non-analytic collider enhancement in the deep squeezed regime. 

For $\mu=4$, the full quartic shape retains sizable oscillatory
deviations from the relativistic prediction over the displayed
momentum range. By contrast, for $\mu=8$, the two complete shapes are
nearly coincident, despite the strong relative enhancement seen when
only the non-analytic channel is retained. This illustrates how the
analytic contributions and the particular solution can mask the
non-analytic enhancement in the full bispectrum. The comparison also
shows that the dependence on $\gamma$ is not restricted to the
oscillatory collider sector; it can produce an additional,
phenomenologically relevant modification of the shape near the
equilateral configuration. However, the real impact on the correlator of this gamma effect is better explained by looking at the equilateral limit.

%%%%%%%%%%%%%%%%%%%%%%%%%%%%%%%%
%%%%%%%%%%%%%%%%%%%%%%%%%%%%%%%%
%%%%%%%%%%%%%%%%%%%%%%%%%%%%%%%%
%%%%%%%%%%%%%%%%%%%%%%%%%%%%%%%%
\subsection{Equilateral Limit}
\label{subsec:equilateral_limit}

\begin{figure}[htbp]
    \centering
    \includegraphics[width=0.85\textwidth]{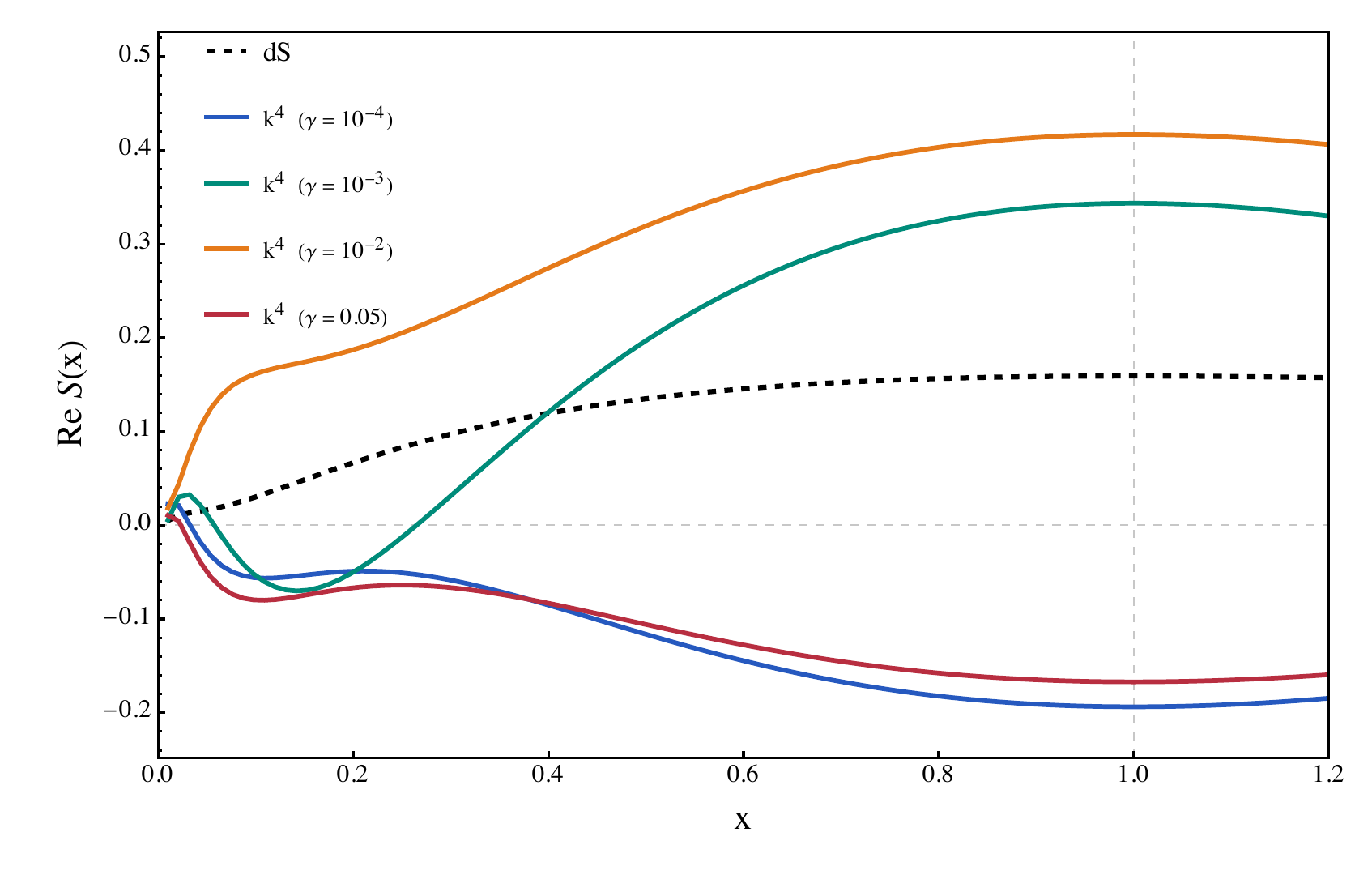}
    \caption{Real part of the full shape $S(x)$ in the isosceles
    configuration, $x=k_3/k_1$, for $\mu=2$. The black dashed line
    denotes the relativistic de Sitter result. The colored curves
    show the quartic ($k^4$) spectator for different values of
    $\gamma$, with the inhomogeneous contribution evaluated by
    Borel--Laplace resummation. The vertical dashed line marks the
    equilateral point $x=1$. Both the magnitude and the sign of the
    equilateral amplitude depend on $\gamma$.}
    \label{fig:full_shape_gamma}
\end{figure}

We now examine the full shape in the equilateral configuration,
$x\equiv k_3/k_1=1$, retaining the analytic and non-analytic
contributions and all momentum channels. Unlike the isolated
non-analytic squeezed signal, the equilateral amplitude depends on
the complete exchange contribution. We begin with the isosceles
shape for fixed $\mu=2$, shown in Fig.~\ref{fig:full_shape_gamma}.

An important feature of Fig.~\ref{fig:full_shape_gamma} is that the
quartic equilateral amplitude can be either positive or negative,
depending on $\gamma$. A change of sign in the equilateral
configuration also occurs in tilted ghost inflation for suitable
values of the tilt parameter $\lambda$~\cite{Ferreira:2026tyj}.
The two models need not share the same mechanism: here we vary the
scale of a purely quartic dispersion relation, whereas in the tilted
model, the relative weight of the quadratic and quartic terms is
modified. Nevertheless, both cases show that the sign of the
full equilateral signal is sensitive to the spectator dispersion.

\begin{figure}[htbp]
    \centering
    \includegraphics[width=0.85\textwidth]{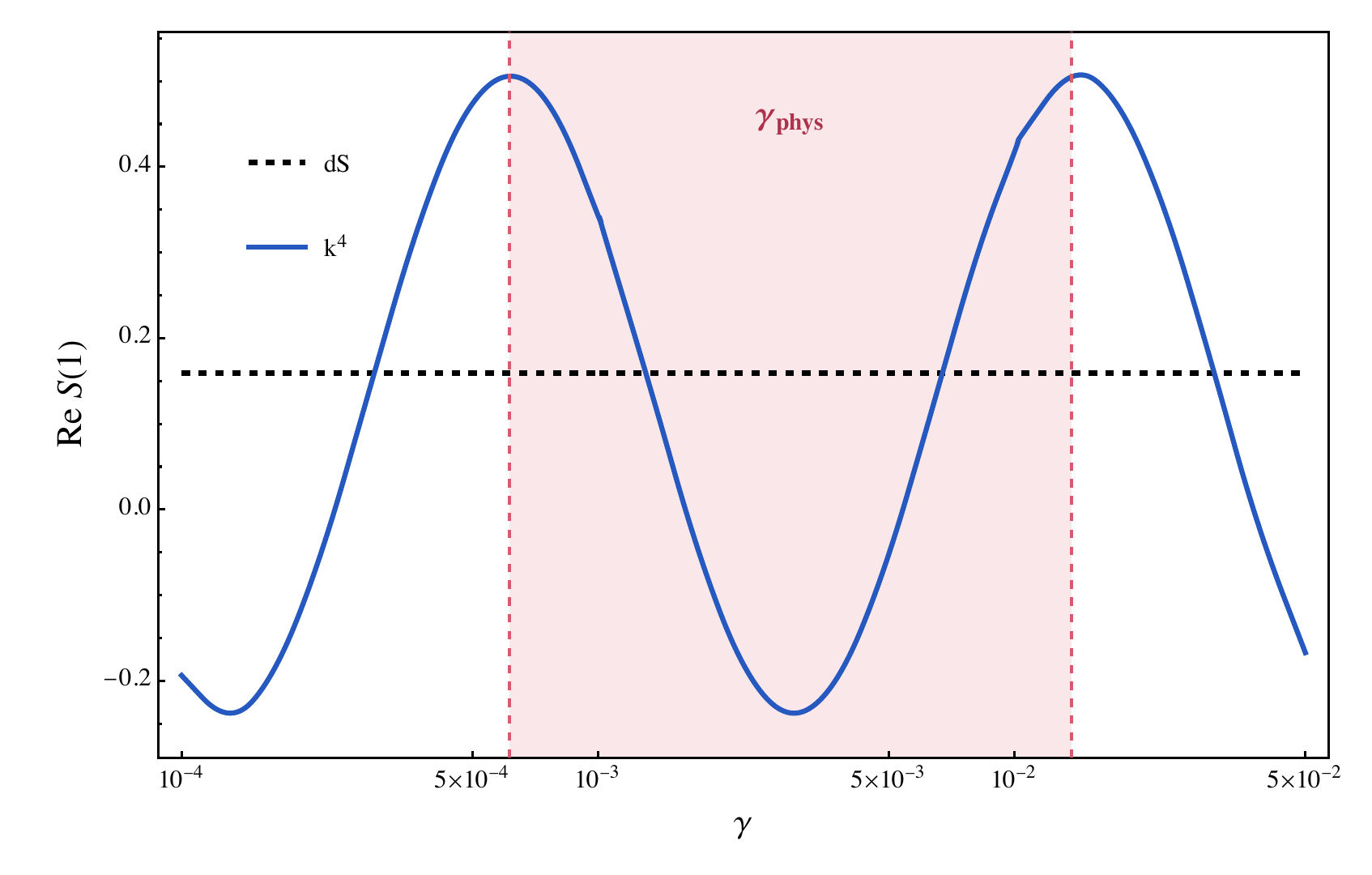}
    \caption{Equilateral full shape, $\operatorname{Re}S(1)$, as a
    function of $\gamma$ at fixed $\mu=2$. The blue curve shows the
    quartic spectator result obtained using Borel--Laplace
    resummation; the horizontal black dashed line is the de Sitter
    result, independent of $\gamma$. The shaded region lies between
    two successive local maxima of the quartic curve and highlights
    its sensitivity to the scale ratio $\gamma=H/M$.}
    \label{fig:equilateral_gamma}
\end{figure}

To examine this sensitivity in more detail, we keep $x=1$ and
$\mu=2$ fixed and scan the higher-derivative parameter $\gamma$.
As shown in Fig.~\ref{fig:equilateral_gamma}, the quartic result
oscillates strongly as a function of $\log\gamma$, crossing the
constant de Sitter prediction and changing sign repeatedly. For the
pure quartic dispersion considered here, $\gamma=H/M$, where $M$
is the scale suppressing the higher-derivative operator. Thus, the
scan can equivalently be interpreted as a dependence on the ratio
$H/M$. The shaded interval, bounded by two consecutive numerical
maxima, illustrates the considerable variation of the equilateral
amplitude over this range of higher-derivative scales. These
boundaries mark maxima of the plotted shape; by themselves, they do
not impose a cutoff constraint.

To understand the origin of the equilateral amplitude, it is useful
to distinguish the time-ordered homogeneous term from the mixed
homogeneous contribution. Schematically, we write
\begin{equation}
    S_{\mathrm{hom}}=S_{\mathrm{ord}}+S_{\mathrm{mix}},
    \qquad
    S_{\mathrm{full}}=S_{\mathrm{hom}}+S_{\mathrm{rem}},
    \label{eq:equilateral_decomposition}
\end{equation}
where $S_{\mathrm{rem}}$ collects the remaining contributions to the
full shape, including the appropriate particular and analytic terms.
To quantify the relative weight of the ordered term \emph{within the
homogeneous sector}, we define the signed ratio
\begin{equation}
    \mathcal{R}_{\mathrm{ord}}^{\mathrm{hom}}
    \equiv
    \frac{\operatorname{Re}S_{\mathrm{ord}}(1)}
         {\operatorname{Re}S_{\mathrm{hom}}(1)}.
    \label{eq:equilateral_ordered_fraction}
\end{equation}
At $\mu=2$ we obtain
\begin{equation}
    \left.\mathcal{R}_{\mathrm{ord}}^{\mathrm{hom}}\right|_{k^4,\,\gamma=10^{-3}}
    \simeq 0.944,
    \qquad
    \left.\mathcal{R}_{\mathrm{ord}}^{\mathrm{hom}}\right|_{\mathrm{dS}}
    \simeq 1.010.
    \label{eq:equilateral_ordered_benchmarks}
\end{equation}
Thus, in both cases, the ordered contribution dominates the
homogeneous sector. The de Sitter ratio slightly exceeds unity
because the mixed term has the opposite sign. Importantly, the
ratios in Eq.~\eqref{eq:equilateral_ordered_benchmarks} are \emph{not}
fractions of the full shape: the additional contributions in
Eq.~\eqref{eq:equilateral_decomposition} must also be included.
The numerical decomposition suggests that the ordered homogeneous
term remains dominant over the mixed homogeneous term as $\mu$
increases; this hierarchy should be checked explicitly for each
mass used in the comparison below.

Finally, we have verified that this behavior persists for
larger masses, with $\mu=2,4,8$. The higher-derivative
modification therefore affects the full bispectrum shape
even in the equilateral configuration, rather than being
restricted to the non-analytic contribution characteristic
of the squeezed limit.

Interestingly, boost-breaking dynamics are known to produce
distinctive signatures even away from the squeezed limit.
In particular, the boostless cosmological collider studied
in Ref.~\cite{PimentelWang_BoostlessCosmologicalColliderBootstrap_2022} exhibits oscillatory
features around nearly equilateral configurations when the
massive mediator propagates with a sufficiently small sound
speed relative to the inflaton. In our setup, boost breaking
arises instead from the quartic dispersion relation of the
massive spectator rather than from a constant reduced sound
speed. Related analyzes have examined the full momentum dependence of boost-breaking signals and the effects of nontrivial features in the correlator equations \cite{Extra:250601555,Extra:251100152}. Although the two mechanisms are different, our
results further illustrate how modified propagation can
affect the full momentum dependence of cosmological
correlators beyond the conventional squeezed regime.

%%%%%%%%%%%%%%%%%%%%%%%%%%%%
%%%%%%%%%%%%%%%%%%%%%%%%%%%%
%%%%%%%%%%%%%%%%%%%%%%%%%%%%
%%%%%%%%%%%%%%%%%%%%%%%%%%%%
\section{Conclusion}\label{sec:conclusion}

We have developed the cosmological bootstrap for the exchange of a massive spectator with quartic dispersion, $\omega^2\propto k^4$. In previous work, we showed that ghost-spectator dynamics can weaken the usual Boltzmann suppression of heavy-particle signals and derived the associated bootstrap equations~\cite{Ferreira:2026tyj}. Here, we have solved these equations, determined the boundary conditions that select the physical solution, and assessed their implications for the complete bispectrum shape.

The quartic dispersion promotes the usual second-order de Sitter seed equation to fourth order and qualitatively changes its singularity structure. The collapsed point $u=0$ becomes an irregular singular point, whereas the de Sitter folded and factorization loci, $u=\pm1$, as well as $u=\infty$, are ordinary points of the homogeneous equation. Near the collapsed endpoint, the four-dimensional homogeneous solution space contains two collider branches, $u^{\ell+5/2\pm i\mu}$, together with two rapidly oscillating WKB branches, $\exp[\pm i/(2\gamma u^2)]$, whose sector-dependent asymptotics give rise to the Stokes phenomenon.

This change has a direct consequence for the choice of vacuum. In the relativistic bootstrap, the Bunch--Davies condition is implemented by requiring the absence of an unphysical folded singularity; in the quartic problem, this criterion is no longer available, and an asymptotic boundary prescription at $u\to0^+$ is required instead. By setting both WKB coefficients to zero and matching the surviving collider branches to Bunch--Davies-normalized Schwinger--Keldysh data, we have constructed a global, WKB-free homogeneous basis. This prescription selects the physical asymptotic sector without requiring an explicit determination of the Stokes multipliers.

For the sourced equations, we have constructed a Borel--Laplace particular solution for general real vertex weights within a specified domain of validity. In terms of $x=1/u$, the fourth-order bootstrap operator is intertwined, under the Laplace transform, with a second-order Bessel operator. The resulting Green kernel defines a convergent integral representation whose asymptotic expansion reproduces the factorially divergent algebraic series near the irregular endpoint, and a kernel identity establishes that this representative satisfies both quartic bootstrap equations simultaneously. Direct Schwinger--Keldysh calculations independently reproduce its leading algebraic collapsed-limit contribution and fix the leading nonanalytic collider coefficients of the homogeneous sector.

The phenomenological analysis reveals two complementary consequences of the modified dispersion. In the squeezed limit, the nonanalytic signal retains its mass-dependent logarithmic frequency, while its amplitude and phase depend on $\gamma=H/M$; the isolated collider signal can be substantially enhanced relative to the relativistic result. Away from this limit, the analytic contributions and the momentum permutations modify the complete shape and may partially mask this enhancement. In particular, the equilateral amplitude is sensitive to $\gamma$ and can change sign; for the parameter choices studied, it also oscillates as a function of $\log\gamma$. Higher-derivative propagation therefore affects not only the nonanalytic squeezed signal but also the full momentum dependence of the bispectrum.

Finally, the Borel--Laplace prescription selects a definite particular solution but does not, by itself, establish its equality with the full time-ordered Schwinger--Keldysh seed at finite momenta. Although the leading collapsed-limit matching is verified, these data alone do not exclude an additional doubly homogeneous contribution. Resolving this global connection problem and determining the associated Stokes data would provide a sharper characterization of the physical boundary conditions for higher-derivative cosmological correlators.

Several directions for future work naturally emerge from our analysis. A direct Mellin--Barnes treatment of the Schwinger--Keldysh integrals could determine whether the Borel--Laplace particular solution requires such a doubly homogeneous contribution at finite momentum. Computing the Stokes matrices of the fourth-order equation would clarify the analytic continuation of the WKB and collider sectors and could provide an intrinsic characterization of the WKB-free boundary prescription. Extending the construction to a mixed quadratic--quartic dispersion relation would shed light on the interpolation between relativistic and ghost-like propagation. More generally, dispersion laws of the form $\omega_{\rm phys}^{2}\propto p^{2z}/M^{2z-2}$ could reveal how the differential order and the singularity structure of the bootstrap equations depend on the dynamical exponent $z$. Finally, extending the analysis to massive fields with spin could uncover new helicity-dependent asymptotic sectors and angular signatures. We leave these questions for future work.

\section*{Acknowledgments}

We would like to thank Guilherme L. Pimentel, Tom Westerdijk,
and Chen Yang for valuable discussions and insights.
We are especially grateful to Guilherme L. Pimentel for
sharing his insights into Borel resummation techniques,
which were particularly valuable in the development of
this work.~We also thank the theoretical physics group at
Scuola Normale Superiore in Pisa for its hospitality
and stimulating research environment.

Preliminary results from this work were presented at
QCD meets Gravity XI (2025), the V Patricio Letelier School
on Mathematical Physics (2026), and the XX Brazilian School
of Cosmology and Gravitation (2026).
We thank the organizers and participants of these events
for stimulating discussions and valuable feedback.

MCF and FTF acknowledge the financial support of the National Scientific and Technological Research Council (CNPq, Brazil) and the Brazilian Federal Agency for Support and Evaluation of Graduate Education (CAPES, Brazil).

\clearpage

\appendix

\section{Generalized hypergeometric functions and their asymptotic limits}
\label{app:generalized_hypergeometric}

In this appendix we collect the definitions and the asymptotic properties of the
generalized hypergeometric functions that are relevant throughout this work. We use
the convention
\begin{equation}
{}_qF_p
\left(
\begin{matrix}
a_1,\ldots,a_q\\
b_1,\ldots,b_p
\end{matrix}
;z
\right)
=
\sum_{n=0}^{\infty}
\frac{(a_1)_n\cdots(a_q)_n}
     {(b_1)_n\cdots(b_p)_n}
\frac{z^n}{n!},
\label{eq:hypergeometric_definition_general}
\end{equation}
where
\begin{equation}
(a)_n=\frac{\Gamma(a+n)}{\Gamma(a)}
\end{equation}
is the Pochhammer symbol. Unless otherwise stated, none of the lower
parameters $b_j$ is assumed to be a non-positive integer.

It is useful to stress that the analytic structure of
${}_qF_p$ depends crucially on the relative number of upper and lower
parameters. We therefore distinguish the cases $q<p$ and $q>p$ below.

\subsection{The case \texorpdfstring{$\boldsymbol{q<p}$}{q < p}}
\label{app:hypergeometric_q_less_p}

For $q<p$, the defining series
\begin{equation}
{}_qF_p
\left(
\begin{matrix}
a_1,\ldots,a_q\\
b_1,\ldots,b_p
\end{matrix}
;z
\right)
=
\sum_{n=0}^{\infty}
\frac{\prod_{i=1}^{q}(a_i)_n}
     {\prod_{j=1}^{p}(b_j)_n}
\frac{z^n}{n!}
\label{eq:q_less_p_definition}
\end{equation}
converges for every finite value of $z$. Therefore, ${}_qF_p$ is an
entire function of its argument. Its small-$z$ expansion is simply
\begin{align}
{}_qF_p(\mathbf a;\mathbf b;z)
={}
1
+
\frac{\displaystyle\prod_{i=1}^{q}a_i}
     {\displaystyle\prod_{j=1}^{p}b_j}\,z + \frac{1}{2}
\frac{\displaystyle\prod_{i=1}^{q}a_i(a_i+1)}
     {\displaystyle\prod_{j=1}^{p}b_j(b_j+1)}
\,z^2
+\mathcal O(z^3).
\label{eq:q_less_p_small_z}
\end{align}
In particular,
\begin{equation}
{}_qF_p(\mathbf a;\mathbf b;0)=1.
\end{equation}

Although there are no finite singularities of this particular solution,
the point $z=\infty$ is an irregular singular point of the corresponding
generalized hypergeometric differential equation. The large-argument
behavior therefore contains exponential sectors.

For later convenience, define
\begin{equation}
\kappa=p-q+1,
\qquad
\nu=
\sum_{i=1}^{q}a_i
-
\sum_{j=1}^{p}b_j
+
\frac{p-q}{2}.
\label{eq:kappa_nu_hypergeometric}
\end{equation}
The exponential asymptotic series can be written as
\begin{equation}
E_{q,p}(z)
=
(2\pi)^{\frac{q-p}{2}}
\kappa^{-\nu-\frac12}
e^{\kappa z^{1/\kappa}}
\sum_{n=0}^{\infty}
c_n
\left(
\kappa z^{1/\kappa}
\right)^{\nu-n},
\qquad
c_0=1.
\label{eq:E_qp_definition}
\end{equation}
Thus, at leading order,
\begin{equation}
E_{q,p}(z)
\sim
(2\pi)^{\frac{q-p}{2}}
\kappa^{-\nu-\frac12}
e^{\kappa z^{1/\kappa}}
\left(
\kappa z^{1/\kappa}
\right)^{\nu}
\left[
1+\mathcal O\!\left(z^{-1/\kappa}\right)
\right].
\label{eq:E_qp_leading}
\end{equation}

The algebraic asymptotic series is
\begin{equation}
H_{q,p}(z)
=
\sum_{m=1}^{q}
\sum_{n=0}^{\infty}
\frac{(-1)^n}{n!}
\Gamma(a_m+n)
\frac{
\displaystyle
\prod_{\substack{i=1\\i\neq m}}^{q}
\Gamma(a_i-a_m-n)
}{
\displaystyle
\prod_{j=1}^{p}
\Gamma(b_j-a_m-n)
}
z^{-a_m-n},
\label{eq:H_qp_definition}
\end{equation}
where the expression is written for generic parameters such that no two
$a_i$ differ by an integer. When this condition is violated, poles
coalesce and logarithmic contributions may appear.

The leading algebraic terms are therefore
\begin{equation}
H_{q,p}(z)
\sim
\sum_{m=1}^{q}
\Gamma(a_m)
\frac{
\displaystyle
\prod_{\substack{i=1\\i\neq m}}^{q}
\Gamma(a_i-a_m)
}{
\displaystyle
\prod_{j=1}^{p}
\Gamma(b_j-a_m)
}
z^{-a_m}
\left[
1+\mathcal O(z^{-1})
\right].
\label{eq:H_qp_leading}
\end{equation}

The case that will be particularly relevant in this work is
\begin{equation}
q=p-1,
\end{equation}
which includes, for example, ${}_2F_3$ and ${}_3F_4$. In this case
$\kappa=2$, and for $z\rightarrow+\infty$ one has
\begin{equation}
\frac{
\displaystyle\prod_{i=1}^{q}\Gamma(a_i)
}{
\displaystyle\prod_{j=1}^{p}\Gamma(b_j)
}
{}_qF_p
\left(
\begin{matrix}
\mathbf a\\
\mathbf b
\end{matrix}
;-z
\right)
\sim
H_{q,p}(z)
+
E_{q,p}\!\left(ze^{-i\pi}\right)
+
E_{q,p}\!\left(ze^{+i\pi}\right).
\label{eq:q_p_minus_one_asymptotic}
\end{equation}
Since
\begin{equation}
\left(ze^{\pm i\pi}\right)^{1/2}
=
\pm i\sqrt{z},
\end{equation}
the two exponential sectors become
\begin{align}
E_{q,p}\!\left(ze^{\pm i\pi}\right)
\sim{}&
\frac{1}{\sqrt{2\pi}}\,
2^{-\nu-\frac12}
e^{\pm 2i\sqrt{z}}
\left(
\pm 2i\sqrt{z}
\right)^\nu
\nonumber\\
&\times
\left[
1+\mathcal O\!\left(z^{-1/2}\right)
\right].
\label{eq:q_p_minus_one_WKB}
\end{align}
Consequently,
\begin{align}
{}_qF_p(\mathbf a;\mathbf b;-z)
\sim{}
\frac{\displaystyle\prod_{j=1}^{p}\Gamma(b_j)}
     {\displaystyle\prod_{i=1}^{q}\Gamma(a_i)}
\Bigg[
H_{q,p}(z) +
\frac{
e^{+2i\sqrt{z}}
(2i\sqrt{z})^\nu
}{
\sqrt{2\pi}\,2^{\nu+\frac12}
}
+
\frac{
e^{-2i\sqrt{z}}
(-2i\sqrt{z})^\nu
}{
\sqrt{2\pi}\,2^{\nu+\frac12}
}
+\cdots
\Bigg].
\label{eq:q_p_minus_one_explicit}
\end{align}
This representation makes explicit the coexistence of algebraic
power-law branches and rapidly oscillating exponential branches at the
irregular singular point $z=\infty$.

For a larger separation, $q\leq p-2$, the dominant asymptotic
representation in the principal sector takes the form
\begin{equation}
\frac{
\displaystyle\prod_{i=1}^{q}\Gamma(a_i)
}{
\displaystyle\prod_{j=1}^{p}\Gamma(b_j)
}
{}_qF_p
\left(
\begin{matrix}
\mathbf a\\
\mathbf b
\end{matrix}
;-z
\right)
\sim
E_{q,p}\!\left(ze^{-i\pi}\right)
+
E_{q,p}\!\left(ze^{+i\pi}\right),
\label{eq:q_less_p_large_gap}
\end{equation}
with additional exponentially subdominant contributions becoming
relevant upon crossing the corresponding Stokes lines. Hence the
precise asymptotic content depends on the sector in the complex
$z$-plane.

\subsection{The case \texorpdfstring{$\boldsymbol{q>p}$}{q > p}}
\label{app:hypergeometric_q_greater_p}

For $q>p$, an important distinction must be made between $q=p+1$ and
$q>p+1$. The defining series
\begin{equation}
{}_qF_p
\left(
\begin{matrix}
a_1,\ldots,a_q\\
b_1,\ldots,b_p
\end{matrix}
;z
\right)
=
\sum_{n=0}^{\infty}
\frac{\prod_{i=1}^{q}(a_i)_n}
     {\prod_{j=1}^{p}(b_j)_n}
\frac{z^n}{n!}
\label{eq:q_greater_p_definition}
\end{equation}
has a nonzero radius of convergence only in the adjacent case
\begin{equation}
q=p+1.
\end{equation}
For $q=p+1$ the radius of convergence is unity,
\begin{equation}
|z|<1,
\end{equation}
and the function outside the unit disk is defined by analytic
continuation. The corresponding differential equation has three
regular singular points,
\begin{equation}
z=0,
\qquad
z=1,
\qquad
z=\infty.
\end{equation}

Around $z=0$ one again finds
\begin{align}
{}_{p+1}F_p(\mathbf a;\mathbf b;z)
={}
1
+
\frac{\displaystyle\prod_{i=1}^{p+1}a_i}
     {\displaystyle\prod_{j=1}^{p}b_j}\,z +
\frac{1}{2}
\frac{\displaystyle\prod_{i=1}^{p+1}a_i(a_i+1)}
     {\displaystyle\prod_{j=1}^{p}b_j(b_j+1)}
z^2
+\mathcal O(z^3).
\label{eq:p_plus_one_p_small_z}
\end{align}

The behavior near $z=1$ is controlled by the parametric excess
\begin{equation}
s
=
\sum_{j=1}^{p}b_j
-
\sum_{i=1}^{p+1}a_i.
\label{eq:parametric_excess}
\end{equation}
For generic noninteger $s$, the local structure contains an analytic
part together with a nonanalytic branch,
\begin{equation}
{}_{p+1}F_p(\mathbf a;\mathbf b;z)
=
\mathcal A(z)
+
\frac{
\displaystyle\prod_{j=1}^{p}\Gamma(b_j)
}{
\displaystyle\prod_{i=1}^{p+1}\Gamma(a_i)
}
\Gamma(-s)
(1-z)^s
\left[
1+\mathcal O(1-z)
\right],
\label{eq:p_plus_one_p_near_one}
\end{equation}
where $\mathcal A(z)$ is analytic at $z=1$. Therefore, when
$\Re(s)<0$, the singular contribution behaves as
\begin{equation}
{}_{p+1}F_p(\mathbf a;\mathbf b;z)
\underset{z\rightarrow1}{\sim}
\frac{
\displaystyle\prod_{j=1}^{p}\Gamma(b_j)
}{
\displaystyle\prod_{i=1}^{p+1}\Gamma(a_i)
}
\Gamma(-s)
(1-z)^s.
\label{eq:p_plus_one_p_singular_one}
\end{equation}

For the zero-balanced case,
\begin{equation}
s=0,
\end{equation}
the power-law branch degenerates into a logarithm,
\begin{equation}
{}_{p+1}F_p(\mathbf a;\mathbf b;z)
=
-
\frac{
\displaystyle\prod_{j=1}^{p}\Gamma(b_j)
}{
\displaystyle\prod_{i=1}^{p+1}\Gamma(a_i)
}
\log(1-z)
+
\mathcal O(1),
\qquad
z\rightarrow1.
\label{eq:p_plus_one_p_zero_balanced}
\end{equation}
More generally, when the parametric excess is an integer, resonances
between the local exponents may generate logarithmic terms. For
$s=m\in\mathbb Z_{\geq0}$, the first nonanalytic contribution has the
form
\begin{equation}
{}_{p+1}F_p(\mathbf a;\mathbf b;z)
\supset
\frac{
\displaystyle\prod_{j=1}^{p}\Gamma(b_j)
}{
\displaystyle\prod_{i=1}^{p+1}\Gamma(a_i)
}
\frac{(-1)^{m+1}}{m!}
(1-z)^m\log(1-z).
\label{eq:p_plus_one_p_integer_excess}
\end{equation}

The large-$z$ behavior is purely algebraic for generic parameters.
Provided that no two upper parameters differ by an integer, the exact
connection formula can be written as
\begin{align}
{}_{p+1}F_p
\left(
\begin{matrix}
a_1,\ldots,a_{p+1}\\
b_1,\ldots,b_p
\end{matrix}
;z
\right)&
={}
\sum_{m=1}^{p+1}
C_m
(-z)^{-a_m}
\nonumber\\
&\times
{}_{p+1}F_p
\left(
\begin{matrix}
a_m,\,
1-b_1+a_m,\ldots,1-b_p+a_m
\\
1-a_1+a_m,\ldots,
\widehat{1-a_m+a_m},
\ldots,
1-a_{p+1}+a_m
\end{matrix}
;\frac1z
\right),
\label{eq:p_plus_one_p_connection_infinity}
\end{align}
where the hat denotes omission of the corresponding entry and
\begin{equation}
C_m
=
\frac{
\displaystyle
\prod_{j=1}^{p}\Gamma(b_j)
\prod_{\substack{i=1\\i\neq m}}^{p+1}
\Gamma(a_i-a_m)
}{
\displaystyle
\prod_{\substack{i=1\\i\neq m}}^{p+1}
\Gamma(a_i)
\prod_{j=1}^{p}
\Gamma(b_j-a_m)
}.
\label{eq:p_plus_one_p_Cm}
\end{equation}
Consequently,
\begin{equation}
{}_{p+1}F_p(\mathbf a;\mathbf b;z)
\underset{|z|\rightarrow\infty}{\sim}
\sum_{m=1}^{p+1}
C_m
(-z)^{-a_m}
\left[
1+\mathcal O(z^{-1})
\right],
\label{eq:p_plus_one_p_large_z}
\end{equation}
with the branches of $(-z)^{-a_m}$ fixed by the chosen analytic
continuation.

Finally, if
\begin{equation}
q>p+1,
\end{equation}
the defining generalized hypergeometric series diverges for every
$z\neq0$ unless at least one upper parameter is a non-positive integer,
in which case the series terminates and reduces to a polynomial.

\section{Direct collapsed-limit Schwinger--Keldysh integrals}
\label{app:sk_collapsed}

The following derivation fixes the four contour components directly at
small exchanged momentum.  It uses the mode convention of
Eq.~\eqref{eq:sigma-mode2}, keeps the time-ordered contact contribution,
and supplies independent boundary data for the WKB-free homogeneous
basis. 

The quartic analysis uses the positive-frequency Bunch--Davies $H^{(1)}$ mode for the pure-$k^4$ dispersion. The choice of positive frequency and the mode normalizations are stated explicitly below. Every phase factor is retained. These are leading soft/internal-momentum formulas, not claims about the full finite-$s$ seed.

Take the following seed function, 
\begin{equation}
\begin{aligned}
 \Ihat_{ab}^{\ell_1\ell_2}
 ={}&-ab\,s^{5+\ell_1+\ell_2}
 \int_{-\infty}^{0}\!\dd\eta_1\,\dd\eta_2\,
 (-\eta_1)^{\ell_1}(-\eta_2)^{\ell_2} \ee^{a\ii k_{12}\eta_1+b\ii k_{34}\eta_2}
 G_{ab}(s,\eta_1,\eta_2).
\end{aligned}
 \label{eq:skcoll_seeddef}
\end{equation}
Introduce $t_j=-\eta_j>0$ and
\begin{equation}
 u=\frac{s}{k_{12}},\qquad v=\frac{s}{k_{34}},\qquad
 p=5+\ell_1+\ell_2.
 \label{eq:skcoll_defs}
\end{equation}
The collapsed limit is $u,v\to0$ with $u/v=k_{34}/k_{12}$ fixed. For the power-series representations below we take $0<u<v$, so $|u/v|<1$; \emph{the ratio need not tend to zero}. The time integrals are understood with the respective Schwinger--Keldysh $\ii\epsilon$ prescriptions. The elementary Fourier transform is
\begin{equation}
 \int_0^\infty\!\dd t\,t^{p-1}\ee^{-a\ii Kt}
 =\ee^{-a\ii\pi p/2}\G(p)K^{-p},\qquad K>0,\quad a=\pm1.
 \label{eq:skcoll_fourier}
\end{equation}
It implies the mass-dependent phase
$\ee^{-a\ii\pi (\ell+\frac52+\sigma\ii\mu)/2}
 =\ee^{-a\ii\pi(\ell+5/2)/2}\ee^{a\sigma\pi\mu/2}$.
For the ordered sectors, $\eta_1>\eta_2$ is equivalent to $t_1<t_2$.

\subsection{Pure quartic dispersion}\label{sec:skcoll_k4}
\paragraph{Mode and small-$s$ expansion.}
Positive frequency is fixed by the canonical adiabatic condition
$\exp[-\ii\int^\eta \omega_s(\eta')\dd\eta']$ with
$\omega_s(\eta)=\gamma s^2(-\eta)>0$ for $\eta<0$.  Since
$-\int^\eta\omega_s\dd\eta'=\gamma s^2\eta^2/2+\mathrm{const.}$,
the positive-frequency phase is $\ee^{+\ii\gamma s^2\eta^2/2}$.
We therefore use, for real $H,\gamma,\mu>0$,
\begin{equation}
 \boxed{\chi_s(-t)=\frac{H\sqrt\pi}{\sqrt8}\,
 \ee^{\ii\pi/4}\ee^{-\pi\mu/4}
 t^{3/2} H^{(1)}_{\ii\mu/2}\!\left(\frac{\gamma s^2t^2}{2}\right).}
 \label{eq:skcoll_k4mode}
\end{equation}
It has the early-time behavior
\begin{equation}
 \chi_s(-t)\underset{t\to\infty}{\sim}
 \frac{H\sqrt{t}}{\sqrt{2\gamma}\,s}\,
 \ee^{+\ii\gamma s^2t^2/2},\qquad
 \chi_s\chi_s^{*\prime}-\chi_s^*\chi_s'=\ii H^2\eta^2,
 \label{eq:skcoll_k4bdcheck}
\end{equation}
where the prime denotes $\partial_\eta$.
For real positive argument, $[H_{\ii\mu/2}^{(1)}]^*
=e^{\pi\mu/2}H_{\ii\mu/2}^{(2)}$.
With $q=\gamma s^2t^2/4$, the leading small-$s$ expansions are
\begin{align}
 \chi_s(-t)&\sim\frac{H\ee^{-\ii\pi/4}}{\sqrt{8\pi}}t^{3/2}
 \left[\ee^{-\pi\mu/4}\G\!\left(\frac{\ii\mu}{2}\right)q^{-\ii\mu/2}
 +\ee^{\pi\mu/4}\G\!\left(-\frac{\ii\mu}{2}\right)q^{\ii\mu/2}\right],\label{eq:skcoll_k4soft}\displaybreak[1]\\
 \chi_s^*(-t)&\sim\frac{H\ee^{\ii\pi/4}}{\sqrt{8\pi}}t^{3/2}
 \left[\ee^{\pi\mu/4}\G\!\left(\frac{\ii\mu}{2}\right)q^{-\ii\mu/2}
 +\ee^{-\pi\mu/4}\G\!\left(-\frac{\ii\mu}{2}\right)q^{\ii\mu/2}\right].\label{eq:skcoll_k4softcc}
\end{align}
The normalized collider powers are already defined by the small-$u$
behavior of $\Phi_\pm^\ell$ in
Eq.~\eqref{eq:Phi_asymptotic_definition}. In this appendix, $\Phi_\pm^\ell$
in the leading-soft expressions below always denotes that leading asymptotic
branch, not an equality to the full global function at finite $u$.
The commutator-like difference entering the time-ordering correction is
\begin{equation}
 \boxed{\chi_s^*(-t_1)\chi_s(-t_2)-\chi_s(-t_1)\chi_s^*(-t_2)
 \sim-\frac{H^2}{2\mu}(t_1t_2)^{3/2}
 \left[\left(\frac{t_1}{t_2}\right)^{\ii\mu}
 -\left(\frac{t_1}{t_2}\right)^{-\ii\mu}\right].}
 \label{eq:skcoll_k4comm}
\end{equation}
Here we have used $\G(\ii x)\G(-\ii x)=\pi/[x\sinh(\pi x)]$.
The $s^{\pm\ii\mu}$ factors cancel in this difference.

\subsection*{The $++$ component}\label{sub:skcoll_k4pp}
For $t_j>0$, the time-ordered propagator is identically
\begin{align}
 G_{++}(-t_1,-t_2)
 &=\chi_s(-t_1)\chi_s^*(-t_2)
 +\Theta(t_1-t_2)\big[\chi_s^*(-t_1)\chi_s(-t_2)
 -\chi_s(-t_1)\chi_s^*(-t_2)\big].\label{eq:skcoll_k4ppdecomp}
\end{align}
Thus the first piece is unrestricted and factorizes; the second is integrated over $t_1>t_2$.
In the first piece, $\chi_s$ at vertex $u$ carries $\ee^{+\sigma\pi\mu/4}$ before the Fourier integral, while $\chi_s^*$ at $v$ carries $\ee^{-\sigma\pi\mu/4}$. Both vertex integrals use $a=+1$ and supply $\ee^{\sigma\pi\mu/2}$. Therefore
\begin{equation}
 \boxed{\begin{aligned}
 \Ihat_{++,\mathrm{fact}}^{\ell_1\ell_2}\big|_{k^4}
 \sim{}&\frac{\ii H^2\ee^{-\ii\pi(\ell_1+\ell_2)/2}}{8\pi}
 \big[\ee^{-3\pi\mu/4}\Phi_-^{\ell_1}(u)
 +\ee^{3\pi\mu/4}\Phi_+^{\ell_1}(u)\big]\\[-2pt]
 &\times\big[\ee^{-\pi\mu/4}\Phi_-^{\ell_2}(v)
 +\ee^{\pi\mu/4}\Phi_+^{\ell_2}(v)\big].
 \end{aligned}}\label{eq:skcoll_k4ppfact}
\end{equation}
For the correction, insert Eq.~\eqref{eq:skcoll_k4comm} into Eq.~\eqref{eq:skcoll_seeddef} and set $t_2=xt_1$:
\begin{align}
 \Ihat_{++,\Theta}^{\ell_1\ell_2}\big|_{k^4}
 &=\frac{H^2s^p}{2\mu}\int_0^\infty\!\dd t_1\int_0^{t_1}\!\dd t_2\,
 t_1^{\ell_1+3/2}t_2^{\ell_2+3/2}
 \ee^{-\ii k_{12}t_1-\ii k_{34}t_2}
 \left[\left(\frac{t_1}{t_2}\right)^{\ii\mu}
 -\left(\frac{t_1}{t_2}\right)^{-\ii\mu}\right]\notag\\
 &=\frac{H^2\G(p)\ee^{-\ii\pi p/2}}{2\mu}u^p
 \int_0^1\!\dd x\,\frac{x^{\ell_2+3/2-\ii\mu}-x^{\ell_2+3/2+\ii\mu}}
 {(1+(u/v)x)^p}.\label{eq:skcoll_k4ppint}
\end{align}
Expanding the $x$-integral for $0<u<v$ gives the analytic fixed-ratio series:
\begin{equation}
 \boxed{\Ihat_{++,\Theta}^{\ell_1\ell_2}\big|_{k^4}
 =H^2\ee^{-\ii\pi(\ell_1+\ell_2)/2}\G(p)u^p
 \sum_{n=0}^\infty\frac{(-u/v)^n(p)_n}
 {n!\big[(n+\ell_2+\tfrac52)^2+\mu^2\big]}.}
 \label{eq:skcoll_k4ppseries}
\end{equation}
The complete leading-soft $++$ integral is Eq.~\eqref{eq:skcoll_k4ppfact} plus Eq.~\eqref{eq:skcoll_k4ppseries}. The latter is analytic in $u/v$ in its disk of convergence; it is not part of the pure collider-branch projection.

\subsection*{The $--$ component}\label{sub:skcoll_k4mm}
The exact $--$ propagator can be written using the same domain $t_1>t_2$:
\begin{equation}
 G_{--}(-t_1,-t_2)=\chi_s^*(-t_1)\chi_s(-t_2)
 +\Theta(t_1-t_2)\big[\chi_s(-t_1)\chi_s^*(-t_2)-\chi_s^*(-t_1)\chi_s(-t_2)\big].
 \label{eq:skcoll_k4mmdecomp}
\end{equation}
The time exponential is now $\ee^{+\ii K_1t_1+\ii K_2t_2}$.
The Fourier factors supply $\ee^{-\sigma\pi\mu/2}$ at each vertex, giving
\begin{equation}
 \boxed{\begin{aligned}
 \Ihat_{--,\mathrm{fact}}^{\ell_1\ell_2}\big|_{k^4}
 \sim{}&-\frac{\ii H^2\ee^{\ii\pi(\ell_1+\ell_2)/2}}{8\pi}
 \big[\ee^{3\pi\mu/4}\Phi_-^{\ell_1}(u)
 +\ee^{-3\pi\mu/4}\Phi_+^{\ell_1}(u)\big]\\[-2pt]
 &\times\big[\ee^{\pi\mu/4}\Phi_-^{\ell_2}(v)
 +\ee^{-\pi\mu/4}\Phi_+^{\ell_2}(v)\big].
 \end{aligned}}\label{eq:skcoll_k4mmfact}
\end{equation}
The bracket in Eq.~\eqref{eq:skcoll_k4mmdecomp} is the negative of
Eq.~\eqref{eq:skcoll_k4comm}. Including the overall $-1$ from the
$--$ seed gives the corresponding analytic series:

\begin{equation}
 \boxed{\Ihat_{--,\Theta}^{\ell_1\ell_2}\big|_{k^4}
 =H^2\ee^{\ii\pi(\ell_1+\ell_2)/2}\G(p)u^p
 \sum_{n=0}^\infty\frac{(-u/v)^n(p)_n}
 {n!\big[(n+\ell_2+\tfrac52)^2+\mu^2\big]}.}
 \label{eq:skcoll_k4mmseries}
\end{equation}
The full leading-soft $--$ contribution is Eqs.~\eqref{eq:skcoll_k4mmfact} and \eqref{eq:skcoll_k4mmseries}. Since $G_{--}=G_{++}^*$, these results obey
$\Ihat_{--}^{\ell_1\ell_2}=(\Ihat_{++}^{\ell_1\ell_2})^*$ for real parameters.

\subsection*{The $+-$ component}\label{sub:skcoll_k4pm}
There is no time-ordering function:
$G_{+-}(s,-t_1,-t_2)=\chi_s^*(-t_1)\chi_s(-t_2)$.
The seed factor is $-ab=+1$, and the Fourier exponentials are
$\ee^{-\ii k_{12}t_1}$ and $\ee^{+\ii k_{34}t_2}$. Thus the mass-dependent factors at the first vertex are $\ee^{-\sigma\pi\mu/4}\ee^{\sigma\pi\mu/2}=\ee^{\sigma\pi\mu/4}$,
whereas those at the second vertex are $\ee^{\sigma\pi\mu/4}\ee^{-\sigma\pi\mu/2}=\ee^{-\sigma\pi\mu/4}$.
Direct use of Eq.~\eqref{eq:skcoll_fourier} gives
\begin{equation}
 \boxed{\begin{aligned}
 \Ihat_{+-}^{\ell_1\ell_2}\big|_{k^4,\mathrm{coll}}
 \sim{}&\frac{H^2\ee^{-\ii\pi(\ell_1-\ell_2)/2}}{8\pi}
 \big[\ee^{-\pi\mu/4}\Phi_-^{\ell_1}(u)
 +\ee^{\pi\mu/4}\Phi_+^{\ell_1}(u)\big]\\[-2pt]
 &\times\big[\ee^{\pi\mu/4}\Phi_-^{\ell_2}(v)
 +\ee^{-\pi\mu/4}\Phi_+^{\ell_2}(v)\big].
 \end{aligned}}\label{eq:skcoll_k4pm}
\end{equation}

\subsection*{The $-+$ component}\label{sub:skcoll_k4mp}
Again, the integral factorizes. Now
$G_{-+}(s,-t_1,-t_2)=\chi_s(-t_1)\chi_s^*(-t_2)$, while the Fourier phases are $\ee^{+\ii k_{12}t_1}$ and $\ee^{-\ii k_{34}t_2}$. The vertex weights reverse those of $+-$:
\begin{equation}
 \boxed{\begin{aligned}
 \Ihat_{-+}^{\ell_1\ell_2}\big|_{k^4,\mathrm{coll}}
 \sim{}&\frac{H^2\ee^{\ii\pi(\ell_1-\ell_2)/2}}{8\pi}
 \big[\ee^{\pi\mu/4}\Phi_-^{\ell_1}(u)
 +\ee^{-\pi\mu/4}\Phi_+^{\ell_1}(u)\big]\\[-2pt]
 &\times\big[\ee^{-\pi\mu/4}\Phi_-^{\ell_2}(v)
 +\ee^{\pi\mu/4}\Phi_+^{\ell_2}(v)\big].
 \end{aligned}}\label{eq:skcoll_k4mp}
\end{equation}
For real parameters $\Ihat_{-+}=(\Ihat_{+-})^*$, providing a direct check on the phases.

The step-function contribution to the directly evaluated ordered
integral is, for $0<u<v$,
\begin{equation}
 \widehat{\mathcal I}_{aa,\Theta}^{\ell_1\ell_2}
 =H^2e^{-ai\pi(\ell_1+\ell_2)/2}\Gamma(p)u^p
 \sum_{n=0}^\infty\frac{(-u/v)^n(p)_n}{n!
 [(n+\ell_2+\frac52)^2+\mu^2]}.
 \label{eq:skcoll_BL_matching}
\end{equation}
This agrees term by term with the leading expression
Eq.~\eqref{eq:k4_bl_collapsed_fixed_ratio}, for the full analytic
fixed-ratio series rather than merely its first term.
The direct time-ordering integral and the Borel--Laplace particular
integral are not thereby identified at finite $s$: equal soft asymptotics
cannot determine possible subleading homogeneous contributions.
Likewise, the normalized factorized products fix the leading collider
coefficients, whereas the global WKB-free completion requires the
separate selection of asymptotic sectors described in
Section~\ref{subsec:k4_homogeneous_solution}.

\section{dS Bootstrap Details}
\label{app:ds_bootstrap_details}

The following appendix contains the detailed calculations moved out of the main de Sitter section: the Frobenius derivation of the homogeneous solutions, the construction of the non-homogeneous soft-series solution, and the folded-limit cancellation check.

\subsection{Homogeneous solutions}

We now solve the homogeneous equation,
\begin{equation}
    \mathcal D_u^\ell\,\mathcal Y^\ell(u)=0.
    \label{eq:hom_bootstrap_eq_usual}
\end{equation}
Since the soft point \(u=0\) is a regular singular point, we use a Frobenius ansatz
\begin{equation}
    \mathcal Y^\ell_\sigma(u)=u^{\rho_\sigma}F_\sigma(u^2),
    \qquad
    \sigma=\pm1,
\end{equation}
with
\begin{equation}
    \rho_\sigma=\ell+\frac52+\sigma\ii\mu .
\end{equation}
We now solve the homogeneous equation
\begin{equation}
    \mathcal D_u^\ell\,\mathcal Y^\ell(u)=0 .
    \label{eq:hom_bootstrap_eq_usualone}
\end{equation}
In the conventions used here, the relevant differential operator is
\begin{equation}
    \mathcal D_u^\ell
    =
    u^2(1-u^2)\frac{d^2}{du^2}
    -2u(\ell+2+u^2)\frac{d}{du}
    +
    \left[
    \left(\ell+\frac52\right)^2+\mu^2
    \right] .
    \label{eq:Dell_operator_usualtwo}
\end{equation}
Thus the homogeneous equation reads
\begin{equation}
    \left\{
    u^2(1-u^2)\frac{d^2}{du^2}
    -2u(\ell+2+u^2)\frac{d}{du}
    +
    \left[
    \left(\ell+\frac52\right)^2+\mu^2
    \right]
    \right\}
    \mathcal Y^\ell(u)=0 .
    \label{eq:hom_eq_expanded_usual}
\end{equation}
The point \(u=0\), which corresponds to the soft point, is a regular singular point of this equation. Therefore, instead of using an ordinary Taylor expansion, we use a Frobenius expansion. The essential idea of the Frobenius method is to allow the leading behavior of the solution to be a non-integer, or even complex, power of \(u\). In its most general form, one would write
\begin{equation}
    \mathcal Y^\ell(u)
    =
    u^\rho
    \sum_{n=0}^{\infty} b_n u^n ,
    \qquad b_0\neq 0 .
    \label{eq:frobenius_general_series}
\end{equation}
The exponent \(\rho\) is not chosen by hand. It is determined by the differential equation itself through the indicial equation. In the present problem, after factoring out the leading power \(u^\rho\), the remaining equation depends naturally on \(u^2\). It is therefore convenient to write the Frobenius ansatz as
\begin{equation}
    \mathcal Y^\ell_\sigma(u)
    =
    u^{\rho_\sigma}
    F_\sigma(u^2),
    \qquad
    \sigma=\pm 1,
    \label{eq:frobenius_ansatz_usual}
\end{equation}
where
\begin{equation}
    F_\sigma(u^2)
    =
    \sum_{n=0}^{\infty} a_n^{(\sigma)} u^{2n}.
    \label{eq:F_sigma_series_u}
\end{equation}
Equivalently, introducing
\begin{equation}
    z=u^2 ,
    \label{eq:z_def}
\end{equation}
we write
\begin{equation}
    F_\sigma(z)
    =
    \sum_{n=0}^{\infty} a_n^{(\sigma)} z^n .
    \label{eq:F_sigma_series_z}
\end{equation}
Thus the Frobenius ansatz is explicitly
\begin{equation}
    \mathcal Y^\ell_\sigma(u)
    =
    u^{\rho_\sigma}
    \sum_{n=0}^{\infty} a_n^{(\sigma)} u^{2n}
    =
    a_0^{(\sigma)}u^{\rho_\sigma}
    +a_1^{(\sigma)}u^{\rho_\sigma+2}
    +a_2^{(\sigma)}u^{\rho_\sigma+4}
    +\cdots .
    \label{eq:Y_frobenius_series_open}
\end{equation}
This is the sense in which the Frobenius ansatz is a series expansion. It differs from an ordinary Taylor series because the leading power \(u^{\rho_\sigma}\) is not fixed to be \(u^0\). Instead, the exponent \(\rho_\sigma\) is determined dynamically by the differential equation.

Let us now derive \(\rho_\sigma\) and the function \(F_\sigma\). Define
\begin{equation}
    L\equiv \ell+\frac52 .
    \label{eq:L_def}
\end{equation}
Using \(z=u^2\), the ansatz
\begin{equation}
    \mathcal Y^\ell(u)=u^\rho F(z)
    \label{eq:ansatz_temp}
\end{equation}
gives
\begin{equation}
    \frac{d\mathcal Y^\ell}{du}
    =
    u^{\rho-1}
    \left[
    \rho F(z)+2zF'(z)
    \right],
    \label{eq:first_derivative_frobenius}
\end{equation}
and
\begin{equation}
    \frac{d^2\mathcal Y^\ell}{du^2}
    =
    u^{\rho-2}
    \left[
    \rho(\rho-1)F(z)
    +2z(2\rho+1)F'(z)
    +4z^2F''(z)
    \right],
    \label{eq:second_derivative_frobenius}
\end{equation}
where primes denote derivatives with respect to \(z\). Substituting these expressions into Eq.~\eqref{eq:hom_eq_expanded_usual}, and dividing by the common factor \(u^\rho\), we obtain
\begin{align}
0
={}&
(1-z)
\left[
\rho(\rho-1)F
+2z(2\rho+1)F'
+4z^2F''
\right]
\nonumber\\
&-2(\ell+2+z)
\left[
\rho F+2zF'
\right]
+
\left[
\left(\ell+\frac52\right)^2+\mu^2
\right]F .
\label{eq:substituted_frobenius}
\end{align}
The leading term near \(z=0\) gives the indicial equation. Keeping only the terms independent of \(z\), we find
\begin{equation}
    \rho(\rho-1)
    -2(\ell+2)\rho
    +
    \left(\ell+\frac52\right)^2+\mu^2
    =
    0 .
    \label{eq:indicial_before_simplification}
\end{equation}
Using \(L=\ell+5/2\), or equivalently \(\ell+2=L-1/2\), this becomes
\begin{equation}
    \rho(\rho-1)
    -2\left(L-\frac12\right)\rho
    +L^2+\mu^2
    =
    0 .
    \label{eq:indicial_L}
\end{equation}
The first two terms simplify as
\begin{equation}
    \rho(\rho-1)
    -2\left(L-\frac12\right)\rho
    =
    \rho^2-2L\rho ,
\end{equation}
and therefore the indicial equation is
\begin{equation}
    \rho^2-2L\rho+L^2+\mu^2=0 .
    \label{eq:indicial_quadratic}
\end{equation}
Equivalently,
\begin{equation}
    (\rho-L)^2+\mu^2=0 .
    \label{eq:indicial_compact}
\end{equation}
Hence the two independent Frobenius exponents are
\begin{equation}
    \rho_\sigma
    =
    L+\sigma i\mu
    =
    \ell+\frac52+\sigma i\mu ,
    \qquad
    \sigma=\pm 1 .
    \label{eq:rho_sigma_result}
\end{equation}
These two roots correspond to the two independent non-analytic soft behaviors
\begin{equation}
    \mathcal Y^\ell_\sigma(u)
    \sim
    u^{\ell+\frac52+\sigma i\mu},
    \qquad
    u\rightarrow 0 .
    \label{eq:soft_behavior_homogeneous}
\end{equation}
We now determine the full function \(F_\sigma(z)\). Once the indicial equation is imposed, the remaining terms in Eq.~\eqref{eq:substituted_frobenius} give
\begin{align}
0
={}&
4z^2(1-z)F_\sigma''
+
2z
\left[
(1-z)(2\rho_\sigma+1)
-2(\ell+2+z)
\right]F_\sigma'
\nonumber\\
&-
z\rho_\sigma(\rho_\sigma+1)F_\sigma .
\label{eq:F_equation_before_division}
\end{align}
Dividing by \(4z\), we find
\begin{equation}
    z(1-z)F_\sigma''
    +
    \frac12
    \left[
    (1-z)(2\rho_\sigma+1)
    -2(\ell+2+z)
    \right]F_\sigma'
    -
    \frac{\rho_\sigma(\rho_\sigma+1)}{4}
    F_\sigma
    =
    0 .
    \label{eq:F_equation_intermediate}
\end{equation}
The coefficient of \(F_\sigma'\) may be simplified using
\begin{equation}
    \ell+2=L-\frac12,
    \qquad
    \rho_\sigma=L+\sigma i\mu .
\end{equation}
Indeed,
\begin{align}
\frac12
\left[
(1-z)(2\rho_\sigma+1)
-2(\ell+2+z)
\right]
&=
\rho_\sigma-\ell-\frac32
-
\left(\rho_\sigma+\frac32\right)z
\nonumber\\
&=
1+\sigma i\mu
-
\left(\rho_\sigma+\frac32\right)z .
\label{eq:Fprime_coefficient_simplified}
\end{align}
Therefore Eq.~\eqref{eq:F_equation_intermediate} becomes
\begin{equation}
    z(1-z)F_\sigma''
    +
    \left[
    1+\sigma i\mu
    -
    \left(\rho_\sigma+\frac32\right)z
    \right]F_\sigma'
    -
    \frac{\rho_\sigma(\rho_\sigma+1)}{4}
    F_\sigma
    =
    0 .
    \label{eq:F_hypergeometric_form_pre}
\end{equation}
This is the Gauss hypergeometric equation,
\begin{equation}
    z(1-z)F''
    +
    \left[
    c-(a+b+1)z
    \right]F'
    -
    abF
    =
    0 ,
    \label{eq:gauss_hypergeometric_equation}
\end{equation}
with parameters
\begin{equation}
    a_\sigma=\frac{\rho_\sigma}{2},
    \qquad
    b_\sigma=\frac{\rho_\sigma+1}{2},
    \qquad
    c_\sigma=1+\sigma i\mu .
    \label{eq:hypergeometric_parameters}
\end{equation}
Indeed,
\begin{equation}
    a_\sigma+b_\sigma+1
    =
    \frac{\rho_\sigma}{2}
    +
    \frac{\rho_\sigma+1}{2}
    +1
    =
    \rho_\sigma+\frac32 ,
\end{equation}
and
\begin{equation}
    a_\sigma b_\sigma
    =
    \frac{\rho_\sigma}{2}
    \frac{\rho_\sigma+1}{2}
    =
    \frac{\rho_\sigma(\rho_\sigma+1)}{4}.
\end{equation}
Let us also see explicitly how the hypergeometric series arises. We insert the Taylor expansion
\begin{equation}
    F_\sigma(z)
    =
    \sum_{n=0}^{\infty}a_n^{(\sigma)}z^n
    \label{eq:F_series_again}
\end{equation}
into Eq.~\eqref{eq:gauss_hypergeometric_equation}. The derivatives are
\begin{equation}
    F_\sigma'(z)
    =
    \sum_{n=1}^{\infty}
    n\,a_n^{(\sigma)}z^{n-1},
    \qquad
    F_\sigma''(z)
    =
    \sum_{n=2}^{\infty}
    n(n-1)a_n^{(\sigma)}z^{n-2}.
    \label{eq:F_series_derivatives}
\end{equation}
Collecting equal powers of \(z\), one obtains the recursion relation
\begin{equation}
    (n+1)(n+c_\sigma)a_{n+1}^{(\sigma)}
    -
    (n+a_\sigma)(n+b_\sigma)a_n^{(\sigma)}
    =
    0 ,
    \qquad n=0,1,2,\ldots .
    \label{eq:hypergeometric_recursion}
\end{equation}
Hence
\begin{equation}
    a_{n+1}^{(\sigma)}
    =
    \frac{(n+a_\sigma)(n+b_\sigma)}
    {(n+1)(n+c_\sigma)}
    a_n^{(\sigma)} .
    \label{eq:recursion_relation_explicit}
\end{equation}
Iterating this relation gives
\begin{equation}
    a_n^{(\sigma)}
    =
    a_0^{(\sigma)}
    \frac{
    (a_\sigma)_n(b_\sigma)_n
    }{
    (c_\sigma)_n\,n!
    },
    \label{eq:an_solution_pochhammer}
\end{equation}
where \((q)_n\) denotes the Pochhammer symbol,
\begin{equation}
    (q)_n
    =
    q(q+1)(q+2)\cdots(q+n-1),
    \qquad
    (q)_0=1 .
    \label{eq:pochhammer_def}
\end{equation}
Therefore
\begin{equation}
    F_\sigma(z)
    =
    a_0^{(\sigma)}
    \sum_{n=0}^{\infty}
    \frac{
    (a_\sigma)_n(b_\sigma)_n
    }{
    (c_\sigma)_n\,n!
    }
    z^n .
    \label{eq:F_hypergeometric_series}
\end{equation}
This is precisely the series representation of the Gauss hypergeometric function,
\begin{equation}
    {}_2F_1(a,b;c;z)
    =
    \sum_{n=0}^{\infty}
    \frac{(a)_n(b)_n}{(c)_n\,n!}z^n .
    \label{eq:gauss_hypergeometric_series}
\end{equation}
Thus
\begin{equation}
    F_\sigma(z)
    =
    a_0^{(\sigma)}
    {}_2F_1
    \left(
    a_\sigma,
    b_\sigma;
    c_\sigma;
    z
    \right).
    \label{eq:F_solution_hypergeometric}
\end{equation}
Using Eq.~\eqref{eq:hypergeometric_parameters}, this becomes
\begin{equation}
    F_\sigma(z)
    =
    a_0^{(\sigma)}
    {}_2F_1
    \left(
    \frac{\rho_\sigma}{2},
    \frac{\rho_\sigma+1}{2};
    1+\sigma i\mu;
    z
    \right).
    \label{eq:F_solution_parameters}
\end{equation}
Returning to \(z=u^2\), the two homogeneous solutions are therefore
\begin{equation}
    \mathcal Y^\ell_\sigma(u)
    =
    a_0^{(\sigma)}
    u^{\rho_\sigma}
    {}_2F_1
    \left(
    \frac{\rho_\sigma}{2},
    \frac{\rho_\sigma+1}{2};
    1+\sigma i\mu;
    u^2
    \right),
    \qquad
    \sigma=\pm1 .
    \label{eq:Y_sigma_solution_rho}
\end{equation}
Substituting
\begin{equation}
    \rho_\sigma=\ell+\frac52+\sigma\ii\mu ,
\end{equation}
we obtain
\begin{equation}
    \mathcal Y^\ell_\sigma(u)
    =
    a_0^{(\sigma)}
    u^{\ell+\frac52+\sigma i\mu}
    {}_2F_1
    \left(
    \frac{\ell+\frac52+\sigma i\mu}{2},
    \frac{\ell+\frac72+\sigma i\mu}{2};
    1+\sigma i\mu;
    u^2
    \right).
    \label{eq:Y_sigma_solution_explicit}
\end{equation}
Equivalently, writing the two branches explicitly as \(\sigma=\pm1\),
\begin{equation}
    \mathcal Y^\ell_\pm(u)
    =
    a_0^{(\pm)}
    u^{\ell+\frac52\pm i\mu}
    {}_2F_1
    \left(
    \frac{\ell+\frac52\pm i\mu}{2},
    \frac{\ell+\frac72\pm i\mu}{2};
    1\pm i\mu;
    u^2
    \right).
    \label{eq:Y_pm_solution_explicit}
\end{equation}
In the normalization convention adapted to the collapsed limit, we choose
\begin{equation}
    a_0^{(\pm)}
    =
    2^{\mp i\mu}i^{\ell+\frac52\pm i\mu},
\end{equation}
or, equivalently, we write the solution as
\begin{equation}
        \mathcal Y^\ell_\pm(u)
    =
    2^{\mp i\mu}
    (iu)^{\ell+\frac52\pm i\mu}
    {}_2F_1
    \left(
    \frac{\ell+\frac52\pm i\mu}{2},
    \frac{\ell+\frac72\pm i\mu}{2};
    1\pm i\mu;
    u^2
    \right)
     .
    \label{eq:Y_pm_normalized_solution}
\end{equation}
Near the soft point \(u=0\), the hypergeometric function tends to one,
\begin{equation}
    {}_2F_1(a,b;c;u^2)
    =
    1+\mathcal O(u^2),
    \qquad u\rightarrow 0 .
\end{equation}
Therefore the two homogeneous solutions behave as
\begin{equation}
    \mathcal Y^\ell_\pm(u)
    =
    2^{\mp i\mu}
    (iu)^{\ell+\frac52\pm i\mu}
    \left[
    1+\mathcal O(u^2)
    \right].
    \label{eq:Y_pm_soft_expansion}
\end{equation}
This makes explicit the two oscillatory soft branches
\begin{equation}
    u^{\ell+\frac52+i\mu},
    \qquad
    u^{\ell+\frac52-i\mu}.
\end{equation}
The Frobenius expansion therefore provides the local soft solution, while the identification with the hypergeometric function gives its closed analytic form.

The appearance of \(u^2\) is a direct consequence of the symmetry of the operator under \(u\to -u\). Substituting into Eq.~\eqref{eq:hom_bootstrap_eq_usual}, one obtains the hypergeometric equation for \(F_\sigma\). The parameters are
\begin{equation}
    a_\sigma=\frac{\ell}{2}+\frac54+\frac{\sigma i\mu}{2},
    \qquad
    b_\sigma=\frac{\ell}{2}+\frac74+\frac{\sigma i\mu}{2},
    \qquad
    c_\sigma=1+\sigma i\mu .
\end{equation}
A convenient normalized basis is
\begin{equation}
\mathcal Y^\ell_\sigma(u)
=
2^{-\sigma i\mu}
 u^{\ell+\frac52+\sigma i\mu}
\Gamma\!\left(\ell+\frac52+\sigma i\mu\right)
\Gamma(-\sigma i\mu)
\,{}_2F_1\!\left(
\begin{matrix}
\frac{\ell}{2}+\frac54+\frac{\sigma i\mu}{2},\
\frac{\ell}{2}+\frac74+\frac{\sigma i\mu}{2}\\
1+\sigma i\mu
\end{matrix}
;u^2
\right).
\label{eq:Y_sigma_usual}
\end{equation}
Finally, we have the solution along the singular regular point ($u \to 0$) of the theory,
\begin{equation}
    \mathcal{Y}^{l}(u) = C_+ \mathcal{Y}^l_+(u) + C_-\mathcal{Y}^l_-(u) . 
\end{equation}
The $C_{\pm}$ is a normalization constant that can be fixed by a proper boundary condition.~Taking the collapsed limit of the four-point function in the full de Sitter theory, this normalization constant can be fixed as follows \cite{ArkaniHamedMaldacena_CosmologicalCollider_2015, ArkaniHamedBaumannLeePimentel_CosmologicalBootstrap_2018},
\begin{equation}
    \lim_{u,v \to 0} \hat{F}(u,v) \propto \mathcal{Y}^{l=-2}_+(u)\mathcal{Y}^{l=-2}_- (v) \bigg|_{u,v \to 0} + \cdots
\end{equation}
where,
\begin{equation}
    \mathcal{Y}_{\pm}^{l}(u)\bigg|_{u \to 0} = 2^{\mp i \mu} u^{l+\frac{5}{2}\pm i \mu } \Gamma \bigg(l+ \frac{5}{2} \pm i \mu \bigg) \Gamma (\mp i \mu).
\end{equation}
The phases associated with the Schwinger--Keldysh branches are kept outside the definition of \(\mathcal Y^\ell_\pm\).~With this convention, the folded-limit cancellation in the Bunch--Davies combination is manifest. If instead one absorbs factors such as \(i^{\ell+5/2\pm i\mu}\) into the basis, the same cancellation still occurs, but the compensating phases appear explicitly in the Schwinger--Keldysh prefactors.

For the Wightman Schwinger--Keldysh seed, the squeezed-limit matching fixes the homogeneous product combination \cite{Qin:2023ejc}
\begin{equation}
\hat{\mathcal I}^{\ell_1\ell_2}_{\pm\mp}(u,v)
=
\frac{e^{\mp i\pi(\ell_1-\ell_2)/2}}{4\pi}
\left[\mathcal Y^{\ell_1}_+(u)+\mathcal Y^{\ell_1}_-(u)\right]
\left[\mathcal Y^{\ell_2}_+(v)+\mathcal Y^{\ell_2}_-(v)\right],
\label{eq:I_pmmp_usual}
\end{equation}
This expression is normalized so that the soft behavior agrees with the product of the two late-time massive-field branches.

\paragraph{Folded-limit check for the homogeneous solution.}
We now verify explicitly that the combination
\begin{equation}
    \mathcal Y^\ell_+(u)+\mathcal Y^\ell_-(u)
\end{equation}
is regular as \(u\to1\). This is the folded point, and in the Bunch--Davies vacuum this singularity must cancel.

For each branch,
\begin{align}
    c_\sigma-a_\sigma-b_\sigma
    &
    =
    1+\sigma i\mu
    -\left(\frac{\ell}{2}+\frac54+\frac{\sigma i\mu}{2}\right)
    -\left(\frac{\ell}{2}+\frac74+\frac{\sigma i\mu}{2}\right)
    \\
    &=-\ell-2 .
\end{align}
Thus each individual branch contains a potentially singular contribution of the form
\begin{equation}
    (1-u^2)^{-\ell-2}.
\end{equation}
Using the continuation formula for the hypergeometric function near \(u^2=1\), the singular part of Eq.~\eqref{eq:Y_sigma_usual} is
\begin{equation}
\left.
\mathcal Y^\ell_\sigma(u)
\right|_{\rm sing}
=
\sigma i\sqrt\pi\,
2^{-1/2}
\Gamma(\ell+2)
\operatorname{csch}(\pi\mu)
(1-u)^{-\ell-2}
+\cdots .
\label{eq:Y_singular_part_usual}
\end{equation}
In deriving this expression we used
\begin{equation}
    1-u^2=(1-u)(1+u)
    \simeq 2(1-u)
    \qquad (u\to1),
\end{equation}
together with the duplication formula for gamma functions. The important point is the relative sign between the two branches:
\begin{equation}
\left.
\mathcal Y^\ell_+(u)
\right|_{\rm sing}
= -
\left.
\mathcal Y^\ell_-(u)
\right|_{\rm sing}.
\end{equation}
Therefore
\begin{align}
\left.
\left[\mathcal Y^\ell_+(u)+\mathcal Y^\ell_-(u)\right]
\right|_{\rm sing}
&=
 i\sqrt\pi\,
2^{-1/2}
\Gamma(\ell+2)\operatorname{csch}(\pi\mu)
(1-u)^{-\ell-2}
\nonumber\\
&\quad
-i\sqrt\pi\,
2^{-1/2}
\Gamma(\ell+2)\operatorname{csch}(\pi\mu)
(1-u)^{-\ell-2}
\nonumber\\
&=0 .
\end{align}
Thus the folded singularity cancels between the two homogeneous branches. This cancellation is the explicit implementation of Bunch--Davies regularity in the Wightman seed.

The same check can be performed at the second vertex, \(v\to1\). Therefore the double-folded limit of the Wightman seed is finite:
\begin{equation}
    \hat{\mathcal I}^{\ell_1\ell_2}_{\pm\mp}(1,1)<\infty .
\end{equation}
For the purpose of this section, the most important result is not the detailed finite constant, but the cancellation of the non-analytic folded contribution. The physical factorization singularity is instead at \(u=-1\), and it is not removed by the Bunch--Davies condition.

\subsection{Non-homogeneous solution}

We now construct the non-homogeneous solution for the \(\pm\pm\) Schwinger--Keldysh branches in the same variables \(u,v\). In the single-folded limit \(v=1\), Eq.~\eqref{eq:dS_bootstrap_source} becomes
\begin{equation}
    \mathcal D_u^{\ell_1}
    \hat{\mathcal I}^{\ell_1\ell_2}_{\pm\pm,\mathrm{inh}}(u,1)
    =
    H^2 e^{\mp i\pi\ell_{12}/2}
    \Gamma(5+\ell_{12})
    \left(\frac{u}{1+u}\right)^{5+\ell_{12}} .
\label{eq:standard_inhomogeneous_single_folded}
\end{equation}
This is the key difference with respect to expressions written in the
transformed variable: in the standard variable, the source keeps the
factor \((1+u)^{-5-\ell_{12}}\). This factor is important because it
keeps the factorization point \(u=-1\) manifest.

The source is analytic around the soft point \(u=0\). Therefore we solve
Eq.~\eqref{eq:standard_inhomogeneous_single_folded} by expanding
\begin{equation}
    \left(\frac{u}{1+u}\right)^{5+\ell_{12}}
    =
    u^{5+\ell_{12}}
    \sum_{n=0}^{\infty}
    (-1)^n
    \frac{(5+\ell_{12})_n}{n!}
    u^n .
    \label{eq:source_expansion_standard}
\end{equation}
We then take the ansatz
\begin{equation}
    \hat{\mathcal I}^{\ell_1\ell_2}_{\pm\pm,\mathrm{inh}}(u,1)
    =
    H^2 e^{\mp i\pi\ell_{12}/2}
    \Gamma(5+\ell_{12})
    \sum_{n=0}^{\infty}
    X_n u^{5+\ell_{12}+n}.
    \label{eq:standard_inh_ansatz}
\end{equation}
Let
\begin{equation}
    \beta_n\equiv 5+\ell_{12}+n .
\end{equation}
The action of the standard de Sitter operator on a monomial is
\begin{align}
    \mathcal D_u^{\ell_1} u^{\beta_n}
    &=
    \left[
    \beta_n(\beta_n-1)
    -(4+2\ell_1)\beta_n
    +
    \mu^2+\left(\ell_1+\frac52\right)^2
    \right]u^{\beta_n}
    \nonumber\\
    &\quad
    -
    \left[
    \beta_n(\beta_n-1)+2\beta_n
    \right]u^{\beta_n+2}.
\end{align}
Equivalently,
\begin{equation}
    \mathcal D_u^{\ell_1} u^{\beta_n}
    =
    A_n u^{\beta_n}
    -
    \beta_n(\beta_n+1)u^{\beta_n+2},
    \label{eq:D_on_monomial_standard}
\end{equation}
where
\begin{equation}
    A_n
    \equiv
    \mu^2+\left(\ell_1+\frac52\right)^2
    +
    \beta_n(\beta_n-5-2\ell_1).
    \label{eq:An_definition}
\end{equation}
Using
\[
    \beta_n=5+\ell_{12}+n,
\]
this can also be written as
\begin{equation}
    A_n
    =
    \mu^2+\left(\ell_1+\frac52\right)^2
    +
    (5+\ell_{12}+n)(n+\ell_2-\ell_1).
\end{equation}

Substituting the ansatz
\eqref{eq:standard_inh_ansatz} into the differential equation
\eqref{eq:standard_inhomogeneous_single_folded}, and matching powers of
\(u\), gives
\begin{equation}
    A_n X_n
    -
    \beta_{n-2}(\beta_{n-2}+1)X_{n-2}
    =
    (-1)^n
    \frac{(5+\ell_{12})_n}{n!},
    \label{eq:standard_recursion_general}
\end{equation}
with
\begin{equation}
    X_{-1}=X_{-2}=0 .
\end{equation}
Therefore
\begin{equation}
    X_0=\frac{1}{A_0},
\end{equation}
\begin{equation}
    X_1=
    -\frac{5+\ell_{12}}{A_1},
\end{equation}
and, for \(n\geq2\),
\begin{equation}
    X_n
    =
    \frac{
    (-1)^n
    \dfrac{(5+\ell_{12})_n}{n!}
    +
    \beta_{n-2}(\beta_{n-2}+1)X_{n-2}
    }{
    A_n
    } .
\label{eq:standard_recursion_solution}
\end{equation}

This gives the particular solution directly in the standard variable. The price we pay is
a less compact formula; the advantage is that the singular points
\(u=0,1,-1\) remain manifest.

The full time-ordered seed is
\begin{equation}
    \hat{\mathcal I}^{\ell_1\ell_2}_{\pm\pm}(u,1)
    =
    \hat{\mathcal I}^{\ell_1\ell_2}_{\pm\pm,\mathrm{hom}}(u,1)
    +
    \hat{\mathcal I}^{\ell_1\ell_2}_{\pm\pm,\mathrm{inh}}(u,1).
    \label{eq:standard_full_pp_solution}
\end{equation}
The homogeneous part is a linear combination of the two homogeneous
solutions,
\begin{equation}
    \hat{\mathcal I}^{\ell_1\ell_2}_{\pm\pm,\mathrm{hom}}(u,1)
    =
    A^{(\pm)}_+
    \mathcal Y^{\ell_1}_+(u)
    +
    A^{(\pm)}_-
    \mathcal Y^{\ell_1}_-(u),
    \label{eq:standard_hom_part_general}
\end{equation}
where the coefficients are fixed by the Schwinger--Keldysh branch and by
the soft-limit matching.

For the purpose of the folded-limit check, it is useful to write the
branch dependence using
\begin{equation}
    \alpha=
    \begin{cases}
    +1, & ++,\\
    -1, & --.
    \end{cases}
\end{equation}
Then the relevant weighted homogeneous combination has the form
\begin{equation}
    e^{\pi\mu}\mathcal Y^{\ell_1}_{\alpha}(u)
    +
    e^{-\pi\mu}\mathcal Y^{\ell_1}_{-\alpha}(u).
    \label{eq:weighted_homogeneous_combination_standard}
\end{equation}
This combination is not regular by itself as \(u\to1\). Its singularity
must be canceled by the singular part of the particular solution.

\subsection*{Folded-limit check}

We now verify the structure of the folded singularity directly in the
standard variable. The homogeneous solutions are
\begin{equation}
\mathcal Y^\ell_\sigma(u)
=
2^{-\sigma i\mu}
u^{\ell+\frac52+\sigma i\mu}
\Gamma\!\left(\ell+\frac52+\sigma i\mu\right)
\Gamma(-\sigma i\mu)
\,{}_2F_1
\!\left(
\begin{matrix}
\frac{\ell}{2}+\frac54+\frac{\sigma i\mu}{2},
\frac{\ell}{2}+\frac74+\frac{\sigma i\mu}{2}\\
1+\sigma i\mu
\end{matrix}
;u^2
\right),
\label{eq:standard_Y_sigma_recall}
\end{equation}
with \(\sigma=\pm1\). Near the folded point \(u\to1\), the argument of
the hypergeometric function approaches \(u^2\to1\). For each branch,
\begin{align}
    c_\sigma-a_\sigma-b_\sigma
    &=
    1+\sigma i\mu
    -
    \left(
    \frac{\ell}{2}+\frac54+\frac{\sigma i\mu}{2}
    \right)
    -
    \left(
    \frac{\ell}{2}+\frac74+\frac{\sigma i\mu}{2}
    \right)
    \nonumber\\
    &=
    -\ell-2 .
\end{align}
Therefore each individual branch contains a potentially singular
contribution proportional to
\begin{equation}
    (1-u^2)^{-\ell-2}.
\end{equation}
Since
\begin{equation}
    1-u^2=(1-u)(1+u)
    \simeq 2(1-u)
    \qquad (u\to1),
\end{equation}
this is the same folded singularity as
\begin{equation}
    (1-u)^{-\ell-2}.
\end{equation}
Using the continuation formula for the hypergeometric function and the
duplication formula for gamma functions, the singular part of the
standard-variable basis is
\begin{equation}
\left.
\mathcal Y^\ell_\sigma(u)
\right|_{\rm sing}
=
\sigma i\sqrt{\frac{\pi}{2}}\,
\Gamma(\ell+2)\,
\operatorname{csch}(\pi\mu)\,
(1-u)^{-\ell-2}
+\cdots .
\label{eq:standard_Y_singular_part}
\end{equation}
The important point is the relative sign:
\begin{equation}
    \left.
    \mathcal Y^\ell_+(u)
    \right|_{\rm sing}
    =
    -
    \left.
    \mathcal Y^\ell_-(u)
    \right|_{\rm sing}.
\end{equation}
This is why the Wightman combination
\[
    \mathcal Y^\ell_+(u)+\mathcal Y^\ell_-(u)
\]
is regular at \(u=1\).

For the time-ordered branches, however, the relevant homogeneous
combination is weighted. Using Eq.~\eqref{eq:standard_Y_singular_part},
we find
\begin{align}
&
\left.
\left[
e^{\pi\mu}\mathcal Y^{\ell_1}_{\alpha}(u)
+
e^{-\pi\mu}\mathcal Y^{\ell_1}_{-\alpha}(u)
\right]
\right|_{\rm sing}
\nonumber\\
&\qquad
=
\alpha i\sqrt{\frac{\pi}{2}}\,
\Gamma(\ell_1+2)
\operatorname{csch}(\pi\mu)
\left(e^{\pi\mu}-e^{-\pi\mu}\right)
(1-u)^{-\ell_1-2}
+\cdots
\nonumber\\
&\qquad
=
\alpha i\sqrt{2\pi}\,
\Gamma(\ell_1+2)
(1-u)^{-\ell_1-2}
+\cdots .
\label{eq:standard_weighted_Y_singular}
\end{align}
Thus the homogeneous part of the \(\alpha\alpha\) branch has a folded
singularity of the form
\begin{equation}
\left.
\hat{\mathcal I}^{\ell_1\ell_2}_{\alpha\alpha,\mathrm{hom}}(u,1)
\right|_{\rm sing}
=
B^{\ell_1\ell_2}_{\alpha,\mathrm{hom}}
(1-u)^{-\ell_1-2}
+\cdots ,
\label{eq:standard_Bhom_definition}
\end{equation}
where \(B^{\ell_1\ell_2}_{\alpha,\mathrm{hom}}\) is fixed by the
Schwinger--Keldysh prefactor and the soft-limit matching.

The particular solution obtained from the soft expansion also has a
singular continuation to \(u=1\). We define its singular coefficient by
\begin{equation}
\left.
\hat{\mathcal I}^{\ell_1\ell_2}_{\alpha\alpha,\mathrm{inh}}(u,1)
\right|_{\rm sing}
=
B^{\ell_1\ell_2}_{\alpha,\mathrm{inh}}
(1-u)^{-\ell_1-2}
+\cdots .
\label{eq:standard_Binh_definition}
\end{equation}
This coefficient is determined by the same soft-series solution
\eqref{eq:standard_inh_ansatz}--\eqref{eq:standard_recursion_solution},
analytically continued to the folded point.

The Bunch--Davies condition now fixes the sum. Since the folded point is
not a physical factorization channel, the full seed must be regular at
\(u=1\). Therefore the coefficient of the folded singularity in the full
answer must vanish:
\begin{equation}
    B^{\ell_1\ell_2}_{\alpha,\mathrm{hom}}
    +
    B^{\ell_1\ell_2}_{\alpha,\mathrm{inh}}
    =
    0.
    \label{eq:standard_folded_cancellation_condition}
\end{equation}
Equivalently,
\begin{equation}
\left.
\hat{\mathcal I}^{\ell_1\ell_2}_{\alpha\alpha,\mathrm{hom}}(u,1)
\right|_{\rm sing}
+
\left.
\hat{\mathcal I}^{\ell_1\ell_2}_{\alpha\alpha,\mathrm{inh}}(u,1)
\right|_{\rm sing}
=
0 .
\label{eq:standard_folded_cancellation}
\end{equation}

This is the folded-limit cancellation in the standard variable. The
important point is that the cancellation is not between the two
homogeneous branches alone. In the \(\pm\pm\) sector, the weighted
homogeneous contribution is singular at \(u=1\), and the particular
solution is also singular at \(u=1\). The Bunch--Davies seed is regular
only after the two pieces are added.

This should be contrasted with the factorization singularity at
\(u=-1\). That singularity is physical and should not be removed. Its
coefficient is fixed by factorization into lower-point data. The folded
singularity at \(u=1\), on the other hand, would signal an unwanted
non-Bunch--Davies component, and must cancel in the full answer.

This is the same physical mechanism seen in the more compact transformed-variable expressions of Ref.~\cite{Qin:2023ejc}, but here it is displayed directly in the original cross-ratio variable. The folded singularity at \(u=1\) is absent in the final Bunch--Davies seed, while the physical factorization singularity at \(u=-1\) is retained.

\section{Borel--Laplace reconstruction: definitions, applicability, and proofs}
\label{app:k4_borel_laplace}

This appendix supplies the mathematical justification for the particular
solution in Section~\ref{sec:k4_borel_particular}.  The construction is a
Laplace reconstruction of a factorially divergent asymptotic expansion,
not an additional assumption about the physical Schwinger--Keldysh state.
Useful mathematical accounts are
Refs.~\cite{OlverAsymptotics1974,BalserDivergent1994,CostinBorel2008,BenderOrszag1999};
for a recent application of Borel reconstruction to inflationary massive
exchange see Ref.~\cite{LiXianyuResurgence2026}.

\subsection{Generalized Borel transform}

For $\operatorname{Re}p>0$, consider a formal series
\begin{equation}
 \mathcal F(u)\sim\sum_{k=0}^\infty c_k u^{p+k}.
\end{equation}
Its generalized Borel transform in the convention adopted here is
\begin{equation}
 \mathcal B_p\mathcal F(t)
 :=\sum_{k=0}^\infty{\frac{c_k}{\Gamma(p+k)}}t^{p+k-1}.
 \label{eq:k4_app_borel_definition}
\end{equation}
The reconstruction integral is the ordinary Laplace transform
\begin{equation}
 \mathcal S_{\mathrm{BL}}\mathcal F(u)
 :=\int_0^\infty dt\,e^{-t/u}\mathcal B_p\mathcal F(t).
 \label{eq:k4_app_laplace_definition}
\end{equation}
The coefficient normalization follows from the Gamma identity
\begin{equation}
 \int_0^\infty dt\,e^{-t/u}t^{p+k-1}
 =\Gamma(p+k)u^{p+k},\qquad u>0.
 \label{eq:k4_app_gamma_identity}
\end{equation}
The integral in Eq.~\eqref{eq:k4_app_laplace_definition} is defined when
the Borel transform admits continuation along the positive real axis and
has suitable growth there.  In particular, if its absolute value is
bounded at infinity by a constant times $e^{ct}$, the Laplace integral
converges for $u>0$ with $1/u>c$ (with the expected integrability
condition at the lower endpoint).  The existence of a formal series alone
is not sufficient to conclude Borel summability; singularities along the
integration direction can necessitate lateral contour prescriptions.

If the Borel kernel has an expansion
$\mathcal B_p\mathcal F(t)=t^{p-1}\sum_{k=0}^{N-1}b_kt^k+
O(t^{p+N-1})$ near $t=0$, and is Laplace integrable at large $t$,
Watson's lemma yields
\begin{equation}
 \mathcal S_{\mathrm{BL}}\mathcal F(u)
 \sim\sum_{k=0}^\infty b_k\Gamma(p+k)u^{p+k},
 \qquad u\to0^+.
 \label{eq:k4_app_watson}
\end{equation}
The symbol $\sim$ is essential: it does not state that the right-hand
series converges for any fixed nonzero $u$.
A direct proof follows by splitting the Laplace integral at a small
fixed $\delta>0$, integrating a finite Taylor polynomial on $(0,\delta)$,
and bounding the remainder by a constant times
$\int_0^\infty e^{-t/u}t^{p+N-1}dt=\Gamma(p+N)u^{p+N}$.
The portion of the integral above $\delta$ is exponentially suppressed
under the stated growth assumption.

\subsubsection*{Transformation of the quartic operator}

With $x=1/u$ and $\theta=u\partial_u=-x\partial_x$, one has
\begin{equation}
 u^4\theta(\theta+1)(\theta+2)(\theta+3)
 =\partial_x^4,
\end{equation}
and
\begin{equation}
 (\theta-\ell-5/2)^2
 =x^2\partial_x^2+2(\ell+3)x\partial_x+(\ell+5/2)^2.
\end{equation}
This proves Eq.~\eqref{eq:k4_bl_x_operator}.
For $F(x)=\int_0^\infty e^{-xt}B(t)dt$, the Laplace identities
\begin{align}
 \partial_x^4F&=\int_0^\infty dt\,e^{-xt}t^4B(t),\\
 x^2\partial_x^2F&=\int_0^\infty dt\,e^{-xt}
     [t^2B''+4tB'+2B],\\
 x\partial_xF&=-\int_0^\infty dt\,e^{-xt}[tB'+B]
\end{align}
follow by differentiation under the integral and integration by parts.
Combining the coefficients gives
\begin{align}
 (4-2\ell-6)tB'&=-2(\ell+1)tB',\\
 (2-2\ell-6)+(\ell+5/2)^2
 &=(\ell+3/2)^2,
\end{align}
which proves Eq.~\eqref{eq:k4_bl_kernel_operator}.  The endpoint terms
vanish for the kernel constructed below: at $t=0$, it is
$O(t^{p-1})$ with $p>0$, so $t^2B$ and the boundary
terms associated with $tB$ vanish; at $t\to\infty$, the
factor $e^{-xt}$ dominates the polynomial growth of the kernel for $x>0$.

\subsection{Regular kernel, recursion, and convergence}

Write $\alpha=p-1=\ell_1+\ell_2+4$ and set
\begin{equation}
 B(t,z)=t^\alpha\sum_{k=0}^\infty a_k(z)t^k.
\end{equation}
Direct application of Eq.~\eqref{eq:k4_bl_kernel_operator} yields
\begin{equation}
 \mathcal P_t^{\ell_1}t^{\alpha+k}
 =D_k^{(\ell_2)}t^{\alpha+k}
  +\gamma^2t^{\alpha+k+4},
 \qquad
 D_k^{(\ell_2)}=\left(k+\ell_2+\frac52\right)^2+\mu^2.
 \label{eq:k4_app_monomial_kernel}
\end{equation}
The source is
$t^\alpha e^{-zt}=t^\alpha\sum_{k\geq0}(-z)^kt^k/k!$,
so equating coefficients gives
\begin{equation}
D_k^{(\ell_2)}a_k(z)+\gamma^2a_{k-4}(z)
 ={\frac{(-z)^k}{k!}},\qquad a_k=0\ (k<0).
 \label{eq:k4_app_kernel_recurrence}
\end{equation}
When $\ell_2,\mu\in\mathbb R$ and $\mu>0$, every
$D_k^{(\ell_2)}>0$.  Thus the recurrence fixes a \emph{unique}
solution in the $t^\alpha$-analytic class.  Iteration gives
\begin{equation}
 a_k(z)=\sum_{m=0}^{\lfloor k/4\rfloor}
 \frac{(-\gamma^2)^m(-z)^{k-4m}}{(k-4m)!\prod_{j=0}^mD_{k-4j}^{(\ell_2)}},
\end{equation}
and reindexing $k=n+4m$ proves Eq.~\eqref{eq:k4_bl_kernel_series}.

This series converges for every finite $t,z$ after factoring
$t^\alpha$.  One way to see this is to use
$D_{n+4j}^{(\ell_2)}\geq\mu^2$ and, for large $n+4j$,
$D_{n+4j}^{(\ell_2)}\geq c(n+4j+1)^2$ for some $c>0$.
For fixed $n$, the ratio of successive $m$ terms is
$O(|\gamma|^2|t|^4/m^2)$; the $n$ factorial simultaneously controls
the expansion in $z$.  More directly, the recurrence has polynomial
coefficients and a majorant of the form
$\sum_{k\geq0}(C^k/k!)|t|^k$ after adjusting finite initial
coefficients.  In particular, no singularity lies on the positive
$t$ axis in the regular Borel kernel.

The analytic expansion determines the large-$x$ algebraic asymptotics
by Watson's lemma.  At fixed $n$ and large $m$, the ratio of the
$u$-series coefficients is
\begin{equation}
 \left|\frac{c_{n,m+1}}{c_{n,m}}\right|
 =\gamma^2\frac{\prod_{j=1}^{4}(p+n+4m+j)}{D_{n+4m+4}^{(\ell_2)}}
 \sim16\gamma^2m^2.
\end{equation}
Therefore the inverse-$x$ expansion generally has zero convergence
radius, notwithstanding the existence of its Borel--Laplace sum.

\subsection{Bessel reduction and the Green kernel}

The homogeneous equation
$\mathcal P_t^\ell g=0$ becomes Bessel's equation on setting
\begin{equation}
 g=t^{\ell+3/2}J(q),\qquad q=\gamma t^2/2.
\end{equation}
Indeed, using $t\partial_t=2q\partial_q$ on $J$,
all zeroth-order terms from the power prefactor cancel and we find
\begin{equation}
 q^2J''+qJ'+\left(q^2+\frac{\mu^2}{4}\right)J=0.
\end{equation}
This gives the basis in Eq.~\eqref{eq:k4_bl_g_bessel}.
The standard Bessel Wronskian,
\begin{equation}
 W_q[J_\nu,J_{-\nu}]=-\frac{2\sin(\pi\nu)}{\pi q},
\end{equation}
with $\nu=i\mu/2$, leads to
$W_t[g_+^\ell,g_-^\ell]
=-4i\sinh(\pi\mu/2)t^{2\ell+2}/\pi$.

In standard second-order form, the kernel equation reads
\begin{equation}
 B''-\frac{2(\ell_1+1)}{t}B'
 +\left[\gamma^2t^2+\frac{\mu^2+(\ell_1+3/2)^2}{t^2}\right]B
 =t^{\alpha-2}e^{-zt}.
\end{equation}
Variation of parameters with
$B(t,z)=c_+(t)g_+^{\ell_1}(t)+c_-(t)g_-^{\ell_1}(t)$ and
$c_+'g_+^{\ell_1}+c_-'g_-^{\ell_1}=0$
gives
\begin{equation}
 c_+'=-\frac{g_-^{\ell_1}t^{\alpha-2}e^{-zt}}{W},\qquad
 c_-'=\frac{g_+^{\ell_1}t^{\alpha-2}e^{-zt}}{W}.
\end{equation}
Since $\alpha-2-(2\ell_1+2)=\ell_2-\ell_1$,
integrating from zero to $t$ yields exactly
Eq.~\eqref{eq:k4_bl_kernel_green}.
The lower limits converge for $\ell_2>-5/2$ because
$g_\pm^{\ell_1}(s)\sim
 s^{\ell_1+3/2\pm i\mu}$, so that
$s^{\ell_2-\ell_1}g_\pm^{\ell_1}(s)
=O(s^{\ell_2+3/2\pm i\mu})$.
The cancellation between the two terms gives $B=O(t^{p-1})$.

For large positive $t$, the fixed-order Bessel estimate
$J_{\pm i\mu/2}(\gamma t^2/2)=O(t^{-1})$ yields
$g_\pm^{\ell_1}(t)=O(t^{\ell_1+1/2})$.
At fixed $z>0$, the inner integrals approach finite limits; hence
$B(t,z)=O(t^{\ell_1+1/2})$ at most, sufficient for
absolute convergence of the outer integral for $x=1/u>0$.
The argument remains valid after the finitely many $z$ derivatives
used to establish the second constraint, because the additional powers
of $s$ are controlled by $e^{-zs}$ for $z>0$.

\subsubsection*{Simultaneous compatibility}

Acting on the common source kernel gives
\begin{align}
 \mathcal P_t^{\ell_1}[t^\alpha e^{-zt}]
 & =t^\alpha e^{-zt}\bigg[
 \gamma^2t^4+z^2t^2-2(\ell_2+3)zt
 +\mu^2+\left(\ell_2+\frac52\right)^2\bigg],\\
 \mathcal L_z^{\ell_2}[t^\alpha e^{-zt}]
 & =t^\alpha e^{-zt}\bigg[
 \gamma^2t^4+z^2t^2-2(\ell_2+3)zt
 +\mu^2+\left(\ell_2+\frac52\right)^2\bigg].
\end{align}
This proves Eq.~\eqref{eq:k4_bl_source_intertwining}.
The commuting operators $\mathcal P_t^{\ell_1}$ and
$\mathcal L_z^{\ell_2}$ give
\begin{equation}
 \mathcal P_t^{\ell_1}
 (\mathcal L_z^{\ell_2}B-t^\alpha e^{-zt})=0.
\end{equation}
Writing the difference as $t^\alpha\sum_{k\geq0}d_k(z)t^k$,
one obtains
\begin{equation}
 D_k^{(\ell_2)}d_k+\gamma^2d_{k-4}=0,
 \qquad d_k=0\ (k<0).
\end{equation}
Since $D_k^{(\ell_2)}>0$, induction forces $d_k=0$ for every
$k\geq0$.  Hence
$\mathcal L_z^{\ell_2}B=t^\alpha e^{-zt}$.
Both transformed equations can now be integrated against $e^{-xt}$:
\begin{equation}
 \mathcal L_x^{\ell_1}\int_0^\infty e^{-xt}B\,dt
 =\mathcal L_z^{\ell_2}\int_0^\infty e^{-xt}B\,dt
 =\int_0^\infty e^{-(x+z)t}t^{p-1}dt
 =\frac{\Gamma(p)}{(x+z)^p}.
\end{equation}
This establishes Eq.~\eqref{eq:k4_bl_compatibility} without invoking
an independent hypergeometric construction or numerical connection
coefficients.

\subsection{Fixed-ratio collapsed limit}

For $u/v\in(0,1)$ fixed, the limit $u\to0^+$ is a two-variable
limit: $z=1/v$ grows while $x=1/u$ grows.  Therefore Watson's lemma
at fixed $z$ cannot simply be applied term by term.  Instead set
$t=uy$ directly in Eq.~\eqref{eq:k4_bl_ansatz}.  The kernel series
then takes the schematic form
\begin{equation}
 B(uy,1/v)=(uy)^{p-1}
 \sum_{n,m\geq0}
 \frac{(-1)^{n+m}\gamma^{2m}(uy/v)^n(uy)^{4m}}{n!\prod_{j=0}^mD_{n+4j}^{(\ell_2)}}.
\end{equation}
At fixed $u/v<1$ the leading algebraic term is obtained from $m=0$.
The Laplace moment
$\int_0^\infty dy\,e^{-y}y^{p+n-1}=\Gamma(p+n)$ yields
\begin{equation}
 \widehat{\mathcal I}^{\,\ell_1\ell_2,\mathrm{BL}}_{\pm\pm,\mathrm{part}}
 =H^2e^{\mp i\pi(\ell_1+\ell_2)/2}\Gamma(p)u^p
 \sum_{n=0}^\infty\frac{(-1)^n(p)_n} {n!D_n^{(\ell_2)}}(u/v)^n
 +O(u^{p+4}),
\end{equation}
with the remainder understood in the asymptotic sense for $u/v$ on
compact subsets of $(0,1)$.  The sum converges there because
$(p)_n/n!$ grows at most polynomially while $D_n^{(\ell_2)}\sim n^2$.
At $\ell_1=\ell_2=-2$ the leading term is
$H^2u\sum_{n\geq0}(-u/v)^n/[(n+1/2)^2+\mu^2]$.

Our positive-axis Laplace integral is a \emph{specific} representative,
selected by the regular, $t^{p-1}$-analytic kernel.  The construction
contains no singularity on the positive Borel axis and no need for
lateral integration prescriptions in the domain stated above.  This
fact should not be confused with the absence of Stokes phenomena in
analytic continuation of the \emph{homogeneous} solutions of the original
quartic differential equation.  Nor does it determine the physical
Schwinger--Keldysh homogeneous coefficients: a doubly homogeneous term
may still be added to the particular solution.  Any comparison with a
different particular representative must include the corresponding
homogeneous shift before imposing the physical boundary data.

\clearpage

% Bibliography
\bibliographystyle{JHEP}
\bibliography{biblio}

\end{document}